\documentclass[acmlarge]{acmart}

\AtBeginDocument{%
  \providecommand\BibTeX{{%
    \normalfont B\kern-0.5em{\scshape i\kern-0.25em b}\kern-0.8em\TeX}}}

\setcopyright{rightsretained}
\acmJournal{IMWUT}
\acmYear{2022} \acmVolume{6} \acmNumber{2} \acmArticle{56} \acmMonth{6} \acmPrice{} \acmDOI{10.1145/3534617}

\newcommand{\m}{\textit{M=}}

\newcommand{\sd}{\textit{SD=}}

\newcommand{\N}{\textit{N=}}

\newcommand{\F}[3]{$F({#1},{#2})={#3}$}

\newcommand{\p}{\textit{p=}}

\newcommand{\pminor}{\textit{p$<$}}

\newcommand\itema{\item[\textbf{RQ~1:}]}
\newcommand\itemb{\item[\textbf{RQ~2:}]}
\newcommand\itemc{\item[\textbf{RQ~3:}]}
\newcommand\itemd{\item[\textbf{RQ~4:}]}

\usepackage{dirtytalk} 
\usepackage[utf8]{inputenc}  
\usepackage{float} 
\usepackage{longtable}
\usepackage{url}
\makeatletter
\g@addto@macro{\UrlBreaks}{\UrlOrds}
\makeatother
\usepackage{multirow} 
\usepackage{color} 
\usepackage{enumitem} 
\usepackage{xspace}
\usepackage{mwe}
\usepackage{xcolor, colortbl} 
\usepackage{soul}
\usepackage{tabularx}
\usepackage{graphicx}

\usepackage{makecell}
\usepackage{hhline}
\usepackage{array}
\usepackage{ragged2e}

\usepackage{soul}
\colorlet{soulyellow}{yellow!40}
\sethlcolor{soulyellow}

\begin{document}

\title[A Design Space for Human Sensor and Actuator Focused In-Vehicle Interaction Based on a Systematic Literature Review]{A Design Space for Human Sensor and Actuator Focused In-Vehicle Interaction Based on a Systematic Literature Review}

\author{Pascal Jansen}
\email{pascal.jansen@uni-ulm.de}
\orcid{0000-0002-9335-5462}
\affiliation{%
  \institution{Institute of Media Informatics, Ulm University}
  \city{Ulm}
  \country{Germany}
}

\author{Mark Colley}
\email{mark.colley@uni-ulm.de}
\orcid{0000-0001-5207-5029}
\affiliation{%
  \institution{Institute of Media Informatics, Ulm University}
  \city{Ulm}
  \country{Germany}
}

\author{Enrico Rukzio}
\email{enrico.rukzio@uni-ulm.de}
\orcid{0000-0002-4213-2226}
\affiliation{%
  \institution{Institute of Media Informatics, Ulm University}
  \city{Ulm}
  \country{Germany}
}
\renewcommand{\shortauthors}{Jansen et al.}

\begin{abstract}
Automotive user interfaces constantly change due to increasing automation, novel features, additional applications, and user demands.
While in-vehicle interaction can utilize numerous promising modalities, no existing overview includes an extensive set of human sensors and actuators and interaction locations throughout the vehicle interior.
We conducted a systematic literature review of 327 publications leading to a design space for in-vehicle interaction that outlines existing and lack of work regarding input and output modalities, locations, and multimodal interaction.
To investigate user acceptance of possible modalities and locations inferred from existing work and gaps unveiled in our design space, we conducted an online study (\N{48}).
The study revealed users' general acceptance of novel modalities (e.g., brain or thermal activity) and interaction with locations other than the front (e.g., seat or table).
Our work helps practitioners evaluate key design decisions, exploit trends, and explore new areas in the domain of in-vehicle interaction.
\end{abstract}



\begin{teaserfigure}
  \includegraphics[width=\textwidth]{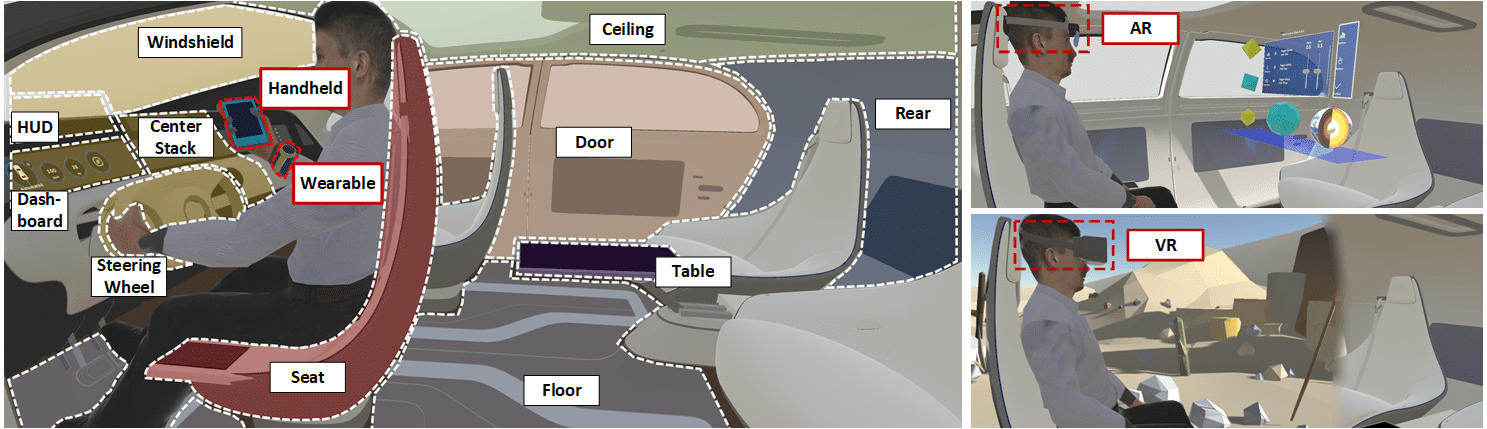}
  \caption{\textit{Anchored} (white outline) and \textit{nomadic} (red outline) interaction locations in a concept vehicle.}
  \label{fig:teaser}
\end{teaserfigure}

\maketitle

\section{Introduction}
With the increasing integration of automation technology into vehicle systems, the scope of in-vehicle interaction is getting broader.
According to the Society of Automotive Engineers (SAE) taxonomy J3016~\cite{sae_levels}, there are six levels of driving automation, ranging from level 0 (no driving automation) to level 5 (full driving automation) in the context of motor vehicles and their operation on roadways.
With automated vehicles (AVs) (SAE levels 3-5) changing the role of the driver, automotive user interfaces (UIs) undergo a paradigm shift~\cite{detjen_how_2021} and the vehicle transforms into a mobile office or living space~\cite{janssen_exploring_2019}.
Accordingly, the driver can perform non-driving related tasks (NDRTs)~\cite{detjen_wizard_2020, WILSON2022150}, such as working, using the smartphone~\cite{davtyan2020effect}, or gaming in virtual reality (VR)~\cite{mcgill2020challenges}.
Hence, the interior design of AVs will adapt, and new input and output locations emerge that are anchored (e.g., door, table, or seat) or nomadic (e.g., handheld or wearable), see~\autoref{fig:teaser}.
For example, future AVs may consist of a 4-seat configuration, where passengers face each other and, like passengers of non-AVs, benefit from UIs located throughout the interior, e.g., rear-seat entertainment~\cite{de2012new,haeling2018car}.
As the interior is a closed space surrounding the passengers, all human senses could be stimulated by output modalities and, to some extent, used as input.
We consider input and output from the human perspective (see~\autoref{fig:sensors_actuators}), i.e., human actuators (e.g., mouth or skin) intentionally generate explicit input modalities (e.g., speech) or subconsciously produce implicit input modalities (e.g., electrodermal activity (EDA)) sensed by vehicle sensors (e.g., microphone), and vehicle actuators (e.g., speaker) generate output modalities (e.g., sound) perceived by human sensors (e.g., ears).

Despite the large body of works, concepts, and prototypes regarding in-vehicle input and output, current automotive UI research does not or only partly consider the full range of input and output modalities (which also contains, e.g., vestibular stimuli, EDA, gustatory stimuli, brain, or heart activity) and novel vehicle interior locations (e.g., rear, floor, or ceiling).
Thus, a new perspective on the in-vehicle interaction space is required, unconstrained from a front-focused design and including such modalities.
Besides, it is partly unknown what modalities and interior locations were already considered in previous works concerning vehicles of any SAE level (0-5).
For manual or assisted driving (SAE~0-2), knowledge about possible input and output modalities and their placement may help in designing interactions with minimal driver distraction~\cite{regan2011driver} and workload levels (physical, visual, and mental)~\cite{burnett2008designing}.
In the context of AVs (SAE~3-5), there exist human factor issues, such as mistrust~\cite{fraedrich2016societal}, loss of control~\cite{frison_why_2019}, or safety concerns~\cite{schoettle2014survey}.
Therefore, it is essential to design in-vehicle interactions that will be accepted~\cite{detjen_how_2021}.
Besides, passengers of AVs can perform NDRTs while interacting with modalities/locations that were previously impractical or dangerous regarding the driving task, e.g., due to sensory overload or reduced takeover readiness.
Still, the usability of such modalities/locations (e.g., swivel seats, or VR) in an AV context is underexplored.

To investigate these problems, we defined the following research questions (RQs):
\begin{itemize} [noitemsep]
    \itema How does current automotive UI research leverage human sensors and actuators for in-vehicle interaction?
    \itemb What vehicle interior locations can be utilized for in-vehicle interaction?
    \itemc What is the design space for in-vehicle interaction, including an extensive set of human sensors and actuators?
    \itemd How do users perceive the usefulness, real-world usage, and comfort of in-vehicle modalities and locations?
\end{itemize}
    
To answer RQs~1-3, we conducted a systematic literature review (SLR).
We gathered a set of keywords for in-vehicle interaction to define the search query for the SLR, which was based on the PRISMA guidelines~\cite{page_prisma_2021,moher2015preferred}.
Our SLR considered vehicles of any SAE level (0-5).
We selected the databases ACM Digital Library (DL)~\cite{ACMDigit45:online}, IEEE Xplore~\cite{IEEEXplo58:online}, and ScienceDirect~\cite{ScienceD38:online}, to search for relevant publications, resulting in an initial set of 2534 publications.
After an abstract screening and a subsequent full-text screening, we considered 327 publications relevant for the synthesis to answer RQs~1-3.

Our SLR shows which approaches for in-vehicle interaction modalities were used by the included publications.
They primarily used visual, auditory, and tactile input and output modalities, while few considered (novel) modalities, such as electrodermal, thermal, olfactory, or cerebral.
Furthermore, our proposed combination matrix for multimodal interaction reveals that most publications utilized visual modalities for multimodal input and output, e.g., in combination with auditory, kinesthetic, or tactile modalities.
Gaps regarding multimodal input containing olfactory modalities and output containing vestibular, electrodermal, or gustatory modalities highlight future multimodal interaction research opportunities.
We then present a design space for in-vehicle interaction that extends~\cite{kern_design_2009} regarding the set of input and output modalities, interior locations, and involved human sensors and actuators.
The design space reveals little to no utilization of thermal, olfactory, gustatory, cerebral, and cardiac input modalities, only a few approaches for vestibular, kinesthetic, and thermal output, and none for electrodermal and gustatory output.
Our design space further shows that publications mainly used the front as output location, e.g., for displays~\cite{broy_evaluating_2015} or vibration~\cite{vo_investigating_2020}, while other locations are not frequently considered, e.g., table, door, rear, floor, or ceiling.
To answer RQ~4, we assessed the feasibility of possible in-vehicle interaction modalities and locations in an online user study (\N{48}) by presenting concept images deduced from related work and gaps in our design space (see~\autoref{fig:input_output_study_concepts} and~\autoref{fig:location_study_concepts}).
The study results reveal that input modalities were more accepted regarding usefulness, usage, and comfort than output modalities.
Besides, well-established input and output modalities in current vehicles, such as auditory or tactile, were generally perceived as more acceptable.
However, novel input and output modalities, e.g., vestibular stimuli, were also perceived as useful.
While participants perceived interaction in some nomadic and anchored interior locations as useful, e.g., handheld, wearable, rear, seat, or table, they deemed other locations less useful, e.g., ceiling, floor, door, or VR.
The results highlight the importance of considering novel modalities and locations in future in-vehicle interaction design.

\textit{Contribution Statement:}
First, we report the results of an SLR on in-vehicle UI research and the analysis of multimodal interaction and utilization of interior locations, leading to a combination matrix for multimodal in-vehicle interaction and visualization of interaction locations accompanied by a self-developed interactive website\footnote{\url{https://in-vehicle-interaction-design-space.onrender.com/} | Interactive tool to support investigation of in-vehicle interaction research and design.}.
Second, we propose a design space for in-vehicle interaction, considering several novel vehicle interior locations and including an extensive set of human sensors and actuators.
Third, we provide the results of an image-based online user study (\N{48}) on perceived usefulness, real-world usage, and comfort of possible in-vehicle interaction approaches deduced from our design space and related work and discuss implications for future interaction design.

\section{Background and Related Work}
Our work is grounded on (1)~a definition of the in-vehicle interaction scheme 
based on a classification of human sensors and actuators and (2)~previous literature reviews on automotive UIs.

\subsection{Human Sensors and Actuators}
\label{sensors and actuators}
\begin{figure*}[ht!]
        \centering
        \includegraphics[width=\textwidth]{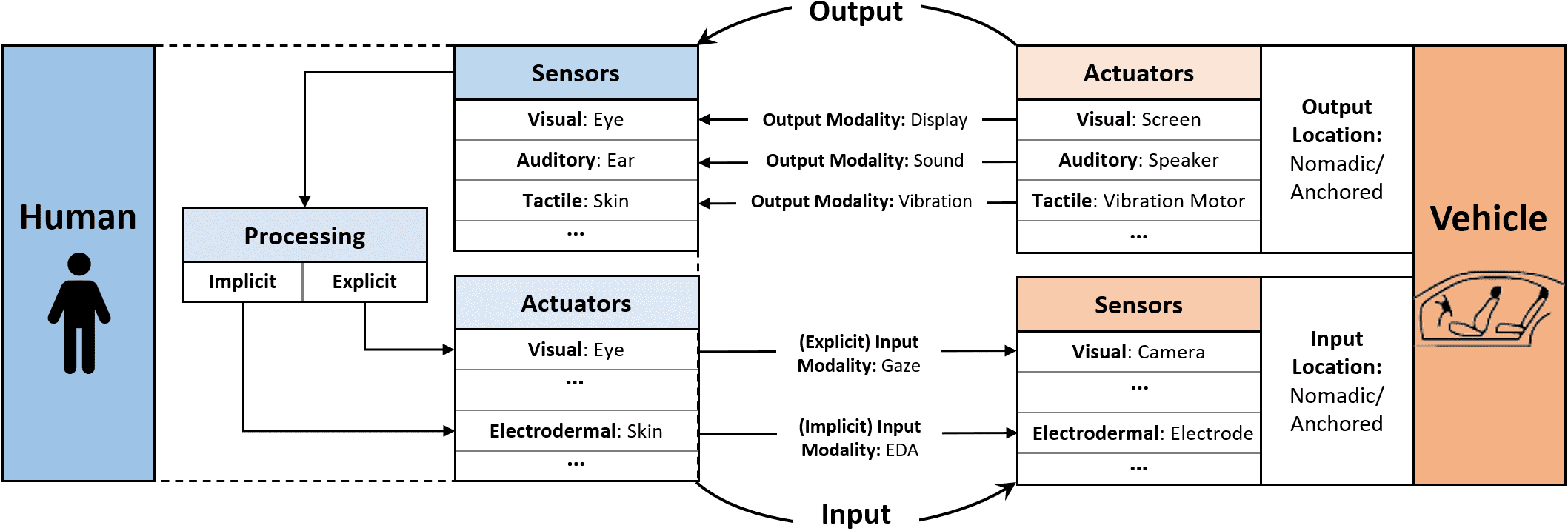}
    \caption{In-vehicle interaction scheme including two agents: a human and a vehicle. A feedback loop is created between both agents, which sense the external world using natural (human) and artificial (vehicle) sensors. Both human and vehicle act upon the environment with their actuators. One possible example is shown after each sensor, actuator, input modality, and output modality.}
    \label{fig:sensors_actuators}
\end{figure*}

Based on~\cite{blattner1996multimodal,benyon2014designing,jaimes2007multimodal,sharma2002toward}, we distinguish seven human sensor/actuator categories: (1)~visual, (2)~auditory, (3)~haptic, (4)~olfactory, (5)~gustatory, (6)~cerebral, and (7)~cardiac.
We follow the definition of sensors and actuators and the respective interaction scheme shown in~\autoref{fig:sensors_actuators}.
The in-vehicle interaction scheme includes two agents: human and vehicle, and describes an input-output feedback loop between both agents~\cite{sharma2002toward}.
A human/the human body intentionally or subconsciously uses actuators such as fingers, brain, or heart to generate input modalities.
In this work, intentionally performed input is considered as \textit{explicit} (e.g., touch, or speech) and subconsciously performed input as \textit{implicit} (e.g., brain or heart activity).
The input modalities are sensed by specific vehicle sensors (e.g., microphones or touchscreens) at vehicle input locations that are either nomadic (e.g., VR, handheld, or wearable) or anchored throughout the interior (e.g., front, seat, door, or rear).
Vehicles use actuator devices, such as screens, speakers, or vibration motors, at nomadic or anchored output locations to generate output modalities (e.g., display, sound, or vibration).
The output modalities are sensed by specific human sensors (e.g., eye, ear, or skin).
In this work, the sensor/actuator categories are used to categorize input and output modalities.
The \textbf{visual} category entails the eye as a sensor enabling passengers to perceive any light-based output modalities and as an actuator that produces explicit input modalities, e.g., gaze, pupil dilation, or blink rate~\cite{sharma2002toward}.
\textbf{Auditory} output modalities are sensed by the ears, while auditory actuators are body parts that can explicitly generate sounds, e.g., mouth for voice or hands/fingers for clapping~\cite{funk_non-verbal_2020}.
According to Benyon~\cite{benyon2014designing} and the ISO standard 9241-910~\cite{ISO_tactile}, we divide \textbf{haptic} into kinesthetic, cutaneous, and vestibular.
We did not consider proprioception, which is the sense of one's body position and movement~\cite{ISO_tactile} as such sensation is already covered by kinesthetic and vestibular sensors/actuators~\cite{jaimes2007multimodal}.
\textbf{Kinesthetic} sensors in the human joints and muscles detect body motion, while kinesthetic actuators generate explicit output modalities, such as muscle activity or body movement.
Such body activity can also be implicitly performed.
\textbf{Cutaneous} is subdivided into electrodermal, tactile, thermal, and pain, which each have specific skin sensors to perceive output modalities such as pressure, temperature, or pain stimuli.
Cutaneous actuators generate skin-related input modalities explicitly via touch and implicitly via EDA or skin temperature.
The \textbf{vestibular} category is adapted from~\cite{benyon2014designing} and describes a sensor that detects balance and general body motion.
However, as the vestibular system is a passive sensor and active body motion is a kinesthetic actuator, there is no vestibular actuator.
Similar to~\cite{benyon2014designing,sharma2002toward}, we include the \textbf{olfactory} and \textbf{gustatory} categories, which each have a dedicated sensor organ (nose and tongue).
However, olfactory actuators are any source of body scent, and gustatory actuators are any source of body flavor, e.g., sweat taste.
Olfactory and gustatory actuators implicitly produce input.
Besides, we include the \textbf{cerebral} and \textbf{cardiac} categories, similar to~\cite{turk2014multimodal,baig2019survey}.
We consider brain activity measured by, e.g., functional near-infrared spectroscopy (fNIRS) or electroencephalography (EEG) as implicit input modality and brain stimuli as output modalities "sensed" by the brain.
Likewise, heart activity (e.g., heart rate) is an implicit input modality, and heart stimuli (e.g., via a defibrillator) are output modalities "sensed" by the heart.

\subsection{Literature Reviews on Automotive User Interfaces}
\label{literature reviews on automotive user interfaces}
There are several reviews on human-vehicle interaction and considerations for the design of automotive UIs.
For example, Lee~\cite{lee2008fifty} analyzed 50 years of driving safety research, including an overview of vehicle technology, and Akamatsu et al.~\cite{akamatsu2013automotive} presented a detailed description of the history of vehicle UIs and related human factors.
Still, these papers give little information on higher automation levels (SAE levels 3, 4, and 5) and future in-vehicle interaction regarding novel modalities.
A review including AVs was conducted by Kun et al.~\cite{kun2016shifting} who identified problem fields for automotive research regarding the transition to higher automation levels.
Similarly, Ayoub et al.~\cite{ayoub_from_2019} identified various trends, e.g., the transition towards AVs or the increasing relevance of NDRTs.
They also summarized a broad range of input and output modalities, including (novel) approaches like augmented reality (AR), VR, or emotion recognition.
An overview of technologies that are being used or developed to perceive user’s intentions for natural and intuitive in-vehicle interaction was presented by Murali et al.~\cite{murali_intelligent_2021}.
They found that novel multimodal sensing devices replace legacy display interfaces and haptic devices such as buttons and knobs.
However, their overview was not based on an SLR, and they did not consider some sensor/actuator categories, e.g., vestibular, olfactory, and gustatory.
Besides, there are reviews regarding human factor-related issues (e.g., distraction, awareness, trust, or acceptance)~\cite{cunningham2015autonomous,jafary2018survey} and technical challenges~\cite{hussain2018autonomous}.
The first design space for driver-based automotive UIs was introduced by Kern and Schmidt~\cite{kern_design_2009} that describes in-vehicle input and output modalities concerning their location in the interior.
However, since driving automation was not yet an omnipresent research topic at the time of publication (2009), their design space focused on a subset of possible in-vehicle modalities (i.e., visual, auditory, and haptic) and locations (i.e., subdivisions of front).
In a later work (2021), Detjen et al.~\cite{detjen_how_2021} discussed the requirements and challenges of interaction with AVs regarding users’ acceptance, namely, security \& privacy, trust \& transparency, safety \& performance, competence \& control, and positive experiences.
They also classified current in-vehicle interaction literature by their contribution to one of the acceptance challenges and used interaction modality.
However, they did not consider novel in-vehicle locations and mainly focused on SAE 3-5 vehicles.

In combination, these works already furnish the direction for future in-vehicle interaction.
However, they are limited due to not including AVs (e.g.,~\cite{lee2008fifty,akamatsu2013automotive}), novel interior locations (e.g.,~\cite{detjen_how_2021}), or considering a subset of possible human sensors and actuators (e.g.,~\cite{kern_design_2009,murali_intelligent_2021}).
In this work, we include in our SLR any SAE level (i.e., from manual to highly automated driving) and approaches for input and output modalities while considering an extensive set of human sensors and actuators.
Besides, we propose a comprehensive design space that extends previous design spaces (such as~\cite{kern_design_2009,detjen_how_2021}) and includes not only the driver but also other passengers as users.

\section{Systematic Literature Review on In-Vehicle Interaction}
\label{systematic literature review on in-vehicle interaction}
\begin{figure*}[ht!]
    \centering
    \includegraphics[width=\textwidth]{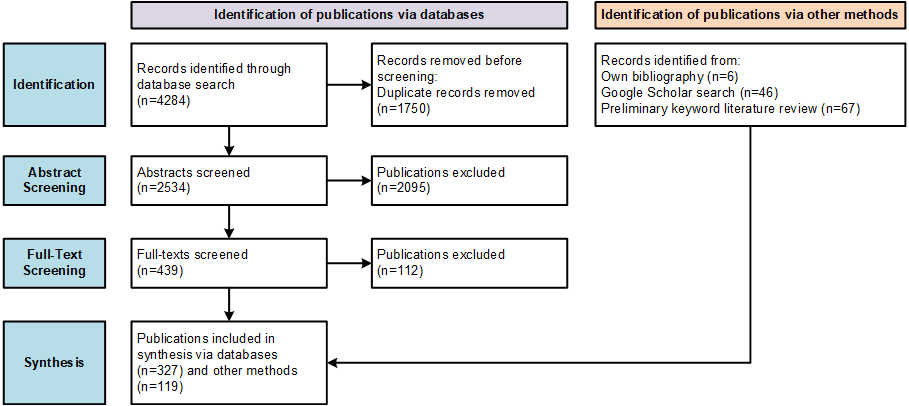}
    \caption{Flow chart of our SLR process based on the PRISMA guidelines proposed by Page et al.~\cite{page_prisma_2021}.
    }
\label{fig:prisma_process}
\end{figure*}

To answer \textit{RQ~1, "How does current automotive UI research leverage human sensors and actuators for in-vehicle interaction?"} and \textit{RQ~2, "What vehicle interior locations can be utilized for in-vehicle interaction?"}, our goal was to elaborate a detailed and comprehensive overview of research on in-vehicle interaction.
Therefore, we employed an SLR.
The process of this SLR is based on the Preferred Reporting Items for Systematic Reviews and Meta-Analysis (PRISMA) proposed by Moher et al.~\cite{moher2009preferred} and Page et al.~\cite{page_prisma_2021}. 
Our multistaged process is depicted in \autoref{fig:prisma_process} and consists of an identification step, a two-step publication screening part, and a synthesis. 

\subsection{Identification}
\label{identification}
To identify publications relevant to our topic, we created a database query using Boolean algebra that combines the keywords of each sensor/actuator category specified in a preparatory step (see {\autoref{tab:keyword_list}} in Appendix).
However, as the used database search engines have keyword limits, we split the query into 12 sub-queries, one for each sensor/actuator category.
For readability, the complete list of used queries is shown in the Appendix (see \autoref{tab:search_queries}).
The query for the auditory category is shown as example:


\vspace{1pt}
\texttt{(vehicle* OR car* OR driver* OR driving OR "in-vehicle") AND (interaction* OR interface*) AND (auditory OR audio OR ear* OR speech OR voice OR verbal OR vocal OR "non-speech" OR sound* OR whistling OR humming OR acoustic OR earcon* OR conversational OR music)}
\vspace{1pt}

We adapted each query to suit the respective search syntax required by the databases (see Supplementary material).
Although the queries were applied independently, the screening process is uniform for all results to prevent differing analyses.
Thus, for example, a publication found in the searches for visual and kinesthetic modalities is coded for both.
If queries yielded identical results, the duplicates were omitted.

We selected the databases ACM DL, IEEE Xplore, and ScienceDirect, covering venues relevant for automotive UI and future mobility research, e.g., Conference on Human Factors in Computing Systems (CHI), Conference on Automotive User Interfaces (AutoUI), or Transportation Research Part F.
Within these databases, we selected a subset of venues (see~\autoref{tab:selected_venues}) based on the 16 most cited HCI venues according to Google Scholar~\cite{GoogleScholarHCI:online} and an additional ten venues relevant for automotive research.
Besides, we defined three search criteria that were applied directly in the database search: A publication had to be (1)~published in the past ten years (2011-2021), (2)~written in English, and (3)~peer-reviewed.
Only publication titles and abstracts were searched, similar to~\cite{pai2004systematic,bargas_old_2011,zawacki2019systematic}, as those parts should contain an overview of all relevant content.

The three databases were queried, and the search results were downloaded on 07/15/2021.
The detailed result counts for each sensor/actuator category are shown in \autoref{tab:search_queries}.
After removing duplicates with Zotero~\cite{zotero:online}, 2534 publications remained (1244 ACM DL (49.1~\%), 1046 IEEE Xplore (41.3~\%), 244 ScienceDirect (9.6~\%)).

\subsection{Screening}
For the publications screening, we employed a two-phase process, consisting of an abstract screening and a subsequent full-text screening.

\subsubsection{Abstract Screening}
\label{abstract screening}
We defined a mandatory inclusion criterion: \textit{in-vehicle interaction}.
The criterion was true when the publication described or employed an interaction or interface between a user and an in-vehicle system in the automotive context.
This includes the use of any input and output modality.
Besides, we defined inclusion criteria that were not explicitly coded to save time but always considered in the decision:
(1)~The publication should be no literature review, taxonomy, or design space.
(2)~It considers a vehicle of any SAE~J3016~\cite{sae_levels} automation level. 
(3)~It describes driver/passenger-vehicle interaction, i.e., no direct passenger-passenger interaction.
(4)~It considers user interaction inside a vehicle, i.e., no external communication (e.g.,~\cite{colley_external_2021}), vehicle-pedestrian interaction, or teleoperation.

Sysrev~\cite{sysrev:online} was used for the screening phase, as it is an online platform that offers tools for collaborative screening of large amounts of publications.
Besides, it provides automatic randomization of the publications to prevent a biased review order.
The Sysrev project is public\footnote{\url{https://sysrev.com/u/5286/p/77887} | This project contains our abstract screening. It provides public access to all 2534 screened publications' metadata, abstracts, and shows our labels. Labeling statistics can be shown, and filters can be used to select publications.} to allow researchers to view our work and enable the use in other research projects.
Two of the authors conducted the abstract screening.
To create a shared understanding of the inclusion criteria, the first 25 publication abstracts were collaboratively reviewed.
Both reviewers then reviewed a subset of 1500 publication abstracts without verbal discussion, and the remaining 1034 publications were rated by one reviewer.
To assess inter-rater agreement, we calculated Cohen's Kappa~\cite{cohen1960coefficient} $\kappa=.83$ ($p0=94.00\%$), which indicates an almost perfect agreement~\cite{landis_the_1977}. 
Ninety-four conflicts were discussed and resolved.

In total, 439 publications were included. 
Based on the results, we defined a code book for the full-text screening, representing our topics of interest for the synthesis.
The code book included the vehicle's SAE level considered in the concept/user study, the sensor/actuator category of the employed input/output modality, multimodal input/output usage, and the interaction location.
We recognize an interaction as multimodal when the user is provided with multiple modes for interacting with the system (e.g., speech and touch)~\cite{turk2014multimodal}.
This includes both sequential, where a user will have to switch between modes of interaction and simultaneous multimodal interaction that allows multiple modes at a time~\cite{raisamo1999multimodal}.
For the interaction locations, we defined two overarching types: (1)~nomadic (e.g., head-mounted display (HMD), smartphone, or wearable) and (2)~anchored (e.g., dashboard, seat, or door).
The abstract screening data set containing all included publications' metadata is provided as supplementary material, enabling researchers to access our raw data for future projects.

\subsubsection{Full-Text Screening}
\label{full-text screening}
We excluded publications in this screening step if at least one of the following exclusion criteria are met:
(1)~The publication considers driver monitoring intended only to measure participants' performance, e.g., cognitive load, drowsiness, or distraction.
(2)~It considers a haptic torque steering support, e.g., for takeover situations.
(3)~It does not meet the inclusion criteria defined in the earlier process steps (see Section~\ref{identification}~and~\ref{abstract screening}).

The two reviewers continued with this full-text screening, which was again conducted in Sysrev and made public\footnote{\url{https://sysrev.com/u/5286/p/78991} | This project contains our full-text screening. It provides public access to all 439 screened publications' labels.}.
In the beginning, 27 publications were rated together to generate a common understanding and test the applicability of the defined code book.
Six of these publications resulted in a conflict and were resolved in a discussion.
The inter-rater agreement was qualitatively checked in a discussion similar to~\cite{tyack_self_2020,robinson_let_2020}.
The reviewers rated the remaining 412 publications individually.

\subsection{Results Overview}
\label{slr_results}
In total, 327 out of 439 publications were marked for inclusion in the synthesis (200 ACM DL (61.1~\%), 68 IEEE Xplore (20.8~\%), 59 ScienceDirect (18.1~\%)).
Within the included publications, we found 306 approaches for input modalities, with the following distribution in the sensor/actuator categories: 35~visual (11.4~\%), 50~auditory (16.3~\%), 101~kinesthetic (33.0~\%), 7~electrodermal (2.3~\%), 79~tactile (25.8~\%), 2~thermal (.7~\%), 10~cerebral (3.3~\%), and 8~cardiac (2.6~\%).
While 171 publications used unimodal input, 38 publications considered multimodal input.
Regarding output modalities, we found 416 approaches: 221~visual (53.1~\%), 110~auditory (26.4~\%), 7~kinesthetic (1.7~\%), 52~tactile (12.5~\%), 4~thermal (.96~\%), and 5~olfactory (1.2~\%).
Multimodal output was employed by 81 publications, while 207 used unimodal output.
Besides, we found 14 publications that utilized a nomadic device (e.g., smartphone or wearable device) for input and 17 publications for output.

\section{Extending the Design Space for In-Vehicle Interaction}
\label{extending_the_design_space}

Although visual, auditory, and haptic modalities dominate current in-vehicle interaction research (see Section~\ref{slr_results}), our SLR reveals various approaches that utilized novel modalities, such as vestibular, thermal, olfactory, cerebral, or cardiac.
This highlights that future automotive UI design should consider such (novel) modalities.
Especially, regarding increasing automation, shifting the driver's role to a passenger requiring adaptive intention recognition that considers different driving situations~\cite{ohn2016looking} and making modalities using physiological, brain, gaze, or emotion interfaces more useful~\cite{detjen_how_2021}.
The results of our SLR further indicate the relevance of answering \textit{RQ~3, "What is the design space for in-vehicle interaction that includes an extensive set of human sensors and actuators?"}.
Therefore, we propose a design space for in-vehicle interaction extending the design space by Kern and Schmidt~\cite{kern_design_2009}.

Two dimensions span our design space: (\textbf{D1}) interaction location and (\textbf{D2}) interaction modality.
\textbf{D1} is defined by two parameters that were selected according to the results of the interaction location coding (see Section~\ref{interaction locations}): \textit{nomadic} and \textit{anchored}.
\textbf{D2} is defined by two parameters: \textit{input modality} and \textit{output modality}.
The parameter nomadic of \textbf{D1} has four levels: AR, VR, handheld, and wearable.
The parameter anchored of \textbf{D1} has 12 values: front, windshield, dashboard, head-up display, center stack, steering wheel, seat, table, door, rear, floor, and ceiling.
The front location is adopted from Kern and Schmidt~\cite{kern_design_2009}.
We decided not to differentiate the interaction locations any further to limit complexity.
However, the set of values can be expanded in the future, e.g., \textit{seat} may be differentiated into seat pan, backrest, and headrest.
Both parameters of \textbf{D2} (input and output) have the same 12 values based on the sensor/actuator categories (see Section~\ref{sensors and actuators}): visual, auditory, vestibular, kinesthetic, electrodermal, tactile, thermal, pain, olfactory, gustatory, cerebral, and cardiac.

We followed a morphological analysis and combined the aforementioned values into a multidimensional matrix, also known as Zwicky box~\cite{zwicky1967morphological}, which is a well-established tool for the creation of design spaces~\cite{ballagas_design_2018,hirzle_design_2019}.
Such a matrix contains all combinations of parameters and helps identify promising families of solutions and a possible lack of solutions by the number of solutions in specific cells.
The resulting matrix containing all approaches for input modalities found in our SLR is shown in~\autoref{tab:design_space_input} and for output modalities in~\autoref{tab:design_space_output}, where \textbf{D1} is positioned on the y-axis and \textbf{D2} on the x-axis.
Grayed-out cells indicate that a solution using the respective modality and location does not make sense due to ethical concerns (e.g., pain stimuli) or technical reasons (e.g., no vestibular actuator or impractical cerebral/cardiac output).
Our proposed design space can be used: (1)~as a tool to classify an existing or one's own work, (2)~as a taxonomy of in-vehicle interaction research to inform a literature search, and (3)~as an ideation tool for the creation of novel interaction concepts utilizing interaction locations throughout the vehicle.

In the following, we describe the found input and output modalities mapped to the sensor/actuator categories in detail (RQ~1).
As the interaction was often multimodal, we also present a combination matrix for multimodal input and output (RQ~1).
We conclude by classifying the found modalities into interaction locations across the vehicle interior and nomadic devices (RQ~2).

\begin{table*}[ht]
\centering
\caption{The matrix obtained by the morphological approach, showing the design space for in-vehicle \textbf{input} modalities. The combinations highlighted in gray make technically no sense or are unethical.}
\resizebox{\textwidth}{!}{%
\begin{tabular}{ccl|l|l|l|l|l|l|l|l|l|l|l|l|}
\cline{4-15}
\multicolumn{1}{l}{} & \multicolumn{1}{l}{} &  & \multicolumn{12}{c|}{\cellcolor[HTML]{FCE5CD}\textbf{D2: (Human Actuated) Input Modality}} \\ \cline{4-15} 
\multicolumn{1}{l}{} & \multicolumn{1}{l}{} &  & \multicolumn{1}{c|}{} & \multicolumn{1}{c|}{} & \multicolumn{6}{c|}{\textbf{Haptic}} & \multicolumn{1}{c|}{} & \multicolumn{1}{c|}{} & \multicolumn{1}{c|}{} & \multicolumn{1}{c|}{} \\ \cline{6-11}
\multicolumn{1}{l}{} & \multicolumn{1}{l}{} &  & \multicolumn{1}{c|}{} & \multicolumn{1}{c|}{} & \multicolumn{1}{c|}{} & \multicolumn{1}{c|}{} & \multicolumn{4}{c|}{\textbf{Cutaneous}} & \multicolumn{1}{c|}{} & \multicolumn{1}{c|}{} & \multicolumn{1}{c|}{} & \multicolumn{1}{c|}{} \\ \cline{8-11}
\multicolumn{1}{l}{} & \multicolumn{1}{l}{} &  & \multicolumn{1}{c|}{\multirow{-3}{*}{\textbf{Visual}}} & \multicolumn{1}{c|}{\multirow{-3}{*}{\textbf{Auditory}}} & \multicolumn{1}{c|}{\multirow{-2}{*}{\textbf{Vestibular}}} & \multicolumn{1}{c|}{\multirow{-2}{*}{\textbf{Kinesthetic}}} & \multicolumn{1}{c|}{\textbf{Electrodermal}} & \multicolumn{1}{c|}{\textbf{Tactile}} & \multicolumn{1}{c|}{\textbf{Thermal}} & \multicolumn{1}{c|}{\textbf{Pain}} & \multicolumn{1}{c|}{\multirow{-3}{*}{\textbf{Olfactory}}} & \multicolumn{1}{c|}{\multirow{-3}{*}{\textbf{Gustatory}}} & \multicolumn{1}{c|}{\multirow{-3}{*}{\textbf{Cerebral}}} & \multicolumn{1}{c|}{\multirow{-3}{*}{\textbf{Cardiac}}} \\ \hline
\multicolumn{1}{|c|}{\cellcolor[HTML]{BDD7EE}} & \multicolumn{1}{c|}{\cellcolor[HTML]{C9DAF8}} & \cellcolor[HTML]{C9DAF8}\textbf{AR} &  &  & \cellcolor[HTML]{D9D9D9} &  &  &  &  & \cellcolor[HTML]{D9D9D9} &  &  &  &  \\ \hhline{|*{1}{>{\arrayrulecolor[HTML]{BDD7EE}}-}|>{\arrayrulecolor[HTML]{C9DAF8}}-|*{13}{>{\arrayrulecolor{black}}-}|} 
\multicolumn{1}{|c|}{\cellcolor[HTML]{BDD7EE}} & \multicolumn{1}{c|}{\cellcolor[HTML]{C9DAF8}} & \cellcolor[HTML]{C9DAF8}\textbf{VR} &  &  & \cellcolor[HTML]{D9D9D9} &  &  &  &  & \cellcolor[HTML]{D9D9D9} &  &  &  &  \\ \hhline{|*{1}{>{\arrayrulecolor[HTML]{BDD7EE}}-}|>{\arrayrulecolor[HTML]{C9DAF8}}-|*{13}{>{\arrayrulecolor{black}}-}|} 
\multicolumn{1}{|c|}{\cellcolor[HTML]{BDD7EE}} & \multicolumn{1}{c|}{\cellcolor[HTML]{C9DAF8}} & \cellcolor[HTML]{C9DAF8}\textbf{Handheld} &  &  & \cellcolor[HTML]{D9D9D9} & \begin{tabular}[c]{@{}l@{}}\textbf{Smartphone Button} \cite{li_evaluation_2018},\\ \textbf{Controller} \cite{vetek_could_2011}\end{tabular} &  & \begin{tabular}[c]{@{}l@{}}\textbf{Smartphone}\\ \textbf{Touch} \cite{munger_simulation_2014,he_texting_2013,williams_affective_2014}\\\cite{perlman_relative_2019,heikkinen_mobile_2013,brumby_empirical_2012,chang_using_2017}\\ \cite{he_does_2018,osswald_hmi_2013,wintersberger_let_2018},\\ \textbf{Tablet Touch}\\ \cite{meschtscherjakov_active_2016,chang_using_2017,capallera_secondary_2019}\end{tabular} &  & \cellcolor[HTML]{D9D9D9} &  &  &  &  \\ \hhline{|*{1}{>{\arrayrulecolor[HTML]{BDD7EE}}-}|>{\arrayrulecolor[HTML]{C9DAF8}}-|*{13}{>{\arrayrulecolor{black}}-}|} 
\multicolumn{1}{|c|}{\cellcolor[HTML]{BDD7EE}} & \multicolumn{1}{c|}{\multirow{-8}{*}{\cellcolor[HTML]{C9DAF8}\rotatebox[origin=c]{90}{\textbf{Nomadic}}}} & \cellcolor[HTML]{C9DAF8}\textbf{Wearable} & \begin{tabular}[c]{@{}l@{}}\textit{Eye-Tracker:}\\ \textbf{Gaze} \cite{benedetto_driver_2011,kim_why_2019,wang_watch_2020},\\ \textbf{Glance} \cite{yao_assessment_2020},\\ \textbf{Blink} \cite{benedetto_driver_2011}, \textbf{Pupil} \cite{benedetto_driver_2011},\\ \textbf{Movement} \cite{pakdamanian_deeptake_2021,dillen_keep_2020,kim_visual_2019}\end{tabular} &  & \cellcolor[HTML]{D9D9D9} & \begin{tabular}[c]{@{}l@{}}\textbf{Smartwatch Acceleration} \cite{kim_why_2019},\\ \textbf{Smart Ring Gesture} \cite{gheran_controls_2020}\end{tabular} & \begin{tabular}[c]{@{}l@{}}\textit{Forearm:}\\ \textbf{EMG} \cite{large_lessons_2019},\\ \textit{Finger:}\\ \textbf{EDA} \cite{wang_drivers_2020},\\ \textit{Finger:}\\ \textbf{GSR} \cite{tobisch_dealing_2020,pakdamanian_deeptake_2021,dillen_keep_2020}\end{tabular} & \begin{tabular}[c]{@{}l@{}}\textbf{Smartwatch}\\ \textbf{Touch} \cite{perlman_relative_2019,williams_towards_2013}\end{tabular} & \begin{tabular}[c]{@{}l@{}}\textit{Finger:} \textbf{Skin}\\ \textbf{Temp.} \cite{wang_drivers_2020}\end{tabular} & \cellcolor[HTML]{D9D9D9} &  &  & \begin{tabular}[c]{@{}l@{}}\textit{Head:} \textbf{EEG}\\ \cite{lu_human_2021,he_brain_2016,hood_use_2012}\\ \cite{bi_head-up_2013,astudillo_cost-efficient_2017,bi_using_2014}\\ \cite{yu_general_2019,martelaro_exploration_2019}\\ \cite{zhu_heavy_2020,chang_using_2017}\end{tabular} & \begin{tabular}[c]{@{}l@{}}\textit{Ear:} \textbf{Heart Rate}\\ \cite{pakdamanian_deeptake_2021,dillen_keep_2020,hernandez_autoemotive_2014},\\ \textit{Forearm:} \textbf{Heart}\\ \textbf{Rate} \cite{large_assessing_2016},\\ \textit{Torso:} \textbf{ECG}\\ \cite{wang_drivers_2020,rony_development_2020}\end{tabular} \\ \hhline{|*{1}{>{\arrayrulecolor[HTML]{BDD7EE}}-}|*{14}{>{\arrayrulecolor{black}}-}|} 
\multicolumn{1}{|c|}{\cellcolor[HTML]{BDD7EE}} & \multicolumn{1}{c|}{\cellcolor[HTML]{D0E0E3}} & \cellcolor[HTML]{D0E0E3}\textbf{Front} & \begin{tabular}[c]{@{}l@{}}\textbf{Gaze} \cite{george_daaria_2012}, \textbf{Glance} \cite{mahajan_exploring_2021},\\ \textbf{Blink} \cite{mahajan_exploring_2021}, \textbf{Pupil} \cite{mahajan_exploring_2021},\\ \textbf{Movement} \cite{wu_eye_2021}\end{tabular} & \begin{tabular}[c]{@{}l@{}}\textbf{Speech} \cite{kim_predicting_2019,nawa_information_2012,lin_adasa_2018}\\ \cite{martelaro_exploration_2019,hackenberg_international_2013,schneeberger_tailoring_2015}\\ \cite{stier_adapting_2020,hofmann_comparison_2014,kim_predicting_2019,large_steering_2017},\\ \textbf{Voice} \cite{kim_visual_2019,truschin_designing_2014,tobisch_dealing_2020}\\ \cite{winzer_intersection_2018,lee_partially_2014,wang_design_2014}\\ \cite{detjen_driving_2021,goulati_user_2011,lin_adasa_2018,williams_affective_2014},\\ \textbf{Humming} \cite{funk_non-verbal_2020},\\ \textbf{Clapping} \cite{funk_non-verbal_2020},\\ \textbf{Snapping} \cite{funk_non-verbal_2020},\\ \textbf{Vocal Affect} \cite{tischler_application_nodate}\end{tabular} & \cellcolor[HTML]{D9D9D9} & \begin{tabular}[c]{@{}l@{}}\textit{Camera-sensed:}\\ \textbf{Body Position} \cite{weyers_driver_2018},\\ \textbf{Body Pose} \cite{rangesh_exploring_2018,heckmann_cora_2019,hernandez_autoemotive_2014},\\ \textbf{Facial Landmarks} \cite{rangesh_exploring_2018},\\ \textbf{Head Pose} \cite{schneeberger_tailoring_2015,truschin_designing_2014,aftab_multimodal_2019,rumelin_free-hand_2013,karatas_namida_2016,reimer_effects_2014},\\ \textbf{Head Movement}\\ \cite{broy_3d_2014,broy_exploring_2014,lauber_content_2015,chang_using_2017,suga_-vehicle_2016,kim_visual_2019,hernandez_autoemotive_2014},\\ \textbf{Head Position} \cite{heckmann_cora_2019},\\ \textbf{Facial Expression} \cite{hernandez_autoemotive_2014,huang_face2multi-modal_2020},\\ \textbf{Emotion} \cite{stier_adapting_2020,sundstrom_gaming_2014,braun_what_2020,zepf_towards_2019,li_designing_2019},\\ \textbf{Gesture} \cite{riener_standardization_2013,detjen_driving_2021,shakeri_bimodal_2017},\\ \textbf{Finger Gesture} \cite{john_real-time_2017}\end{tabular} &  &  &  & \cellcolor[HTML]{D9D9D9} &  &  &  &  \\ \hhline{|*{1}{>{\arrayrulecolor[HTML]{BDD7EE}}-}|>{\arrayrulecolor[HTML]{D0E0E3}}-|*{13}{>{\arrayrulecolor{black}}-}|} 
\multicolumn{1}{|c|}{\cellcolor[HTML]{BDD7EE}} & \multicolumn{1}{c|}{\cellcolor[HTML]{D0E0E3}} & \cellcolor[HTML]{D0E0E3}\textbf{Windshield} & \textbf{Gaze} \cite{riegler_gaze-based_2020,aftab_multimodal_2019,rumelin_free-hand_2013} &  & \cellcolor[HTML]{D9D9D9} & \begin{tabular}[c]{@{}l@{}}\textit{Camera-sensed:}\\ \textbf{Finger Pointing} \cite{rumelin_free-hand_2013,aftab_multimodal_2019}\end{tabular} &  &  &  & \cellcolor[HTML]{D9D9D9} &  &  &  &  \\ \hhline{|*{1}{>{\arrayrulecolor[HTML]{BDD7EE}}-}|>{\arrayrulecolor[HTML]{D0E0E3}}-|*{13}{>{\arrayrulecolor{black}}-}|} 
\multicolumn{1}{|c|}{\cellcolor[HTML]{BDD7EE}} & \multicolumn{1}{c|}{\cellcolor[HTML]{D0E0E3}} & \cellcolor[HTML]{D0E0E3}\textbf{Dashboard} & \textbf{Gaze} \cite{clark_directability_2019,kim_cascaded_2020}, \textbf{Glance} \cite{fridman_what_2017} &  & \cellcolor[HTML]{D9D9D9} &  &  & \textbf{Touch} \cite{voinescu_utility_2020} &  & \cellcolor[HTML]{D9D9D9} &  &  &  &  \\ \hhline{|*{1}{>{\arrayrulecolor[HTML]{BDD7EE}}-}|>{\arrayrulecolor[HTML]{D0E0E3}}-|*{13}{>{\arrayrulecolor{black}}-}|} 
\multicolumn{1}{|c|}{\cellcolor[HTML]{BDD7EE}} & \multicolumn{1}{c|}{\cellcolor[HTML]{D0E0E3}} & \cellcolor[HTML]{D0E0E3}\textbf{HUD} &  &  & \cellcolor[HTML]{D9D9D9} &  &  &  &  & \cellcolor[HTML]{D9D9D9} &  &  &  &  \\ \hhline{|*{1}{>{\arrayrulecolor[HTML]{BDD7EE}}-}|>{\arrayrulecolor[HTML]{D0E0E3}}-|*{13}{>{\arrayrulecolor{black}}-}|} 
\multicolumn{1}{|c|}{\cellcolor[HTML]{BDD7EE}} & \multicolumn{1}{c|}{\cellcolor[HTML]{D0E0E3}} & \cellcolor[HTML]{D0E0E3}\textbf{\begin{tabular}[c]{@{}l@{}}Center\\ Stack\end{tabular}} & \textbf{Gaze} \cite{roider_just_2018}, \textbf{Glance} \cite{ulahannan_using_2021} &  & \cellcolor[HTML]{D9D9D9} & \begin{tabular}[c]{@{}l@{}}\textbf{Button} \cite{reimer_effects_2014,kidd_considering_2017,forster_your_2016,manawadu_multimodal_2017,walch_click_2018},\\ \textbf{Rotary Controller} \cite{alvarez_help_2015,lauber_youve_2014,grane_driving_2013}\\\cite{kidd_considering_2017,mitsopoulos-rubens_effects_2011,jeon_effects_2015,doring_gestural_2011,schartmuller_type-o-steer_2019,chang_dont_2016,large_evaluating_2019},\\ \textbf{Keypad} \cite{vieira_kansei_2017,tardieu_sonification_2015,hofmann_comparison_2014},\\ \textbf{Pointing Device} \cite{ros_scribble_2018},\\ \textbf{Lever} \cite{manawadu_multimodal_2017}, \textbf{Sidestick} \cite{panzirsch_3dof-sidestick_2015},\\ \textbf{Shape Changing} \cite{winzer_intersection_2018},\\ \textit{Camera-sensed:}\\ \textbf{Gesture} \cite{choi_designing_2018,winzer_intersection_2018,quintal_hapwheel_2021,gupta_towards_2016,ohn-bar_hand_2012,may_designing_2017},\\ \textbf{Finger Gesture} \cite{deo_-vehicle_2016,shakeri_novel_2017,gupta_towards_2016,may_designing_2017},\\ \textbf{Finger Pointing} \cite{lauber_what_2014,lauber_youve_2014,ahmad_selection_2018,ahmad_interactive_2014},\\ \textit{Ultrasound-sensed:} \\ \textbf{Finger Pointing} \cite{riener_natural_2011},\\ \textbf{Gesture} \cite{georgiou_haptic_2017}\end{tabular} & \begin{tabular}[c]{@{}l@{}}\textbf{Body}\\ \textbf{Conductivity} \cite{chang_dont_2016}\end{tabular} & \begin{tabular}[c]{@{}l@{}}\textbf{Touch} \cite{lee_partially_2014,detjen_driving_2021,goulati_user_2011}\\\cite{jung_voicetactile_2020,lin_adasa_2018,jeon_effects_2015,kopinski_touch_2016}\\ \cite{ng_evaluation_2017,heikkinen_mobile_2013,pichen_stuck_2019,manawadu_multimodal_2017},\\ \textbf{Pressure} \cite{ng_evaluation_2017,huber_towards_2016},\\ \textbf{Pen} \cite{ishiguro_leadingdisplay_2019},\\ \textbf{Indirect Touch} \cite{berger_tactile_2019},\\ \textbf{Shape Changing} \cite{winzer_intersection_2018}\end{tabular} &  & \cellcolor[HTML]{D9D9D9} &  &  &  &  \\ \hhline{|*{1}{>{\arrayrulecolor[HTML]{BDD7EE}}-}|>{\arrayrulecolor[HTML]{D0E0E3}}-|*{13}{>{\arrayrulecolor{black}}-}|} 
\multicolumn{1}{|c|}{\cellcolor[HTML]{BDD7EE}} & \multicolumn{1}{c|}{\cellcolor[HTML]{D0E0E3}} & \cellcolor[HTML]{D0E0E3}\textbf{\begin{tabular}[c]{@{}l@{}}\textbf{Steering}\\ \textbf{Wheel}\end{tabular}} &  &  & \cellcolor[HTML]{D9D9D9} & \begin{tabular}[c]{@{}l@{}}\textbf{Button} \cite{reimer_effects_2014,li_evaluation_2018,nawa_information_2012,wang_design_2014,lauber_youve_2014,forster_your_2016}\\ \cite{jeon_effects_2015,murer_exploring_2012,schartmuller_type-o-steer_2019},\\ \textbf{Lever} \cite{hornberger_evaluation_2018}, \textbf{Switch} \cite{vo_investigating_2020},\\ \textbf{Steering Impedance} \cite{bhardwaj_whos_2020},\\ \textbf{On-Wheel Gesture} \cite{lee_-wheel_2015},\\ \textbf{Keyboard} \cite{hofmann_comparison_2014,schartmuller_type-o-steer_2019}, \textbf{Mouse} \cite{jakus_user_2015},\\ \textit{Camera-sensed:}\\ \textbf{Hand Position} \cite{weyers_driver_2018},\\ \textit{Electric-field-sensed:}\\ \textbf{Finger Gesture} \cite{hackenberg_international_2013,mahr_determining_2011}\end{tabular} & \textbf{GSR} \cite{hernandez_autoemotive_2014} & \begin{tabular}[c]{@{}l@{}}\textbf{Touch} \cite{hernandez_autoemotive_2014,nawa_information_2012,doring_gestural_2011}\\\cite{murer_exploring_2012,heikkinen_mobile_2013,schartmuller_type-o-steer_2019}\\ \cite{large_twist_2016,huber_force-enabled_2017},\\ \textbf{Pressure} \cite{huber_force-enabled_2017}\end{tabular} & \begin{tabular}[c]{@{}l@{}}\textbf{Skin Temp.}\\ \cite{hernandez_autoemotive_2014}\end{tabular} & \cellcolor[HTML]{D9D9D9} &  &  &  & \begin{tabular}[c]{@{}l@{}}\textbf{Heart}\\ \textbf{Rate} \cite{tobisch_dealing_2020}\end{tabular} \\ \hhline{|*{1}{>{\arrayrulecolor[HTML]{BDD7EE}}-}|>{\arrayrulecolor[HTML]{D0E0E3}}-|*{13}{>{\arrayrulecolor{black}}-}|} 
\multicolumn{1}{|c|}{\cellcolor[HTML]{BDD7EE}} & \multicolumn{1}{c|}{\cellcolor[HTML]{D0E0E3}} & \cellcolor[HTML]{D0E0E3}\textbf{Seat} &  &  & \cellcolor[HTML]{D9D9D9} & \textbf{Body Pressure} \cite{sundstrom_gaming_2014,kim_predicting_2019,tobisch_dealing_2020,nandi_design_2020} &  &  &  & \cellcolor[HTML]{D9D9D9} &  &  &  &  \\ \hhline{|*{1}{>{\arrayrulecolor[HTML]{BDD7EE}}-}|>{\arrayrulecolor[HTML]{D0E0E3}}-|*{13}{>{\arrayrulecolor{black}}-}|} 
\multicolumn{1}{|c|}{\cellcolor[HTML]{BDD7EE}} & \multicolumn{1}{c|}{\cellcolor[HTML]{D0E0E3}} & \cellcolor[HTML]{D0E0E3}\textbf{Table} &  &  & \cellcolor[HTML]{D9D9D9} & \textbf{Pen} \cite{ishiguro_leadingdisplay_2019} &  & \textbf{Touch} \cite{chang_using_2017} &  & \cellcolor[HTML]{D9D9D9} &  &  &  &  \\ \hhline{|*{1}{>{\arrayrulecolor[HTML]{BDD7EE}}-}|>{\arrayrulecolor[HTML]{D0E0E3}}-|*{13}{>{\arrayrulecolor{black}}-}|} 
\multicolumn{1}{|c|}{\cellcolor[HTML]{BDD7EE}} & \multicolumn{1}{c|}{\cellcolor[HTML]{D0E0E3}} & \cellcolor[HTML]{D0E0E3}\textbf{Door} &  &  & \cellcolor[HTML]{D9D9D9} &  &  & \textbf{Touch} \cite{berger_ar-enabled_2021,hakkila_exploring_2014} &  & \cellcolor[HTML]{D9D9D9} &  &  &  &  \\ \hhline{|*{1}{>{\arrayrulecolor[HTML]{BDD7EE}}-}|>{\arrayrulecolor[HTML]{D0E0E3}}-|*{13}{>{\arrayrulecolor{black}}-}|} 
\multicolumn{1}{|c|}{\cellcolor[HTML]{BDD7EE}} & \multicolumn{1}{c|}{\cellcolor[HTML]{D0E0E3}} & \cellcolor[HTML]{D0E0E3}\textbf{Rear} &  &  & \cellcolor[HTML]{D9D9D9} &  &  &  &  & \cellcolor[HTML]{D9D9D9} &  &  &  &  \\ \hhline{|*{1}{>{\arrayrulecolor[HTML]{BDD7EE}}-}|>{\arrayrulecolor[HTML]{D0E0E3}}-|*{13}{>{\arrayrulecolor{black}}-}|} 
\multicolumn{1}{|c|}{\cellcolor[HTML]{BDD7EE}} & \multicolumn{1}{c|}{\cellcolor[HTML]{D0E0E3}} & \cellcolor[HTML]{D0E0E3}\textbf{Floor} &  &  & \cellcolor[HTML]{D9D9D9} & \textbf{Foot Position} \cite{rangesh_exploring_2018} &  &  &  & \cellcolor[HTML]{D9D9D9} &  &  &  &  \\  \hhline{|*{1}{>{\arrayrulecolor[HTML]{BDD7EE}}-}|>{\arrayrulecolor[HTML]{D0E0E3}}-|*{13}{>{\arrayrulecolor{black}}-}|} 
\multicolumn{1}{|c|}{\multirow{-50}{*}{\cellcolor[HTML]{BDD7EE}\rotatebox[origin=c]{90}{\textbf{D1: Input Location}}}} & \multicolumn{1}{c|}{\multirow{-40}{*}{\cellcolor[HTML]{D0E0E3}\rotatebox[origin=c]{90}{\textbf{Anchored}}}} & \cellcolor[HTML]{D0E0E3}\textbf{Ceiling} &  &  & \cellcolor[HTML]{D9D9D9} & \textbf{Pen} \cite{ishiguro_leadingdisplay_2019} &  &  &  & \cellcolor[HTML]{D9D9D9} &  &  &  &  \\ \hline
\end{tabular}%
}
\label{tab:design_space_input}
\end{table*}

\subsection{Input Modalities}
\label{input modalities}
Users explicitly or implicitly act using their actuators on in-vehicle systems that are equipped with sensors, e.g., cameras or touchscreens.
In the following, we present examples for each human actuator category of the found input modalities. 
However, we did not find any publications included in our synthesis that considered olfactory or gustatory input modalities.

\textbf{Visual:} 
Our SLR revealed that the eyes were mainly used in gaze-based interaction.
For example, Riegler et al.~\cite{riegler_gaze-based_2020} employed gaze-based interaction with windshield displays and evaluated the dwell time for object selection.
Roider et al.~\cite{roider_just_2018} investigated gaze-activated voice input to reduce visual distraction, and they enabled drivers to utilize gaze input to perform a secondary task~\cite{roider_effects_2017}.
Gaze was also used implicitly to infer users' interest in objects.
For example, by analyzing the gaze direction to determine the referred object in an interaction~\cite{wang_watch_2020} or to identify referenced objects outside the vehicle~\cite{aftab_multimodal_2019}.
An in-vehicle system could also detect glances, which occur when a user briefly looks at something, e.g., at secondary screens while driving~\cite{smith2005methodology}.
In this context, Ulahannan et al.~\cite{ulahannan_using_2021} used glance behavior to inform adaptive UIs in AVs.
A 6-second sequence of glances was used by Fridman et al.~\cite{fridman_what_2017} to predict the driver state.
Likewise, Mahajan et al.~\cite{mahajan_exploring_2021} analyzed glance behavior (e.g., frequency) regarding situation awareness before takeover requests (TORs).
Besides, longer eye movement sequences can be used, which we found to be mainly implicit input for predictive algorithms.
For example, eye movements before a TOR were used to predict takeover behavior to enable driver context adaptive UIs~\cite{pakdamanian_deeptake_2021}, and to predict subsequent driving performance~\cite{wu_eye_2021}.
Other (often implicit) visual input modalities are eye blinks and pupil characteristics.
For example, Benedetto et al.~\cite{benedetto_driver_2011} measured pupil diameter, blink rate, and blink duration to assess the driver's mental workload and inform accident prevention systems.
Likewise, Mahajan et al.~\cite{mahajan_exploring_2021} measured pupil diameter and blink frequency to assess the changed driver alertness due to automation and the TOR performance.
However, we did not find any driver assistance systems already built into vehicles, such as drowsiness detection using visual input, which could be due to the set time range (2011-2021).

\textbf{Auditory:}
All sounds that are produced by human actuators, such as the vocal tract, can be used as auditory input for in-vehicle systems.
Our synthesis showed that speech-based inputs were mainly used in conversational/dialog systems, which can reply via computer-generated speech.
For example, dialog systems by Nawa et al.~\cite{nawa_information_2012} and Lin et al.~\cite{lin_adasa_2018} retrieve speech input to anticipate the dialog context and response relevant information.
Similarly, dialog systems were evaluated regarding usability and driver distraction~\cite{hofmann_comparison_2014}, and design guidelines for dialog UIs were defined to address issues such as mental workload or task-related fatigue~\cite{large_lessons_2019}.
While dialog systems require context, simpler forms of speech-based input are voice commands.
For example, Winzer et al.~\cite{winzer_intersection_2018} utilized voice commands like "left" or "right" for control of a traffic light assistant.
Lee et al.~\cite{lee_partially_2014} enabled users to change the automated driving state via, e.g., "auto drive" or "manual drive".
We also found approaches that used voice commands for communication with a robot assistant~\cite{williams_affective_2014} and interaction with an automated cab~\cite{large_please_2019}.
However, auditory input is more than using natural language in a dialog or voice command.
We found in-vehicle systems that utilized subtle messages conveyed by the human voice.
Tischler et al.~\cite{tischler_application_nodate} described emotion recognition (inference of valence and arousal) using changes of pitch, intensity, and energy of the speech signal.
Furthermore, non-speech sounds generated by the vocal tract or body parts were used.
For example, Funk et al.~\cite{funk_non-verbal_2020} enabled nonverbal auditory input (humming, snapping, and clapping) for interaction with an assistant while driving.

\textbf{Kinesthetic:}
Kinesthetic input comprises the position and movement of body parts and activity of muscles and joints~\cite{ISO_tactile}, to which we also attribute the facial muscle activity needed for emotion and facial expressions.
For example, body position classifiers were used to recognize driver distraction~\cite{weyers_driver_2018}, and posture tracking allows inference of situation awareness and reaction times~\cite{lotz_response_2019}.
The body pressure applied to the seat was also analyzed to inform optimal seat vibration points~\cite{kim_seated_2020}.
Apart from whole body characteristics, body parts such as the head can provide more specific input.
For example, head pose tracking was used to analyze driver behavior and monitor attention~\cite{borghi_embedded_2017}, and head movement to adapt the rendering of stereoscopic 3D displays~\cite{broy_exploring_2014}.
We also found a work that measured the distance between feet and pedals to estimate situation awareness~\cite{rangesh_exploring_2018}.
Besides, facial expressions were used to measure the driver's stress level~\cite{hernandez_autoemotive_2014}, to adapt the personality of a voice assistant~\cite{braun_at_2019}, and facial landmarks were utilized to classify situation awareness~\cite{rangesh_exploring_2018}.
Another commonly found kinesthetic modality was gesture input.
For example, Jiang et al.~\cite{jiang_givs_2020} investigated display control via hand gestures, and Ahmad et al.~\cite{ahmad_you_2016} used hand pointing to predict the display item a user intends to select.
Some approaches considered finger gestures and pointing, e.g., to control in-vehicle functions such as volume~\cite{john_real-time_2017} or for contactless operation of a screen using mid-air finger movements~\cite{riener_natural_2011}.
We also found various approaches to recognize kinesthetic input via hardware such as buttons~\cite{kidd_considering_2017}, levers~\cite{hornberger_evaluation_2018}, switches~\cite{vo_investigating_2020}, and rotary controllers~\cite{alvarez_help_2015}.
Schartmüller et al.~\cite{schartmuller_type-o-steer_2019} reimagined the steering wheel for NDRTs by attaching a keyboard, while Jakus et al.~\cite{jakus_user_2015} attached a computer mouse to the steering wheel backside.
There were also approaches leveraging shape-changing input, e.g., Heijboer et al.~\cite{heijboer_physical_2019} imagined a shape-changing dashboard that users can actively deform.

\textbf{Electrodermal:}
In contrast to, e.g., visual, auditory, or kinesthetic input, EDA is exclusively implicit input.
For example, Wang et al.~\cite{wang_carpacio_2017} used body electric conductivity to distinguish driver and passenger touches on screens.
Madrid et al.~\cite{madrid_biometric_2018} presented a biometric UI for stress and awareness detection using driver's galvanic skin response (GSR), which was also used by Pakdamanian et al.~\cite{pakdamanian_deeptake_2021} to predict takeover performance.
Besides, electromyography of the user's forearm muscles was used to recognize micro-gestures performed on the steering wheel~\cite{angelini_opportunistic_2013}.

\textbf{Tactile:}
We mostly found approaches that recognized tactile input via touch-sensitive surfaces, e.g.,~\cite{detjen_driving_2021,ng_evaluation_2017,large_please_2019,huber_force-enabled_2017}.
However, Berger et al.~\cite{berger_tactile_2019} investigated an absolute indirect touch interaction concept to enable passenger interaction with an infotainment system.
Apart from touch-based input, we also found publications that considered pressure-based inputs.
For example, Swette et al.~\cite{swette_comparing_2013} presented pressure-based buttons on a touch surface, and finger pressure input deforms the shape-changing dashboard by Heijboer et al.~\cite{heijboer_physical_2019}.

\textbf{Thermal:}
The temperature dispensed via the skin can be utilized as a thermal input.
However, as humans do not actively control their body temperature, we only found approaches that used implicit thermal inputs, e.g., to assess driver emotion~\cite{wang_drivers_2020}, or stress~\cite{hernandez_autoemotive_2014}.

\textbf{Cerebral:}
We found that cerebral input modalities were mainly used for implicit interactions.
For example, Zhu et al.~\cite{zhu_heavy_2020} employed EEG for truck driver drowsiness detection and early warning system.
However, some approaches utilized brain signals for explicit control of vehicle functions.
He et al.~\cite{he_brain_2016} proposed an interface between drivers and in-vehicle devices using brain signals that can be used while performing primary driving tasks.
Likewise, in a concept by Hood et al.~\cite{hood_use_2012}, a brain-computer interface enabled drivers to control vehicle functions, including acceleration and steering.

\textbf{Cardiac:}
In contrast to cerebral input, we did not find any approach for explicit cardiac input.
Instead, the heart was only used as an actuator for implicit input.
For example, Madrid et al.~\cite{madrid_biometric_2018} assessed the driver's heart rate to detect stress and enable real-time feedback so that drivers can take corrective actions on time.
Similarly, Hayashi et al.~\cite{hayashi_development_2020} developed an abnormal sign detection system based on driver's heart pulse measures.
Electrocardiogram measures were also used by Wang et al.~\cite{wang_drivers_2020} to improve driver emotion recognition.


\begin{table*}[ht]
\centering
\caption{The matrix obtained by the morphological approach, showing the design space for in-vehicle \textbf{output} modalities. The combinations highlighted in gray make technically no sense or are unethical.}
\resizebox{\textwidth}{!}{%
\begin{tabular}{ccl|l|l|l|l|l|l|l|l|l|l|l|l|}
\cline{4-15}
\multicolumn{1}{l}{} & \multicolumn{1}{l}{} &  & \multicolumn{12}{c|}{\cellcolor[HTML]{FCE5CD}\textbf{D2: (Human Sensed) Output Modality}} \\ \cline{4-15} 
\multicolumn{1}{l}{} & \multicolumn{1}{l}{} &  & \multicolumn{1}{c|}{} & \multicolumn{1}{c|}{} & \multicolumn{6}{c|}{\textbf{Haptic}} & \multicolumn{1}{c|}{} & \multicolumn{1}{c|}{} & \multicolumn{1}{c|}{} & \multicolumn{1}{c|}{} \\ \cline{6-11}
\multicolumn{1}{l}{} & \multicolumn{1}{l}{} &  & \multicolumn{1}{c|}{} & \multicolumn{1}{c|}{} & \multicolumn{1}{c|}{} & \multicolumn{1}{c|}{} & \multicolumn{4}{c|}{\textbf{Cutaneous}} & \multicolumn{1}{c|}{} & \multicolumn{1}{c|}{} & \multicolumn{1}{c|}{} & \multicolumn{1}{c|}{} \\ \cline{8-11}
\multicolumn{1}{l}{} & \multicolumn{1}{l}{} &  & \multicolumn{1}{c|}{\multirow{-3}{*}{\textbf{Visual}}} & \multicolumn{1}{c|}{\multirow{-3}{*}{\textbf{Auditory}}} & \multicolumn{1}{c|}{\multirow{-2}{*}{\textbf{Vestibular}}} & \multicolumn{1}{c|}{\multirow{-2}{*}{\textbf{Kinesthetic}}} & \multicolumn{1}{c|}{\textbf{Electrodermal}} & \multicolumn{1}{c|}{\textbf{Tactile}} & \multicolumn{1}{c|}{\textbf{Thermal}} & \multicolumn{1}{c|}{\textbf{Pain}} & \multicolumn{1}{c|}{\multirow{-3}{*}{\textbf{Olfactory}}} & \multicolumn{1}{c|}{\multirow{-3}{*}{\textbf{Gustatory}}} & \multicolumn{1}{c|}{\multirow{-3}{*}{\textbf{Cerebral}}} & \multicolumn{1}{c|}{\multirow{-3}{*}{\textbf{Cardiac}}} \\ \hline
\multicolumn{1}{|c|}{\cellcolor[HTML]{BDD7EE}} & \multicolumn{1}{c|}{\cellcolor[HTML]{C9DAF8}} & \cellcolor[HTML]{C9DAF8}\textbf{AR} & \begin{tabular}[c]{@{}l@{}}\textbf{AR} \cite{korthauer_watch_2020,wiegand_incarar_2019,lauber_-your-face_2014},\\ \textbf{AR Smartglasses} \cite{he_does_2018}\end{tabular} &  &  &  &  &  &  & \cellcolor[HTML]{D9D9D9} &  &  & \cellcolor[HTML]{D9D9D9} & \cellcolor[HTML]{D9D9D9} \\ \hhline{|*{1}{>{\arrayrulecolor[HTML]{BDD7EE}}-}|>{\arrayrulecolor[HTML]{C9DAF8}}-|*{13}{>{\arrayrulecolor{black}}-}|} 
\multicolumn{1}{|c|}{\cellcolor[HTML]{BDD7EE}} & \multicolumn{1}{c|}{\cellcolor[HTML]{C9DAF8}} & \cellcolor[HTML]{C9DAF8}\textbf{VR} & \begin{tabular}[c]{@{}l@{}}\textbf{See-through Cockpit} \cite{lindemann_examining_2017},\\ \textbf{VR HMD} \cite{mcgill_i_2017,li_exploration_2020,hock_carvr_2017}\end{tabular} &  &  &  &  &  &  & \cellcolor[HTML]{D9D9D9} &  &  & \cellcolor[HTML]{D9D9D9} & \cellcolor[HTML]{D9D9D9} \\ \hhline{|*{1}{>{\arrayrulecolor[HTML]{BDD7EE}}-}|>{\arrayrulecolor[HTML]{C9DAF8}}-|*{13}{>{\arrayrulecolor{black}}-}|} 
\multicolumn{1}{|c|}{\cellcolor[HTML]{BDD7EE}} & \multicolumn{1}{c|}{\cellcolor[HTML]{C9DAF8}} & \cellcolor[HTML]{C9DAF8}\textbf{Handheld} & \begin{tabular}[c]{@{}l@{}}\textbf{Smartphone} \cite{he_texting_2013,perlman_relative_2019,brumby_empirical_2012,heikkinen_mobile_2013}\\ \cite{he_does_2018,wang_carpacio_2017,osswald_hmi_2013,wintersberger_let_2018},\\ \textbf{Tablet} \cite{meschtscherjakov_active_2016,capallera_secondary_2019,oliveira_evaluating_2018}\end{tabular} & \begin{tabular}[c]{@{}l@{}}\textbf{Smartphone Sound} \cite{gable_advanced_2013},\\ \textbf{Smartphone Voice} \cite{gable_advanced_2013}\end{tabular} &  &  &  &  &  & \cellcolor[HTML]{D9D9D9} &  &  & \cellcolor[HTML]{D9D9D9} & \cellcolor[HTML]{D9D9D9} \\ \hhline{|*{1}{>{\arrayrulecolor[HTML]{BDD7EE}}-}|>{\arrayrulecolor[HTML]{C9DAF8}}-|*{13}{>{\arrayrulecolor{black}}-}|} 
\multicolumn{1}{|c|}{\cellcolor[HTML]{BDD7EE}} & \multicolumn{1}{c|}{\multirow{-8}{*}{\cellcolor[HTML]{C9DAF8}\rotatebox[origin=c]{90}{\textbf{Nomadic}}}} & \cellcolor[HTML]{C9DAF8}\textbf{Wearable} & \textbf{Smartwatch} \cite{perlman_relative_2019,li_combined_2018} &  &  &  &  & \begin{tabular}[c]{@{}l@{}}\textbf{Smart Ring} \cite{gheran_controls_2020},\\ \textbf{Vibration Wristband} \cite{politis_beep_2015},\\ \textbf{Vibration Belt} \cite{kruger_tactile_2021,kruger_approach_2018},\\ \textbf{Tactor Belt} \cite{politis_evaluating_2014}\end{tabular} &  & \cellcolor[HTML]{D9D9D9} &  &  & \cellcolor[HTML]{D9D9D9} & \cellcolor[HTML]{D9D9D9} \\ \hhline{|*{1}{>{\arrayrulecolor[HTML]{BDD7EE}}-}|*{14}{>{\arrayrulecolor{black}}-}|} 
\multicolumn{1}{|c|}{\cellcolor[HTML]{BDD7EE}} & \multicolumn{1}{c|}{\cellcolor[HTML]{D0E0E3}} & \cellcolor[HTML]{D0E0E3}\textbf{Front} &  & \begin{tabular}[c]{@{}l@{}}\textbf{Sound}\\ \cite{kutchek_takeover_2019,fagerlonn_graded_2012,wang_using_2017,petermeijer_take-over_2017,schieben_evaluation_2014}\\ \cite{reinmueller_adaptive_2018,kim_exploring_2013,houtenbos_concurrent_2017,seppelt_keeping_2019},\\ \textbf{Speech}\\ \cite{baldwin_loudness_2011,lin_voice_2018,kim_predicting_2019,martelaro_exploration_2019,hackenberg_international_2013}\\ \cite{stier_adapting_2020,jeon_effects_2015,large_steering_2017,chang_dont_2016},\\ \textbf{Voice}\\ \cite{guo_can_2021,braun_at_2019,mahajan_exploring_2021,hayashi_development_2020,di_campli_san_vito_investigation_2018,mitsopoulos-rubens_effects_2011},\\ \textbf{Music} \cite{burnett_altering_2017,chen_manipulating_2021,hernandez_autoemotive_2014},\\ \textbf{Earcon}\\ \cite{ovcharova_effectiveness_2012,gang_dont_2018,jeon_multimodal_2019,roider_effects_2017,schneider_increasing_2021,shakeri_novel_2017}\end{tabular} &  &  &  &  &  & \cellcolor[HTML]{D9D9D9} & \begin{tabular}[c]{@{}l@{}}\textbf{Scent}\\ \cite{dmitrenko_comparison_2016,dmitrenko_caroma_2020}\\ \cite{dmitrenko_what_2017,dmitrenko_towards_2019}\\ \cite{dmitrenko_i_2018,schartmuller_sick_2020}\end{tabular} &  & \cellcolor[HTML]{D9D9D9} & \cellcolor[HTML]{D9D9D9} \\ \hhline{|*{1}{>{\arrayrulecolor[HTML]{BDD7EE}}-}|>{\arrayrulecolor[HTML]{D0E0E3}}-|*{13}{>{\arrayrulecolor{black}}-}|} 
\multicolumn{1}{|c|}{\cellcolor[HTML]{BDD7EE}} & \multicolumn{1}{c|}{\cellcolor[HTML]{D0E0E3}} & \cellcolor[HTML]{D0E0E3}\textbf{Windshield} & \begin{tabular}[c]{@{}l@{}}\textbf{AR} \cite{phan_enhancing_2016,kim_are_2017,bark_personal_2014,kunze_augmented_2018,schneider_field_2019}\\ \cite{lindemann_exploring_2019,oliveira_influence_2020,wiegand_early_2018,feierle_head-up_2019,morra_building_2019},\\ \textbf{Ambient Light} \cite{locken_increasing_2020,langlois_adas_2013,wang_designing_2017,meschtscherjakov_chase_2020}\\ \cite{schneider_increasing_2021,van_den_beukel_driving_2016,van_den_beukel_supporting_2016}\end{tabular} &  &  &  &  &  &  & \cellcolor[HTML]{D9D9D9} &  &  & \cellcolor[HTML]{D9D9D9} & \cellcolor[HTML]{D9D9D9} \\ \hhline{|*{1}{>{\arrayrulecolor[HTML]{BDD7EE}}-}|>{\arrayrulecolor[HTML]{D0E0E3}}-|*{13}{>{\arrayrulecolor{black}}-}|} 
\multicolumn{1}{|c|}{\cellcolor[HTML]{BDD7EE}} & \multicolumn{1}{c|}{\cellcolor[HTML]{D0E0E3}} & \cellcolor[HTML]{D0E0E3}\textbf{Dashboard} & \begin{tabular}[c]{@{}l@{}}\textbf{Display} \cite{kim_usability_2011,kohlhaas_anticipatory_2011,steinberger_coastmaster_2016,stromberg_driver_2011}\\ \cite{rittger_measuring_2017,ma_investigating_2021,jung_displayed_2015,oliveira_influence_2020,weber_service_2019,ulahannan_designing_2020},\\ \textbf{Peripheral Light} \cite{kunze_evaluation_2018,sadeghian_borojeni_feel_2018},\\ \textbf{Robotic Companion} \cite{row_dooboo_2016},\\ \textbf{LED} \cite{lees_cross-modal_2012}\\ \textbf{Shutter Display Glasses} \cite{broy_3d_2014,broy_exploring_2014},\\ \textbf{Ambient Light} \cite{borojeni_assisting_2016,borojeni_reading_2018,clark_directability_2019,shakeri_novel_2017,shakeri_bimodal_2017}\end{tabular} &  &  &  &  &  &  & \cellcolor[HTML]{D9D9D9} &  &  & \cellcolor[HTML]{D9D9D9} & \cellcolor[HTML]{D9D9D9} \\ \hhline{|*{1}{>{\arrayrulecolor[HTML]{BDD7EE}}-}|>{\arrayrulecolor[HTML]{D0E0E3}}-|*{13}{>{\arrayrulecolor{black}}-}|} 
\multicolumn{1}{|c|}{\cellcolor[HTML]{BDD7EE}} & \multicolumn{1}{c|}{\cellcolor[HTML]{D0E0E3}} & \cellcolor[HTML]{D0E0E3}\textbf{HUD} & \begin{tabular}[c]{@{}l@{}}\textbf{HUD} \cite{yan_development_2016,phan_enhancing_2016,haeuslschmid_recognition_2017,hauslschmid_supportingtrust_2017,schartmuller_text_2019}\\ \cite{beck_perceived_2018,lauber_-your-face_2014,schneider_field_2019,van_der_heiden_visual_2019,feierle_head-up_2019},\\ \textbf{Stereoscopic 3D} \cite{wiegand_early_2018}\end{tabular} &  &  &  &  &  &  & \cellcolor[HTML]{D9D9D9} &  &  & \cellcolor[HTML]{D9D9D9} & \cellcolor[HTML]{D9D9D9} \\ \hhline{|*{1}{>{\arrayrulecolor[HTML]{BDD7EE}}-}|>{\arrayrulecolor[HTML]{D0E0E3}}-|*{13}{>{\arrayrulecolor{black}}-}|} 
\multicolumn{1}{|c|}{\cellcolor[HTML]{BDD7EE}} & \multicolumn{1}{c|}{\cellcolor[HTML]{D0E0E3}} & \cellcolor[HTML]{D0E0E3}\textbf{\begin{tabular}[c]{@{}l@{}}\textbf{Center}\\ \textbf{Stack}\end{tabular}} & \begin{tabular}[c]{@{}l@{}}\textbf{Display} \cite{korthauer_watch_2020,reimer_exploratory_2012,steinberger_coastmaster_2016,broy_evaluating_2015,yan_spatial_2019}\\ \cite{lindemann_exploring_2019,weber_service_2019,colley_effects_2021,miglani_compatibility_2016,miglani_compatibility_2016},\\ \textbf{Robotic Companion}\\ \cite{lee_autonomous_2019,kraus_human_2016,karatas_namida_2016,williams_affective_2014,williams_towards_2013},\\ \textbf{Robotic Display} \cite{yang_affective_2013},\\ \textbf{LED} \cite{zhu_heavy_2020,loehmann_heartbeat_2014,heijboer_physical_2019}\end{tabular} & \begin{tabular}[c]{@{}l@{}}\textbf{Robotic}\\ \textbf{Companion Voice} \cite{yang_affective_2013}\end{tabular} &  & \begin{tabular}[c]{@{}l@{}}\textbf{Shape Changing}\\ \cite{zimmermann_i_2014,cockburn_reducing_2018,heijboer_physical_2019}\end{tabular} &  & \begin{tabular}[c]{@{}l@{}}\textbf{Pin-Array} \cite{jung_voicetactile_2020},\\ \textbf{Vibration}\\ \cite{vo_investigating_2020,ng_evaluation_2017,ng_investigating_2016,loehmann_heartbeat_2014,ng_evaluation_2017-1,huber_towards_2016},\\ \textbf{Ultrasound}\\ \cite{brown_ultrahapticons_2020,korres_mid-air_2020,large_feel_2019,shakeri_may_2018,georgiou_haptic_2017,harrington_exploring_2018},\\ \textbf{Shape Changing}\\ \cite{zimmermann_i_2014,cockburn_reducing_2018,heijboer_physical_2019},\\ \textbf{Texture} \cite{zimmermann_i_2014},\\ \textbf{Piezo Haptic} \cite{heijboer_physical_2019}\end{tabular} & \begin{tabular}[c]{@{}l@{}}\textbf{Peltier}\\ \textbf{Element}\\ \cite{di_campli_san_vito_investigation_2018,heijboer_physical_2019}\end{tabular} & \cellcolor[HTML]{D9D9D9} &  &  & \cellcolor[HTML]{D9D9D9} & \cellcolor[HTML]{D9D9D9} \\ \hhline{|*{1}{>{\arrayrulecolor[HTML]{BDD7EE}}-}|>{\arrayrulecolor[HTML]{D0E0E3}}-|*{13}{>{\arrayrulecolor{black}}-}|} 
\multicolumn{1}{|c|}{\cellcolor[HTML]{BDD7EE}} & \multicolumn{1}{c|}{\cellcolor[HTML]{D0E0E3}} & \cellcolor[HTML]{D0E0E3}\textbf{\begin{tabular}[c]{@{}l@{}}\textbf{Steering}\\ \textbf{Wheel}\end{tabular}} & \begin{tabular}[c]{@{}l@{}}\textbf{Display} \cite{wilfinger_wheels_2013,kunze_evaluation_2018,heikkinen_mobile_2013,mok_reinventing_2017},\\ \textbf{LED} \cite{johns_looking_2017,huang_research_2015,madrid_biometric_2018},\\ \textbf{Ambient Light} \cite{sun_improvement_2021,van_den_beukel_supporting_2016,mok_reinventing_2017},\\ \textbf{Projection} \cite{mok_reinventing_2017},\\ \textbf{Shape Changing} \cite{mok_reinventing_2017}\end{tabular} &  &  & \begin{tabular}[c]{@{}l@{}}\textbf{Steering Impedance}\\ \cite{bhardwaj_whos_2020,huang_research_2015},\\ \textbf{Shape Changing}\\ \cite{mok_reinventing_2017,mok_actions_2017,kerschbaum_transforming_2015}\end{tabular} &  & \begin{tabular}[c]{@{}l@{}}\textbf{Haptic Tic} \cite{schieben_evaluation_2014},\\ \textbf{Vibration}\\ \cite{reinmueller_adaptive_2018,nawa_information_2012,sucu_haptic_2013,vo_investigating_2020,dmitrenko_towards_2019,quintal_hapwheel_2021},\\ \textbf{Cutaneous Push}\\ \cite{di_campli_san_vito_haptic_2019,shakeri_novel_2017,shakeri_bimodal_2017}\end{tabular} & \begin{tabular}[c]{@{}l@{}}\textbf{Peltier}\\ \textbf{Element}\\ \cite{di_campli_san_vito_haptic_2019}\end{tabular} & \cellcolor[HTML]{D9D9D9} &  &  & \cellcolor[HTML]{D9D9D9} & \cellcolor[HTML]{D9D9D9} \\ \hhline{|*{1}{>{\arrayrulecolor[HTML]{BDD7EE}}-}|>{\arrayrulecolor[HTML]{D0E0E3}}-|*{13}{>{\arrayrulecolor{black}}-}|} 
\multicolumn{1}{|c|}{\cellcolor[HTML]{BDD7EE}} & \multicolumn{1}{c|}{\cellcolor[HTML]{D0E0E3}} & \cellcolor[HTML]{D0E0E3}\textbf{Seat} & \begin{tabular}[c]{@{}l@{}}\textbf{Display (headrest)} \cite{sundstrom_gaming_2014},\\ \textbf{Display (backrest)} \cite{berger_ar-enabled_2021,berger_tactile_2019}\end{tabular} &  & \begin{tabular}[c]{@{}l@{}}\textbf{Seat}\\ \textbf{Rotation} \cite{sun_shaping_2021},\\ \textbf{Seat}\\ \textbf{Tilt} \cite{winner_effects_2016}\end{tabular} & \begin{tabular}[c]{@{}l@{}}\textbf{Shape Changing} \cite{grah_dorsal_2015},\\ \textbf{Seat Rotation} \cite{sun_shaping_2021},\\ \textbf{Seat Movement} \cite{pfleging_multimodal_2012},\\ \textbf{Brake Pulse} \cite{gaspar_examining_2015}\end{tabular} &  & \begin{tabular}[c]{@{}l@{}}\textbf{Vibration}\\ \cite{petermeijer_take-over_2017,geitner_comparison_2019,riener_subliminal_2012,kunze_preliminary_2018,capallera_convey_2019}\\ \cite{kim_seated_2020,diwischek_tactile_2015,gaspar_examining_2015,schneider_increasing_2021,van_den_beukel_driving_2016},\\ \textbf{Shape Changing} \cite{grah_dorsal_2015},\\ \textbf{Seat Belt Tension} \cite{gaspar_examining_2015},\\ \textbf{Tactor} \cite{lees_cross-modal_2012},\\ \textbf{LED Strip} \cite{politis_evaluation_2018}\end{tabular} &  & \cellcolor[HTML]{D9D9D9} &  &  & \cellcolor[HTML]{D9D9D9} & \cellcolor[HTML]{D9D9D9} \\ \hhline{|*{1}{>{\arrayrulecolor[HTML]{BDD7EE}}-}|>{\arrayrulecolor[HTML]{D0E0E3}}-|*{13}{>{\arrayrulecolor{black}}-}|} 
\multicolumn{1}{|c|}{\cellcolor[HTML]{BDD7EE}} & \multicolumn{1}{c|}{\cellcolor[HTML]{D0E0E3}} & \cellcolor[HTML]{D0E0E3}\textbf{Table} & \begin{tabular}[c]{@{}l@{}}\textbf{Robotic Display} \cite{ishiguro_leadingdisplay_2019},\\ \textbf{Display} \cite{oliveira_evaluating_2018}\end{tabular} &  &  &  &  &  &  & \cellcolor[HTML]{D9D9D9} &  &  & \cellcolor[HTML]{D9D9D9} & \cellcolor[HTML]{D9D9D9} \\  \hhline{|*{1}{>{\arrayrulecolor[HTML]{BDD7EE}}-}|>{\arrayrulecolor[HTML]{D0E0E3}}-|*{13}{>{\arrayrulecolor{black}}-}|} 
\multicolumn{1}{|c|}{\cellcolor[HTML]{BDD7EE}} & \multicolumn{1}{c|}{\cellcolor[HTML]{D0E0E3}} & \cellcolor[HTML]{D0E0E3}\textbf{Door} & \begin{tabular}[c]{@{}l@{}}\textbf{Display} \cite{berger_ar-enabled_2021},\\ \textbf{Door Window AR} \cite{berger_ar-enabled_2021,hakkila_exploring_2014}\end{tabular} &  &  &  &  &  &  & \cellcolor[HTML]{D9D9D9} &  &  & \cellcolor[HTML]{D9D9D9} & \cellcolor[HTML]{D9D9D9} \\ \hhline{|*{1}{>{\arrayrulecolor[HTML]{BDD7EE}}-}|>{\arrayrulecolor[HTML]{D0E0E3}}-|*{13}{>{\arrayrulecolor{black}}-}|} 
\multicolumn{1}{|c|}{\cellcolor[HTML]{BDD7EE}} & \multicolumn{1}{c|}{\cellcolor[HTML]{D0E0E3}} & \cellcolor[HTML]{D0E0E3}\textbf{Rear} &  & \textbf{Music} \cite{burnett_altering_2017}, \textbf{Earcon} \cite{gang_dont_2018} &  &  &  &  &  & \cellcolor[HTML]{D9D9D9} &  &  & \cellcolor[HTML]{D9D9D9} & \cellcolor[HTML]{D9D9D9} \\ \hhline{|*{1}{>{\arrayrulecolor[HTML]{BDD7EE}}-}|>{\arrayrulecolor[HTML]{D0E0E3}}-|*{13}{>{\arrayrulecolor{black}}-}|} 
\multicolumn{1}{|c|}{\cellcolor[HTML]{BDD7EE}} & \multicolumn{1}{c|}{\cellcolor[HTML]{D0E0E3}} & \cellcolor[HTML]{D0E0E3}\textbf{Floor} &  & \textbf{Earcon} \cite{gang_dont_2018} &  & \textbf{Haptic Pedal} \cite{corno_design_2013,henzler_are_2015} &  &  &  & \cellcolor[HTML]{D9D9D9} &  &  & \cellcolor[HTML]{D9D9D9} & \cellcolor[HTML]{D9D9D9} \\ \hhline{|*{1}{>{\arrayrulecolor[HTML]{BDD7EE}}-}|>{\arrayrulecolor[HTML]{D0E0E3}}-|*{13}{>{\arrayrulecolor{black}}-}|} 
\multicolumn{1}{|c|}{\multirow{-50}{*}{\cellcolor[HTML]{BDD7EE}\rotatebox[origin=c]{90}{\textbf{D1: Output Location}}}} & \multicolumn{1}{c|}{\multirow{-40}{*}{\cellcolor[HTML]{D0E0E3}\rotatebox[origin=c]{90}{\textbf{Anchored}}}} & \cellcolor[HTML]{D0E0E3}\textbf{Ceiling} & \begin{tabular}[c]{@{}l@{}}\textbf{Robotic Display} \cite{ishiguro_leadingdisplay_2019},\\ \textbf{Display} \cite{sun_shaping_2021}\end{tabular} & \textbf{Earcon} \cite{gang_dont_2018} &  &  &  &  &  & \cellcolor[HTML]{D9D9D9} &  &  & \cellcolor[HTML]{D9D9D9} & \cellcolor[HTML]{D9D9D9} \\ \hline
\end{tabular}%
}
\label{tab:design_space_output}
\end{table*}

\subsection{Output Modalities}
\label{output modalities}
Human sensors perceive any output modality generated by vehicle actuators, e.g., speakers or vibration motors.
In the following, we present examples for each human sensor category of found output modalities. 
However, we did not find any publications that considered electrodermal, pain, gustatory, cerebral, or cardiac output modalities.

\textbf{Visual:}
The simplest form of visual output is light.
For example, Johns et al.~\cite{johns_looking_2017} used a steering wheel LED strip to create a virtual steering wheel that moves in advance while the steering wheel is locked.
In a concept by Zhu et al.~\cite{zhu_heavy_2020}, a single LED was used to warn in case of driver drowsiness.
Besides, we found approaches for peripheral light feedback.
For example, peripheral LED strips acting as chase lights to influence the driver's perception of speed~\cite{meschtscherjakov_chase_2020} or using peripheral light flashes behind the steering wheel to indicate a TOR~\cite{sadeghian_borojeni_feel_2018}.
Some approaches utilize ambient light, e.g., to highlight pedestrian trajectories~\cite{locken_increasing_2020}.
Most approaches in our synthesis used screens of various sizes and functions, e.g.,~\cite{kim_usability_2011,korthauer_watch_2020,wilfinger_wheels_2013,sundstrom_gaming_2014}.
However, we also found different visualization approaches, e.g., head-up displays (HUDs)~\cite{schneider_field_2019}, stereoscopic displays~\cite{wiegand_early_2018}, active shutter glasses~\cite{broy_3d_2014}, robotic companion displays~\cite{yang_affective_2013}, projections~\cite{mok_reinventing_2017}, and shape changing devices~\cite{mok_reinventing_2017}.
Regarding HUDs, Beck and Park~\cite{beck_perceived_2018} investigated the perceived importance of different HUD information items, and Häuslschmid et al.~\cite{haeuslschmid_recognition_2017} evaluated the recognition of stimuli on a windshield HUD.
Wiegand et al.~\cite{wiegand_early_2018} envisioned a stereoscopic 3D display to show information for early takeover preparation.
A concept by Broy et al.~\cite{broy_3d_2014} used shutter glasses to visualize a 3D effect on a dashboard display.
Yang et al.~\cite{yang_affective_2013} presented a robotic companion display that automatically pans and tilts.
Mok et al.~\cite{mok_reinventing_2017} projected the vehicle's automation status on the steering wheel while also changing its shape to visualize a TOR.
We also found various approaches for visual output that utilized AR.
For example, Lauber and Butz~\cite{lauber_-your-face_2014} displayed driver warnings via AR HMD.
There are also approaches without HMDs, e.g., Akash et al.~\cite{akash_toward_2020} augmented the windshield with AR-highlights of environment objects, or Lindemann et al.~\cite{lindemann_exploring_2019} explored the use of windshield AR for driver assistance in short-notice takeovers.
Similar to windshield AR, Häkkilä et al.~\cite{hakkila_exploring_2014} proposed a concept for AR-enabled door windows.
Besides, we found in-vehicle VR concepts.
For example, McGill et al.~\cite{mcgill_i_2017} and Li et al.~\cite{li_exploration_2020} investigated the use of VR HMDs for in-vehicle applications.

\textbf{Auditory:}
We found various forms of simple auditory output ranging from sounds, e.g.,~\cite{fagerlonn_graded_2012} to earcons that are brief, distinctive, often melodic sounds to convey a message similar to an icon, e.g.,~\cite{gang_dont_2018}. 
Besides, we found approaches that produce artificial speech and interact with the user in a dialog system, e.g.,~\cite{kim_predicting_2019}.
However, some concepts used text-to-speech systems primarily used in non-conversational interaction, e.g.,~\cite{guo_can_2021}.
While most publications that considered auditory output focused on speech and sound, we found approaches that utilized music.
For example, Burnett et al.~\cite{burnett_altering_2017} presented a concept to alter the speed perception of passengers through subliminal music adaptions.
Similarly, Hernandez et al.~\cite{hernandez_autoemotive_2014} used adaptive music (e.g., relaxing songs) if the driver is stressed, and Chen~\cite{chen_manipulating_2021} manipulated background music using blended sonification to communicate reliability in automated driving.

\textbf{Vestibular:}
Vestibular output modalities address the human sensor of balance and often occur coupled with kinesthetic output as the body is moved by an exterior force to stimulate the vestibular system in the inner ear~\cite{benyon2014designing}.
In our synthesis, we found a concept by Winner and Wachenfeld~\cite{winner_effects_2016} that automatically adapts the seat's tilt angle when an automated vehicle is driving in a curve to normalize the passenger's vestibular experience.

\textbf{Kinesthetic:}
Kinesthetic output consists of any movement sensed over muscles and joints, making it a versatile way to provide feedback to users.
For example, the shape-changing steering wheel by Mok et al.~\cite{mok_reinventing_2017} provided kinesthetic stimuli to hands and arms to indicate the automation state and TORs.
Bhardwaj et al.~\cite{bhardwaj_whos_2020} employed a system that adapts the steering wheel impedance to ensure a safe transition of control between vehicle and driver.
A swivel seat can also be utilized, as shown by, e.g., Sun et al.~\cite{sun_shaping_2021}.
Besides, whole-body kinesthetic output was investigated by Gaspar et al.~\cite{gaspar_examining_2015} who conveyed forward collision warnings via brake pulses.
We also found approaches for haptic pedals that, e.g., provided force feedback to guide drivers toward applying ideal throttle~\cite{corno_design_2013} or a pressure point such that the pedal is pressed no further than the recommended angle for ecological driving~\cite{henzler_are_2015}.

\textbf{Tactile:}
Most publications in the synthesis employed vibration motors, e.g.,~\cite{vo_investigating_2020,van_den_beukel_driving_2016,loehmann_heartbeat_2014,kim_seated_2020}.
However, some approaches used tactors.
For example, in the form of a cutaneous push on the steering wheel for navigation~\cite{di_campli_san_vito_haptic_2020}, a tactor to convey warning cues for driver inattention~\cite{lees_cross-modal_2012}, a pin array that creates tactile patterns for fingers and hand~\cite{jung_voicetactile_2020}, a silicone touchscreen cover foil to increase tactile feedback~\cite{zimmermann_i_2014}, or a 3D printed stencil that makes underlying touchscreen controls tangible~\cite{cockburn_reducing_2018}.
Besides, instead of skin contact with the tactile device, \citet{shakeri_may_2018} and Harrington et al.~\cite{harrington_exploring_2018} proposed mid-air ultrasonic tactile feedback for gesture interaction.

\textbf{Thermal:}
In our synthesis, we only found approaches for thermal output via skin contact.
For example, \citet{di_campli_san_vito_investigation_2018} used a Peltier device to apply thermal feedback to the driver's finger, where the temperature indicated the desired lane change direction.
They also attached a Peltier element to the steering wheel to provide thermal feedback indicating navigation cues~\cite{di_campli_san_vito_haptic_2019}.
Besides, \citet{heijboer_physical_2019} envisioned a dashboard that conveys information via temperature changes.

\textbf{Olfactory:}
Schartmüller et al.~\cite{schartmuller_sick_2020} applied essential oils on a piece of felt attached to the center mirror to investigate non-invasive motion sickness mitigation in AVs.
Similarly, Dmitrenko et al.~\cite{dmitrenko_caroma_2020} utilized essential oils to produce pleasant scents to promote safer driving, better mood, and improved well-being in drivers.
They also compared scent output methods and their usability for in-vehicle olfactory interactions regarding distance, volume, and speed~\cite{dmitrenko_comparison_2016}.


\subsection{Interaction Locations}
\label{interaction locations}
According to our code book (see Section~\ref{abstract screening}), we distinguish between two types of interaction locations: (1)~nomadic and (2)~anchored. The locations are visualized in~\autoref{fig:teaser} and on our interactive website\footnote{\url{https://in-vehicle-interaction-design-space.onrender.com/} | Interactive tool to support investigation of in-vehicle interaction research and design.}. 

\textbf{Nomadic:}
The nomadic interaction location contains devices unbound within the vehicle interior such as HMDs (e.g., AR or VR), handheld devices (e.g., smartphone or laptop), and wearable devices (e.g., smartwatch or smart ring), see~\autoref{fig:teaser}.
We found that input sensed via nomadic devices was mainly kinesthetic and tactile.
Kinesthetic input was performed on handheld devices, such as buttons on a smartphone~\cite{li_evaluation_2018}, gaming controller~\cite{hock_carvr_2017}, or wearables, e.g., smartwatch~\cite{hald_using_2018} and smart ring~\cite{gheran_controls_2020}.
Similarly, tactile input was performed on touchscreen handheld devices, e.g., smartphone~\cite{munger_simulation_2014} or tablet~\cite{meschtscherjakov_active_2016}, and a wearable smartwatch~\cite{perlman_relative_2019}.
Besides, physiological input modalities were measured using wearable sensors at nomadic locations, e.g., EDA on the forearm~\cite{angelini_opportunistic_2013}, skin temperature on the fingers~\cite{wang_drivers_2020}, brain activity on the head\cite{bi_using_2014}, and heart activity on ear~~\cite{pakdamanian_deeptake_2021} or torso~\cite{rony_development_2020}.
Regarding output, we found various approaches that utilized nomadic displays, such as AR~\cite{korthauer_watch_2020}, VR~\cite{mcgill_i_2017}, smartphone~\cite{he_does_2018}, tablet~\cite{oliveira_evaluating_2018}, or smartwatch~\cite{li_combined_2018}.
However, our SLR also revealed some approaches that provided tactile output at wearable locations via, e.g., smart ring~\cite{gheran_controls_2020}, vibration wristband~\cite{politis_beep_2015}, vibration belt~\cite{kruger_approach_2018}, or tactor belt~\cite{politis_evaluating_2014}.

\textbf{Anchored:}
The anchored interaction location contains specific locations throughout the vehicle interior: front, rear, door, ceiling, floor, table, and seat (see~\autoref{fig:teaser}).
To account for the large number of publications focusing on front-based in-vehicle interaction, similar to~\cite{kern_design_2009}, we further differentiate the front into windshield, dashboard, head-up display, center stack, and steering wheel.
Our synthesis showed that input modalities were mainly sensed via vehicle sensors located in the front.
For example, cameras for visual input, such as gaze~\cite{george_daaria_2012}, blinking~\cite{mahajan_exploring_2021}, or pupil dilation~\cite{mahajan_exploring_2021}, or kinesthetic input, such as gesture~\cite{detjen_driving_2021}, facial expression~\cite{huang_face2multi-modal_2020}, or head movement~\cite{robbins_comparing_2019}.
Similarly, microphones were placed in the front to sense auditory input (e.g., speech~\cite{martelaro_exploration_2019} or humming~\cite{funk_non-verbal_2020}).
Other input modalities often require direct contact with an anchored interior location.
For example, kinesthetic input at the center stack~\cite{manawadu_multimodal_2017}, steering wheel~\cite{lee_-wheel_2015}, seat~\cite{kim_seated_2020}, table~\cite{ishiguro_leadingdisplay_2019}, or ceiling~\cite{ishiguro_leadingdisplay_2019}.
We also found approaches for tactile input at similar locations, e.g., dashboard~\cite{voinescu_utility_2020}, center stack~\cite{ng_evaluation_2017}, steering wheel~\cite{pfleging_multimodal_2012}, table~\cite{oliveira_evaluating_2018}, or door~\cite{hakkila_exploring_2014}.
Only a few publications considered novel input modalities at anchored locations, such as EDA~\cite{hernandez_autoemotive_2014}, skin temperature~\cite{hernandez_autoemotive_2014}, or heart rate~\cite{madrid_biometric_2018} (all located on the steering wheel).
Regarding output, we found that out of the 12 anchored locations, only rear and floor were not utilized for visual output.
Besides, output was mainly provided using auditory modalities via speakers in the front (e.g.,~\cite{petermeijer_take-over_2017}), and via kinesthetic modalities at the center stack (e.g.,~\cite{heijboer_physical_2019}), steering wheel (e.g.,~\cite{huang_research_2015}), seat (e.g.,~\cite{grah_dorsal_2015}), or floor~\cite{henzler_are_2015}.
Similarly, tactile output was provided at the center stack (e.g.,~\cite{vo_investigating_2020}), steering wheel (e.g.,~\cite{schieben_evaluation_2014}), or seat (e.g.,~\cite{gaspar_examining_2015}).
Only a few publications considered other output modalities at anchored locations, e.g., vestibular output at the seat~\cite{winner_effects_2016}, scent output in the front~\cite{schartmuller_sick_2020}, or thermal output on the center stack~\cite{di_campli_san_vito_investigation_2018} or steering wheel~\cite{di_campli_san_vito_haptic_2019}.

\subsection{Multimodal Interaction}
\label{multimodal interaction}
Multimodal interfaces provide multiple modes of interaction between passengers and vehicle.
We do not distinguish between sequential multimodal, requiring users to switch between modes, and simultaneous multimodal that allows users to use multiple modes at a time.
Due to visualization complexity, we do not consider modality changes, e.g., when a passenger uses gaze input and the vehicle responds with speech output.
Instead, our interactive website can filter publications that considered such modality changes from our publication data set. 
To visualize the found approaches for multimodal interaction, we created a square matrix of order 12 representing combinations between modalities of any two sensor/actuator categories (see~\autoref{tab:multimodal_interaction}).
Both dimensions contain the sensor/actuator categories: visual, auditory, vestibular, kinesthetic, electrodermal, tactile, thermal, pain, olfactory, gustatory, cerebral, and cardiac.
We further differentiate between specific modalities within these categories, e.g., speech, sound, and music within the auditory category.
As in Section~\ref{input modalities}~and~\ref{output modalities}, we consider input and output from the human perspective, i.e., input is produced by human actuators (e.g., hands), and output is perceived by human sensors (e.g., skin).
A matrix cell contains a combination of specific modalities, e.g., display $\times$ sound or gesture $\times$ touch $\times$ lever, and an empty cell indicates that we found no approach for the respective multimodal interaction in our SLR.
Gray-colored cells highlight that combinations may not be feasible due to ethical (pain) or technical reasons (vestibular input, gustatory input, brain stimuli, and heart stimuli).
\begin{table*}[ht]
\centering
\caption{Combination matrix for multimodal in-vehicle interaction. Green-colored cells show combinations of multimodal input, and yellow-colored cells show multimodal output. The combinations highlighted in gray make technically no sense or are unethical.}
\resizebox{\textwidth}{!}{%
\begin{tabular}{ccc|l|l|l|l|l|l|l|l|l|l|l|l|}
\cline{4-15}
\multicolumn{1}{l}{} & \multicolumn{1}{l}{} & \multicolumn{1}{l|}{} & \multicolumn{1}{c|}{} & \multicolumn{1}{c|}{} & \multicolumn{6}{c|}{\textbf{Haptic}} & \multicolumn{1}{c|}{} & \multicolumn{1}{c|}{} & \multicolumn{1}{c|}{} & \multicolumn{1}{c|}{} \\ \cline{6-11}
\multicolumn{1}{l}{} & \multicolumn{1}{l}{} & \multicolumn{1}{l|}{} & \multicolumn{1}{c|}{} & \multicolumn{1}{c|}{} & \multicolumn{1}{c|}{} & \multicolumn{1}{c|}{} & \multicolumn{4}{c|}{\textbf{Cutaneous}} & \multicolumn{1}{c|}{} & \multicolumn{1}{c|}{} & \multicolumn{1}{c|}{} & \multicolumn{1}{c|}{} \\ \cline{8-11}
\multicolumn{1}{l}{} & \multicolumn{1}{l}{} & \multicolumn{1}{l|}{} & \multicolumn{1}{c|}{\multirow{-3}{*}{\textbf{Visual}}} & \multicolumn{1}{c|}{\multirow{-3}{*}{\textbf{Auditory}}} & \multicolumn{1}{c|}{\multirow{-2}{*}{\textbf{Vestibular}}} & \multicolumn{1}{c|}{\multirow{-2}{*}{\textbf{Kinesthetic}}} & \multicolumn{1}{c|}{\textbf{Electrodermal}} & \multicolumn{1}{c|}{\textbf{Tactile}} & \multicolumn{1}{c|}{\textbf{Thermal}} & \multicolumn{1}{c|}{\textbf{Pain}} & \multicolumn{1}{c|}{\multirow{-3}{*}{\textbf{Olfactory}}} & \multicolumn{1}{c|}{\multirow{-3}{*}{\textbf{Gustatory}}} & \multicolumn{1}{c|}{\multirow{-3}{*}{\textbf{Cerebral}}} & \multicolumn{1}{c|}{\multirow{-3}{*}{\textbf{Cardiac}}} \\ \hline
\multicolumn{3}{|c|}{\textbf{Visual}} & \cellcolor[HTML]{666666} & \cellcolor[HTML]{FFF2CC}\begin{tabular}[c]{@{}l@{}}\textbf{Speech x Display} \cite{reimer_effects_2014,chang_using_2017,large_please_2019,macek_mostly_2013}\\ \cite{cohen-lazry_impact_2020,williams_affective_2014,tchankue_impact_2011,walch_click_2018,walch_cooperative_2019,walch_touch_2017,lin_adasa_2018,choi_designing_2018}\\ \cite{clark_directability_2019,schneeberger_tailoring_2015,sykes_human_2020,forster_increasing_2017,hofmann_comparison_2014,jeon_multimodal_2019},\\ \textbf{Speech x Robotic Companion}\\ \cite{lee_autonomous_2019,kraus_human_2016,williams_affective_2014,yang_affective_2013,karatas_namida_2016},\\ \textbf{Voice x Display} \cite{kim_visual_2019,mitsopoulos-rubens_effects_2011,braun_what_2020},\\ \textbf{Speech x LED} \cite{politis_evaluation_2018,shakeri_may_2018,sun_improvement_2021,clark_directability_2019},\\ \textbf{Earcon x Display} \cite{shakeri_novel_2017,gang_dont_2018,jeon_multimodal_2019},\\ \textbf{Earcon x LED} \cite{shakeri_novel_2017},\\ \textbf{Display x Music} \cite{angelini_gesturing_2014,wang_drivers_2020},\\ \textbf{LED x Sound} \cite{shakeri_bimodal_2017,shakeri_may_2018,lees_cross-modal_2012}\\ \cite{huang_research_2015,van_den_beukel_driving_2016,van_den_beukel_supporting_2016,sadeghian_borojeni_feel_2018,borojeni_assisting_2016,borojeni_reading_2018,dmitrenko_towards_2019},\\ \textbf{Display x Sound} \cite{shakeri_may_2018,doring_gestural_2011,walch_click_2018}\\ \cite{ng_investigating_2016,walch_towards_2016,wang_watch_2020,rumelin_free-hand_2013,houtenbos_concurrent_2017,politis_beep_2015,topliss_establishing_2018,di_lena_-vehicle_2017}\\ \cite{van_den_beukel_driving_2016,forster_increasing_2017,van_den_beukel_supporting_2016},\\ \textbf{Speech x HUD} \cite{jakus_user_2015,li_evaluation_2018}\\ \cite{wang_design_2014,clark_directability_2019,politis_beep_2015,li_evaluation_2019},\\ \textbf{Sound x HUD} \cite{lees_cross-modal_2012,horberry_human-centered_2021},\\\textbf{ Windshield AR x Sound} \cite{topliss_establishing_2018,schewe_ecological_2020}\end{tabular} & \cellcolor[HTML]{FFF2CC} & \cellcolor[HTML]{FFF2CC}\begin{tabular}[c]{@{}l@{}}\textbf{LED x Shape Changing}\\ \cite{heijboer_physical_2019,mok_reinventing_2017},\\ \textbf{Display x Seat} \\ \textbf{Movement} \cite{pfleging_multimodal_2012,sun_shaping_2021},\\ \textbf{LED x Steering Torque} \cite{huang_research_2015},\\ \textbf{Display x Haptic Pedal} \cite{henzler_are_2015},\\ \textbf{Shape Changing} \cite{mok_actions_2017,kerschbaum_transforming_2015}\end{tabular} & \cellcolor[HTML]{FFF2CC} & \cellcolor[HTML]{FFF2CC}\begin{tabular}[c]{@{}l@{}}\textbf{Display x Vibration}\\ \cite{politis_evaluation_2018,quintal_hapwheel_2021,eriksson_rolling_2019,ng_investigating_2016,ng_evaluation_2017-1,huber_towards_2016,van_den_beukel_driving_2016,van_den_beukel_supporting_2016},\\ \textbf{LED x Vibration}\\ \cite{politis_evaluation_2018,loehmann_heartbeat_2014,van_den_beukel_driving_2016,van_den_beukel_supporting_2016,dmitrenko_towards_2019},\\ \textbf{Display x Cutanous Push} \cite{shakeri_novel_2017},\\ \textbf{LED x Cutaneous Push}\\ \cite{shakeri_novel_2017,shakeri_bimodal_2017,lees_cross-modal_2012},\\ \textbf{Display x Ultrasound}\\ \cite{shakeri_may_2018,georgiou_haptic_2017},\\ \textbf{LED x Ultrasound} \cite{shakeri_may_2018},\\ \textbf{Windshield AR x Vibration} \cite{eriksson_rolling_2019},\\ \textbf{LED x Shape Changing} \cite{heijboer_physical_2019,mok_reinventing_2017},\\ \textbf{HUD x Cutaneous Push} \cite{lees_cross-modal_2012,politis_evaluating_2014},\\ \textbf{HUD x Vibration} \cite{politis_beep_2015,horberry_human-centered_2021},\\ \textbf{Shape Changing} \cite{mok_actions_2017,kerschbaum_transforming_2015}\end{tabular} & \cellcolor[HTML]{FFF2CC}\begin{tabular}[c]{@{}l@{}}\textbf{LED x Peltier}\\ \textbf{Element} \cite{heijboer_physical_2019},\\ \textbf{Display x Seat}\\ \textbf{Heating} \cite{pfleging_multimodal_2012}\end{tabular} & \cellcolor[HTML]{D9D9D9} & \cellcolor[HTML]{FFF2CC}\begin{tabular}[c]{@{}l@{}}\textbf{Display x Air}\\ \textbf{Conditioning}\\ \cite{pfleging_multimodal_2012},\\ \textbf{HUD x}\\ \textbf{Scent} \cite{dmitrenko_i_2018},\\ \textbf{LED x}\\ \textbf{Scent} \cite{dmitrenko_towards_2019}\end{tabular} & \cellcolor[HTML]{FFF2CC} & \cellcolor[HTML]{D9D9D9} & \cellcolor[HTML]{D9D9D9} \\ \hline
\multicolumn{3}{|c|}{\textbf{Auditory}} & \cellcolor[HTML]{D9EAD3}\begin{tabular}[c]{@{}l@{}}\textbf{Gaze x Voice} \cite{clark_directability_2019}, \textbf{Pupil}\\ \textbf{Diameter x Eye Blink x}\\ \textbf{Glance x Voice} \cite{mahajan_exploring_2021}, \textbf{Gaze x}\\ \textbf{Speech} \cite{roider_just_2018,aftab_multimodal_2019,wang_watch_2020,kim_cascaded_2020}\end{tabular} & \cellcolor[HTML]{666666} & \cellcolor[HTML]{FFF2CC} & \cellcolor[HTML]{FFF2CC}\begin{tabular}[c]{@{}l@{}}\textbf{Sound x Steering Torque} \cite{huang_research_2015},\\ \textbf{Sound x Seat Rotation} \cite{sun_shaping_2021}\end{tabular} & \cellcolor[HTML]{FFF2CC} & \cellcolor[HTML]{FFF2CC}\begin{tabular}[c]{@{}l@{}}\textbf{Cutaneous Push}\\ \textbf{x Earcon} \cite{shakeri_novel_2017},\\ \textbf{x Sound} \cite{shakeri_bimodal_2017,lees_cross-modal_2012,politis_evaluating_2014},\\ \textbf{x Speech} \cite{jung_voicetactile_2020},\\ \textbf{Ultrasound}\\ \textbf{x Sound x Speech} \cite{shakeri_may_2018},\\ \textbf{Vibration x Sound}\\ \cite{ng_investigating_2016,petermeijer_take-over_2017,horberry_human-centered_2021,van_den_beukel_driving_2016,geitner_comparison_2019,dmitrenko_towards_2019},\\ \textbf{Vibration x Speech} \cite{politis_beep_2015}\end{tabular} & \cellcolor[HTML]{FFF2CC} & \cellcolor[HTML]{D9D9D9} & \cellcolor[HTML]{FFF2CC}\begin{tabular}[c]{@{}l@{}}\textbf{Scent x}\\ \textbf{Sound} \cite{dmitrenko_towards_2019}\end{tabular} & \cellcolor[HTML]{FFF2CC} & \cellcolor[HTML]{D9D9D9} & \cellcolor[HTML]{D9D9D9} \\ \hline
\multicolumn{1}{|c|}{} & \multicolumn{2}{c|}{\textbf{Vestibular}} & \cellcolor[HTML]{D9D9D9} & \cellcolor[HTML]{D9D9D9} & \cellcolor[HTML]{666666} & \cellcolor[HTML]{FFF2CC} & \cellcolor[HTML]{FFF2CC} & \cellcolor[HTML]{FFF2CC} & \cellcolor[HTML]{FFF2CC} & \cellcolor[HTML]{D9D9D9} & \cellcolor[HTML]{FFF2CC} & \cellcolor[HTML]{FFF2CC} & \cellcolor[HTML]{D9D9D9} & \cellcolor[HTML]{D9D9D9} \\ \cline{2-15} 
\multicolumn{1}{|c|}{} & \multicolumn{2}{c|}{\textbf{Kinesthetic}} & \cellcolor[HTML]{D9EAD3}\begin{tabular}[c]{@{}l@{}}\textbf{Gaze x Finger Pointing x}\\ \textbf{Head Pose} \cite{aftab_multimodal_2019,rumelin_free-hand_2013,aftab_you_2020},\\ \textbf{Gaze x Button} \cite{wang_watch_2020,kim_cascaded_2020,lotz_response_2019},\\ \textbf{Gaze x Gesture} \cite{kim_cascaded_2020,rumelin_free-hand_2013,ahmad_predictive_2018},\\ \textbf{Eye x Head Movement} \cite{robbins_comparing_2019},\\ \textbf{Gaze x Hand Pointing} \cite{gomaa_studying_2020},\\ \textbf{Gaze x Head Pose} \cite{amadori_decision_2020,lotz_response_2019},\\ \textbf{Eye x Facial Landmarks x}\\ \textbf{Body Pose x Foot Pose} \cite{rangesh_exploring_2018}\end{tabular} & \cellcolor[HTML]{D9EAD3}\begin{tabular}[c]{@{}l@{}}\textbf{Speech x Button}\\ \cite{reimer_effects_2014,li_evaluation_2018,pfleging_multimodal_2012,wang_design_2014,wang_watch_2020,kim_cascaded_2020},\\ \textbf{Voice x Gesture} \cite{winzer_intersection_2018},\\ \textbf{Speech x Gesture} \cite{pfleging_multimodal_2012,kim_cascaded_2020},\\ \textbf{Speech x Emotion} \cite{zepf_towards_2019},\\ \textbf{Speech x Head Pose x}\\ \textbf{Finger Pointing} \cite{aftab_multimodal_2019}\end{tabular} & \cellcolor[HTML]{D9D9D9} & \cellcolor[HTML]{666666} & \cellcolor[HTML]{FFF2CC} & \cellcolor[HTML]{FFF2CC} & \cellcolor[HTML]{FFF2CC}\begin{tabular}[c]{@{}l@{}}\textbf{Peltier Element}\\ \textbf{x Shape}\\ \textbf{Changing} \cite{heijboer_physical_2019}\end{tabular} & \cellcolor[HTML]{D9D9D9} & \cellcolor[HTML]{FFF2CC} & \cellcolor[HTML]{FFF2CC} & \cellcolor[HTML]{D9D9D9} & \cellcolor[HTML]{D9D9D9} \\ \cline{2-15} 
\multicolumn{1}{|c|}{} & \multicolumn{1}{c|}{} & \textbf{\begin{tabular}[c]{@{}c@{}}\textbf{Electro-}\\ \textbf{dermal}\end{tabular}} & \cellcolor[HTML]{D9EAD3}\begin{tabular}[c]{@{}l@{}}\textbf{Eye Movement x}\\ \textbf{GSR} \cite{pakdamanian_deeptake_2021,dillen_keep_2020}\end{tabular} & \cellcolor[HTML]{D9EAD3} & \cellcolor[HTML]{D9D9D9} & \cellcolor[HTML]{D9EAD3}\textbf{Body Posture x GSR} \cite{madrid_biometric_2018} & \cellcolor[HTML]{666666} & \cellcolor[HTML]{FFF2CC} & \cellcolor[HTML]{FFF2CC} & \cellcolor[HTML]{D9D9D9} & \cellcolor[HTML]{FFF2CC} & \cellcolor[HTML]{FFF2CC} & \cellcolor[HTML]{D9D9D9} & \cellcolor[HTML]{D9D9D9} \\ \cline{3-15} 
\multicolumn{1}{|c|}{} & \multicolumn{1}{c|}{} & \textbf{Tactile} & \cellcolor[HTML]{D9EAD3}\textbf{Gaze x Touch} \cite{kim_cascaded_2020} & \cellcolor[HTML]{D9EAD3}\begin{tabular}[c]{@{}l@{}}\textbf{Speech x Touch} \\ \cite{pfleging_multimodal_2012,jung_voicetactile_2020,large_please_2019,goulati_user_2011,perlman_relative_2019,kim_cascaded_2020}\end{tabular} & \cellcolor[HTML]{D9D9D9} & \cellcolor[HTML]{D9EAD3}\begin{tabular}[c]{@{}l@{}}\textbf{Rotary Controller x Touch x}\\ \textbf{Button} \cite{schartmuller_type-o-steer_2019}, \textbf{Hand Pointing x}\\ \textbf{Touch x Button} \cite{ahmad_selection_2018,kim_cascaded_2020},\\ \textbf{Gesture x Touch} \cite{aslan_leap_2015,kim_cascaded_2020},\\ \textbf{Gesture x Touch x Lever} \cite{manawadu_multimodal_2017},\\ \textbf{Finger Movement x Touch} \cite{lauber_what_2014}\end{tabular} & \cellcolor[HTML]{D9EAD3} & \cellcolor[HTML]{666666} & \cellcolor[HTML]{FFF2CC}\begin{tabular}[c]{@{}l@{}}\textbf{Peltier Element}\\ \textbf{x Shape}\\ \textbf{Changing} \cite{heijboer_physical_2019}\end{tabular} & \cellcolor[HTML]{D9D9D9} & \cellcolor[HTML]{FFF2CC}\begin{tabular}[c]{@{}l@{}}\textbf{Scent x}\\ \textbf{Vibration} \cite{dmitrenko_towards_2019}\end{tabular} & \cellcolor[HTML]{FFF2CC} & \cellcolor[HTML]{D9D9D9} & \cellcolor[HTML]{D9D9D9} \\ \cline{3-15} 
\multicolumn{1}{|c|}{} & \multicolumn{1}{c|}{} & \textbf{Thermal} & \cellcolor[HTML]{D9EAD3} & \cellcolor[HTML]{D9EAD3} & \cellcolor[HTML]{D9D9D9} & \cellcolor[HTML]{D9EAD3} & \cellcolor[HTML]{D9EAD3}\begin{tabular}[c]{@{}l@{}}\textbf{EDA x}\\ \textbf{Skin Temp.} \cite{wang_drivers_2020}\end{tabular} & \cellcolor[HTML]{D9EAD3} & \cellcolor[HTML]{666666} & \cellcolor[HTML]{D9D9D9} & \cellcolor[HTML]{FFF2CC} & \cellcolor[HTML]{FFF2CC} & \cellcolor[HTML]{D9D9D9} & \cellcolor[HTML]{D9D9D9} \\ \cline{3-15} 
\multicolumn{1}{|c|}{\multirow{-20}{*}{\rotatebox[origin=c]{90}{\textbf{Haptic}}}} & \multicolumn{1}{c|}{\multirow{-10}{*}{\rotatebox[origin=c]{90}{\textbf{Cutaneous}}}} & \textbf{Pain} & \cellcolor[HTML]{D9D9D9} & \cellcolor[HTML]{D9D9D9} & \cellcolor[HTML]{D9D9D9} & \cellcolor[HTML]{D9D9D9} & \cellcolor[HTML]{D9D9D9} & \cellcolor[HTML]{D9D9D9} & \cellcolor[HTML]{D9D9D9} & \cellcolor[HTML]{666666} & \cellcolor[HTML]{D9D9D9} & \cellcolor[HTML]{D9D9D9} & \cellcolor[HTML]{D9D9D9} & \cellcolor[HTML]{D9D9D9} \\ \hline
\multicolumn{3}{|c|}{\textbf{Olfactory}} & \cellcolor[HTML]{D9EAD3} & \cellcolor[HTML]{D9EAD3} & \cellcolor[HTML]{D9D9D9} & \cellcolor[HTML]{D9EAD3} & \cellcolor[HTML]{D9EAD3} & \cellcolor[HTML]{D9EAD3} & \cellcolor[HTML]{D9EAD3} & \cellcolor[HTML]{D9D9D9} & \cellcolor[HTML]{666666} & \cellcolor[HTML]{FFF2CC} & \cellcolor[HTML]{D9D9D9} & \cellcolor[HTML]{D9D9D9} \\ \hline
\multicolumn{3}{|c|}{\textbf{Gustatory}} & \cellcolor[HTML]{D9D9D9} & \cellcolor[HTML]{D9D9D9} & \cellcolor[HTML]{D9D9D9} & \cellcolor[HTML]{D9D9D9} & \cellcolor[HTML]{D9D9D9} & \cellcolor[HTML]{D9D9D9} & \cellcolor[HTML]{D9D9D9} & \cellcolor[HTML]{D9D9D9} & \cellcolor[HTML]{D9D9D9} & \cellcolor[HTML]{666666} & \cellcolor[HTML]{D9D9D9} & \cellcolor[HTML]{D9D9D9} \\ \hline
\multicolumn{3}{|c|}{\textbf{Cerebral}} & \cellcolor[HTML]{D9EAD3} & \cellcolor[HTML]{D9EAD3} & \cellcolor[HTML]{D9D9D9} & \cellcolor[HTML]{D9EAD3}Head \textbf{Movement x EEG} \cite{li_combined_2018} & \cellcolor[HTML]{D9EAD3} & \cellcolor[HTML]{D9EAD3} & \cellcolor[HTML]{D9EAD3} & \cellcolor[HTML]{D9D9D9} & \cellcolor[HTML]{D9EAD3} & \cellcolor[HTML]{D9D9D9} & \cellcolor[HTML]{666666} & \cellcolor[HTML]{D9D9D9} \\ \hline
\multicolumn{3}{|c|}{\textbf{Cardiac}} & \cellcolor[HTML]{D9EAD3}\begin{tabular}[c]{@{}l@{}}\textbf{Eye Movement x}\\ \textbf{Heart Rate} \cite{pakdamanian_deeptake_2021,dillen_keep_2020}\end{tabular} & \cellcolor[HTML]{D9EAD3}\textbf{Speech x Heart Rate} \cite{large_assessing_2016} & \cellcolor[HTML]{D9D9D9} & \cellcolor[HTML]{D9EAD3}\begin{tabular}[c]{@{}l@{}}\textbf{Body Posture x Heart Rate}\\ \cite{madrid_biometric_2018,hayashi_development_2020}, \textbf{Head Pose x Heart}\\ \textbf{Rate} \cite{hayashi_development_2020}\end{tabular} & \cellcolor[HTML]{D9EAD3}\begin{tabular}[c]{@{}l@{}}\textbf{EDA x ECG} \cite{wang_drivers_2020},\\ \textbf{Heart Rate x}\\ \textbf{GSR} \cite{madrid_biometric_2018,pakdamanian_deeptake_2021,dillen_keep_2020}\end{tabular} & \cellcolor[HTML]{D9EAD3} & \cellcolor[HTML]{D9EAD3}\textbf{Skin Temp. x ECG} \cite{wang_drivers_2020} & \cellcolor[HTML]{D9D9D9} & \cellcolor[HTML]{D9EAD3} & \cellcolor[HTML]{D9D9D9} & \cellcolor[HTML]{D9EAD3} & \cellcolor[HTML]{666666} \\ \hline
\end{tabular}%
}
\label{tab:multimodal_interaction}
\end{table*}

In our synthesis, we found that most publications utilized visual modalities for \textbf{multimodal input}.
Gaze was combined with auditory inputs, e.g., speech~\cite{roider_just_2018} and voice~\cite{clark_directability_2019}, with kinesthetic inputs, e.g., finger pointing~\cite{aftab_multimodal_2019}, hand gestures~\cite{gomaa_studying_2020}, head pose~\cite{amadori_decision_2020}, and body pose~\cite{rangesh_exploring_2018}, or with tactile touch input~\cite{kim_cascaded_2020}.
Besides, some publications combined eye movement with electrodermal (GSR~\cite{dillen_keep_2020}) or with cardiac modalities (heart rate~\cite{pakdamanian_deeptake_2021}).
Moreover, 30 publications used auditory, kinesthetic, or tactile modalities for multimodal input.
Regarding auditory input, speech was used dominantly in combination with, e.g., button~\cite{li_evaluation_2018}, gesture~\cite{pfleging_multimodal_2012}, emotion~\cite{zepf_towards_2019}, touch~\cite{perlman_relative_2019}, or heart rate~\cite{large_assessing_2016}.
Kinesthetic modalities were used diversely, e.g., combined with GSR~\cite{madrid_biometric_2018}, touch~\cite{ahmad_selection_2018}, EEG~\cite{li_combined_2018}, or heart rate~\cite{hayashi_development_2020}.
Only four publications utilized electrodermal or thermal input modalities for multimodal interaction.
For example, skin temperature was used with EDA~\cite{wang_drivers_2020}, or in combination with electrocardiogram~\cite{wang_drivers_2020}.
We did not find any publications that considered multimodal input using olfactory modalities.

Regarding \textbf{multimodal output}, we found that most publications include visual modalities, mainly because their approaches required a display to visualize a task or an UI.
Accordingly, display was combined with various output modalities, e.g., with auditory (speech~\cite{tchankue_impact_2011}, sound~\cite{ng_investigating_2016}, music~\cite{angelini_gesturing_2014}), kinesthetic (haptic pedal~\cite{henzler_are_2015}, seat movement~\cite{pfleging_multimodal_2012}), or tactile (vibration~\cite{van_den_beukel_driving_2016}, ultrasound~\cite{georgiou_haptic_2017}, cutaneous push~\cite{shakeri_novel_2017}).
Other visual output modalities used were LED and HUD.
For example, LED was combined with speech~\cite{sun_improvement_2021}, sound~\cite{sadeghian_borojeni_feel_2018}, shape changing~\cite{heijboer_physical_2019}, vibration~\cite{loehmann_heartbeat_2014}, ultrasound~\cite{shakeri_may_2018}, Peltier element~\cite{heijboer_physical_2019}, or scent~\cite{dmitrenko_towards_2019}.
HUD was combined with, e.g., speech~\cite{politis_beep_2015}, sound~\cite{horberry_human-centered_2021}, vibration~\cite{politis_beep_2015}, cutaneous push~\cite{lees_cross-modal_2012}, or scent~\cite{dmitrenko_i_2018}.
Besides, publications used multimodal output that combined windshield AR with sound~\cite{schewe_ecological_2020} or vibration~\cite{eriksson_rolling_2019}.
We also found approaches for multimodal output that did not use visual output.
For example, auditory combined with kinesthetic output (sound $\times$ steering torque~\cite{huang_research_2015} and sound $\times$ seat rotation~\cite{sun_shaping_2021}) or with tactile output~\cite{jung_voicetactile_2020}.
We found that only four publications utilized thermal or olfactory modalities for multimodal output.
Peltier elements were used along shape-changing output~\cite{heijboer_physical_2019}, and scent was combined with sound~\cite{dmitrenko_towards_2019} or vibration~\cite{dmitrenko_towards_2019}.
We did not find any publications that considered multimodal output containing vestibular, electrodermal, or gustatory modalities.
\section{User Acceptance of In-Vehicle Interaction}
The SLR results and our design space highlight many opportunities for novel in-vehicle interaction approaches other than visual, auditory, and tactile.
However, it is unclear whether experimental modalities (e.g., electrodermal, cerebral, or cardiac) are feasible.
For instance, passengers might perceive modalities as invasive (e.g., cerebral) or unpleasant (e.g., thermal) and, therefore, reject usage.
Moreover, our SLR and design space revealed different interaction locations throughout the interior, with many opportunities for novel interaction locations other than the front.
Still, it is unknown whether users perceive interaction at such unfamiliar locations as practical, e.g., rear, floor, or ceiling.
To gain initial insights, we designed and conducted a within-subject online study with \N{48} participants.
The independent variables were \textit{interaction type} with two levels: input and output, \textit{modality} with 11 levels: visual, auditory, vestibular (only output), kinesthetic, electrodermal, tactile, thermal, olfactory, gustatory (only output), cerebral (only input), and cardiac (only input), and \textit{interaction location} with ten levels: AR, VR, handheld, wearable, seat, door, floor, rear, table, and ceiling.
Following our design space, we excluded impractical modalities (e.g., pain) and removed the front location, as user acceptance of this location is already well known to research.
The following RQ guided the exploratory study:
\textit{RQ~4: How do users perceive the (1)~usefulness, (2)~real-world usage, and (3)~comfort of different input/output modalities and interaction locations within a conceptual vehicle?}


\subsection{Study Procedure and Measurements}
\begin{figure}
        \centering
        \includegraphics[width=\textwidth]{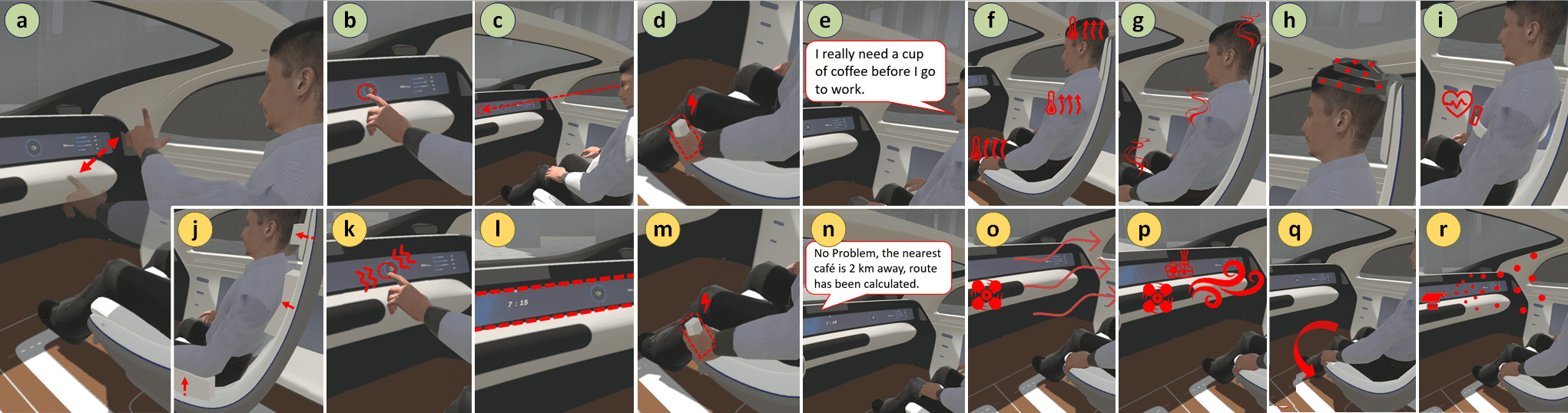}
    \caption{Concept images used in the online study. Input modalities: a)~gesture, b)~touch, c)~gaze, d)~electrodermal, e)~speech, f)~thermal, g)~body odor, h)~cerebral, and i)~cardiac. Output modalities: j)~shape changing, k)~vibration, l)~display, m)~electrodermal, n)~speech, o)~thermal, p)~scent, q)~seat rotation, and r)~gustatory.}
\label{fig:input_output_study_concepts}
\end{figure}

We opted for an image-based survey for two reasons.
First, we aimed for meta-level insights covering the mere concept of an interaction, disregarding profound functionalities of each modality.
Second, it is impractical to develop working prototypes for all 18 input/output modalities, while also using all 10 locations.
The study consisted of a questionnaire, starting with an introduction of aim and scenario.
In the following, there were two parts:
\textit{Part 1} showed 18 images (9 input and 9 output) of a passenger on the front seat using possible interaction modalities in a concept vehicle (e.g., see~\autoref{fig:input_output_study_concepts}).
\textit{Part 2} showed 20 images (10 input and 10 output) of a passenger on the front seat interacting at possible interaction locations within a concept vehicle (e.g., see~\autoref{fig:location_study_concepts}).
The full versions of the study images are visible in the Appendix~\autoref{fig:study_images_full_input},~\autoref{fig:study_images_full_output},~\autoref{fig:study_images_full_location_in}, and~\autoref{fig:study_images_full_location_out}.
For the interaction locations in \textit{Part 2}, we always showed gesture input, and for output, we used a display.
However, this was exemplary, and any other input or output modality from \textit{Part 1} may also be feasible.
\textit{Part 1} was introduced first to ensure that all concepts for input and output modalities were already known to the participants in \textit{Part 2}.
The images per part were presented in randomized order for each participant, and after each part, participants could provide open feedback.
Finally, there was a demographic and a concluding questionnaire.
On average, a session lasted 17 min and participants were compensated with £2.
\begin{figure}
        \centering
        \includegraphics[width=\textwidth]{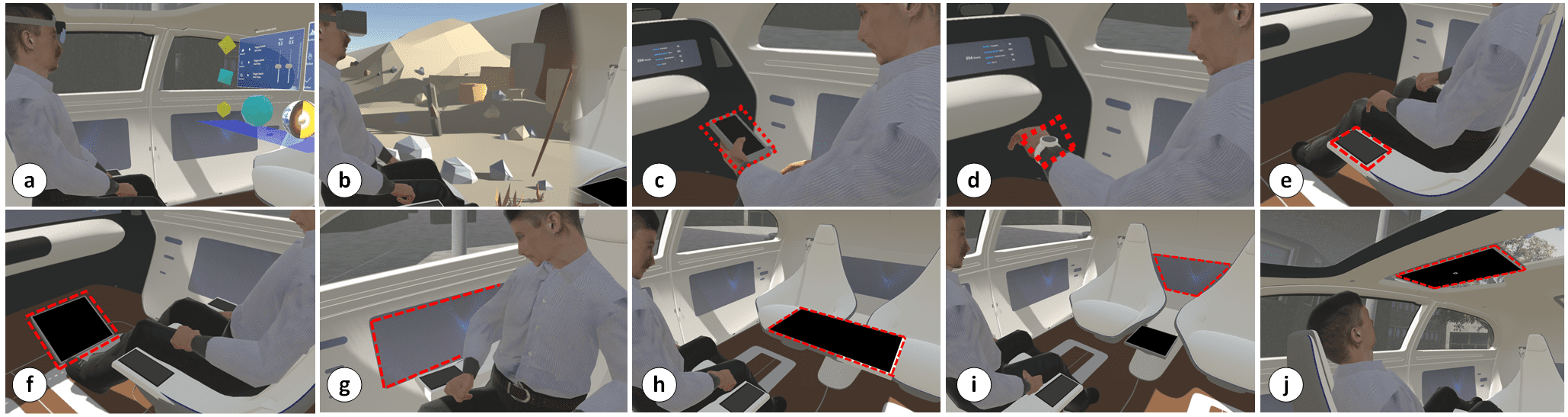}
    \caption{Concept images used in the online study. Nomadic interaction locations: a)~AR, b)~VR, c)~handheld, and d)~wearable. Anchored interaction locations: e)~seat, f)~floor, g)~door, h)~table, i)~rear, and j)~ceiling.}
\label{fig:location_study_concepts}
\end{figure}

After each image in \textit{Part 1}, a self-developed 7-point scale (\textit{1=strongly disagree} to \textit{7=strongly agree}) was used to assess the participants' perceived usefulness of the interaction ("I consider the presented interaction to be useful."), whether they could imagine real-world usage ("I would use the presented interaction in a real vehicle."), and the anticipated comfort of use ("I would feel comfortable using the presented interaction.").
For \textit{Part 2}, three adapted questions were shown after each image: "I consider the depicted interaction location to be useful.", "I would use the depicted interaction location in a real vehicle.", and "I would feel comfortable using the depicted interaction location.".

\subsection{Results}
\label{study_results}
Before the study, we computed the required sample size via a priori power analysis using G*Power~\cite{faul_gpower_2009}.
To achieve a power of 0.95 with $\alpha=0.05$, 37 participants should result in a small effect size (0.15~\cite{funder_effect_size_2019}) in a within-subjects ANOVA.
Originally, 62 individuals were recruited via \href{https://www.prolific.co/}{Prolific} (a platform to find participants~\cite{PALAN201822}, already used in automotive research, e.g.,~\cite{colley_prolific_2021}), but 14 had to be excluded due to failed attention checks (e.g., "Check the fourth option."). 
We recruited US citizens only to avoid confounding variables such as driving laws (e.g., left and right traffic).
Of the remaining 48 participants, 29 were male, 18 female, and one non-binary.
Participants were, on average, \m{34.25} (\sd{8.83}, between 19 and 50) years old and, in total, 44 participants had a driver license.
R in version 4.1.2, RStudio in version 2021.09.0 and, for non-parametric data (checked via normality distribution and homogeneity of variance assumption), the non-parametric ANOVA (NPAV) by Lüpsen~\cite{luepsen2020r} was employed.
All packages were up-to-date in January 2022.

\subsubsection{Input and Output Modalities}
\begin{figure*}[ht!]
    \centering
    \begin{minipage}{.33\textwidth}
    \centering
    \includegraphics[width=0.99\textwidth]{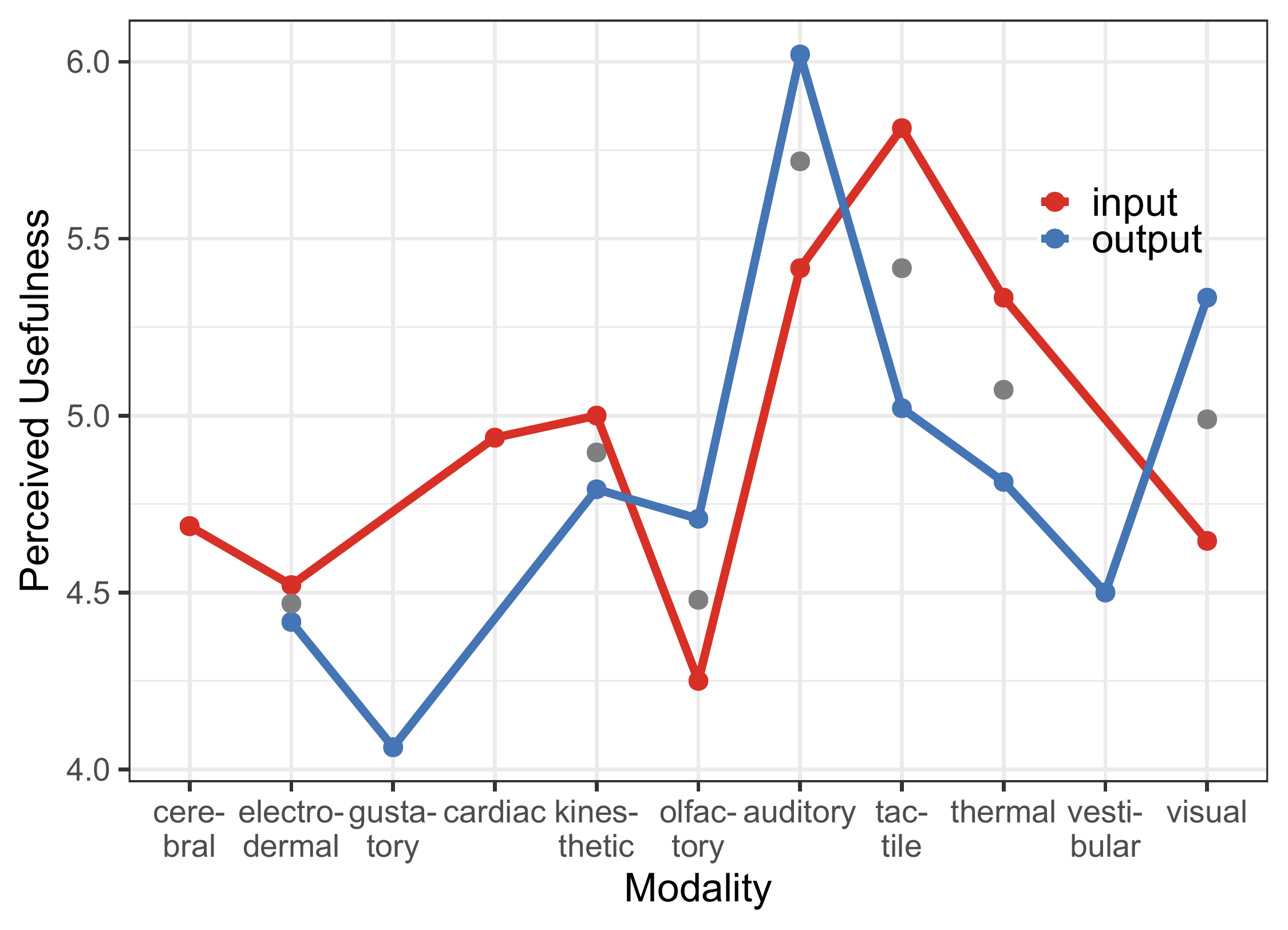}
    \caption{IE on perceived usefulness.}
    \label{fig:pu_mod}
    \end{minipage}
        \begin{minipage}{.33\textwidth}
        \centering
    \includegraphics[width=0.99\textwidth]{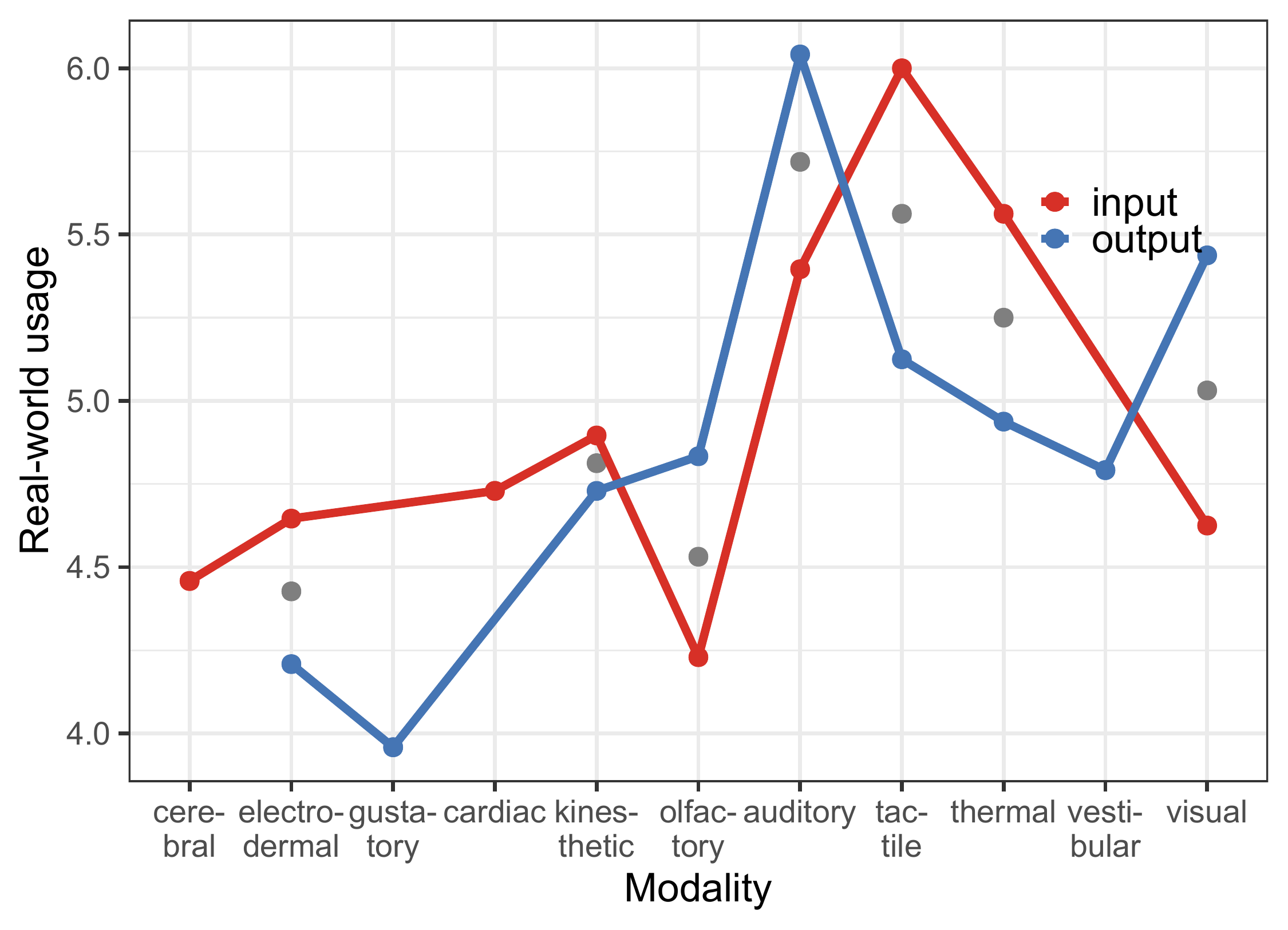}
    \caption{IE on real-world usage.}
    \label{fig:ru_mod}
    \end{minipage}
    \begin{minipage}{.33\textwidth}
    \centering
    \includegraphics[width=0.99\textwidth]{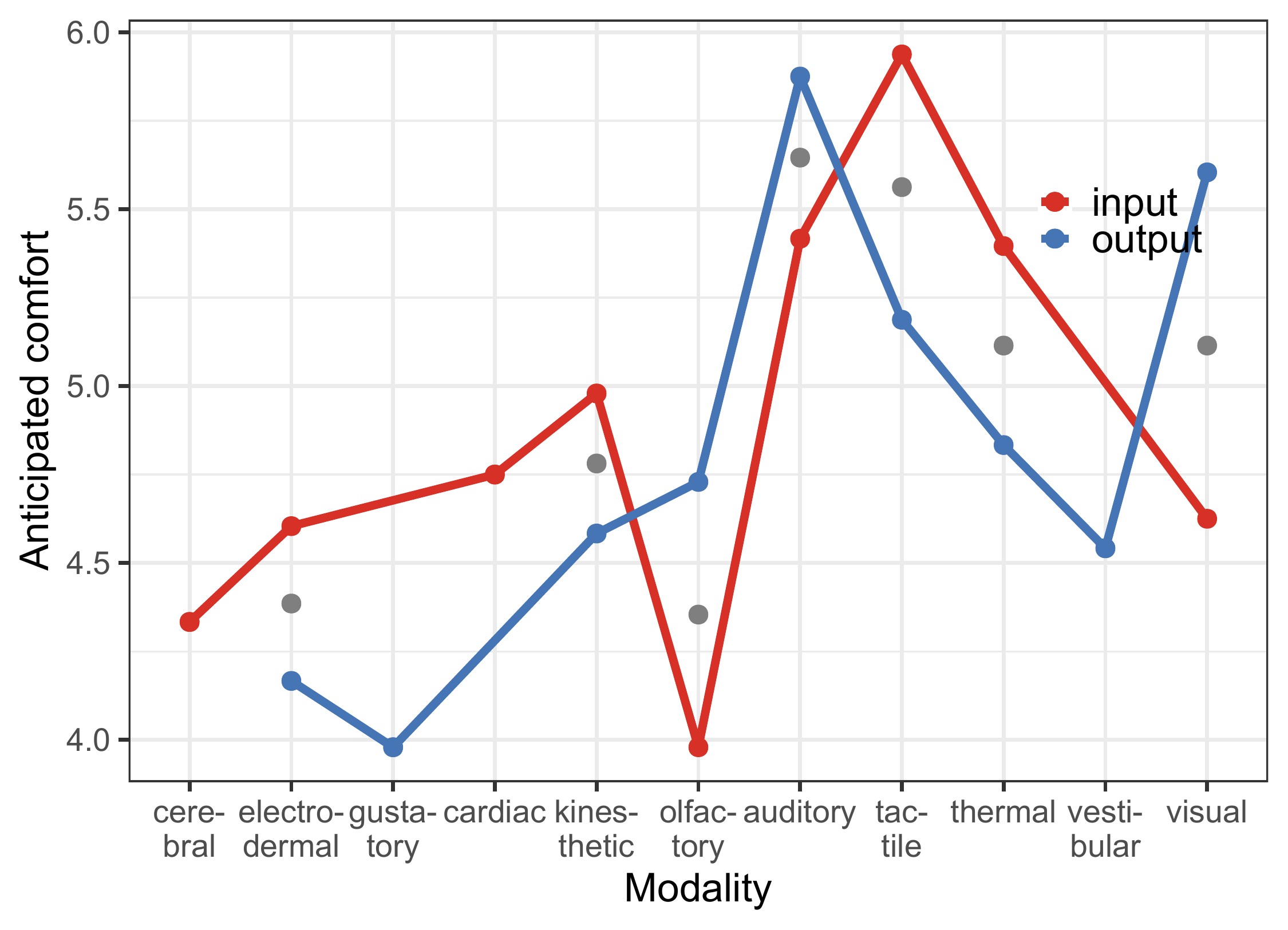}
    \caption{IE on anticipated comfort.}
    \label{fig:ac_mod}
    \end{minipage}
\end{figure*}

\autoref{fig:pu_mod} shows the average ratings of the perceived usefulness of a \textit{modality}.
The NPAV found a significant main effect of \textit{modality} on perceived usefulness (\F{10}{470}{8.45}, \pminor{0.001}).
The NPAV also found a significant interaction effect (IE) of \textit{interaction type} (input/output) $\times$ \textit{modality} on perceived usefulness (\F{6}{282}{6.16}, \pminor{0.001}).
\autoref{fig:ru_mod} shows the average ratings of the anticipated real-world usage of a \textit{modality}.
The NPAV found a significant main effect of \textit{modality} on real-world usage (\F{10}{470}{9.03}, \pminor{0.001}).
The NPAV additionally found a significant IE of \textit{interaction type} $\times$ \textit{modality} on real-world usage (\F{6}{282}{4.25}, \pminor{0.001}).
\autoref{fig:ac_mod} shows the average ratings of the anticipated comfort of a \textit{modality}.
The NPAV found a significant main effect of \textit{modality} on anticipated comfort (\F{10}{470}{10.62}, \pminor{0.001}).
Besides, the NPAV found a significant IE of \textit{interaction type} $\times$ \textit{modality} on anticipated comfort (\F{6}{282}{6.25}, \pminor{0.001}).

\subsubsection{Interaction Locations}
\begin{figure*}[ht!]
    \centering
        \begin{minipage}{.33\textwidth}
    \centering
    \includegraphics[width=0.95\textwidth]{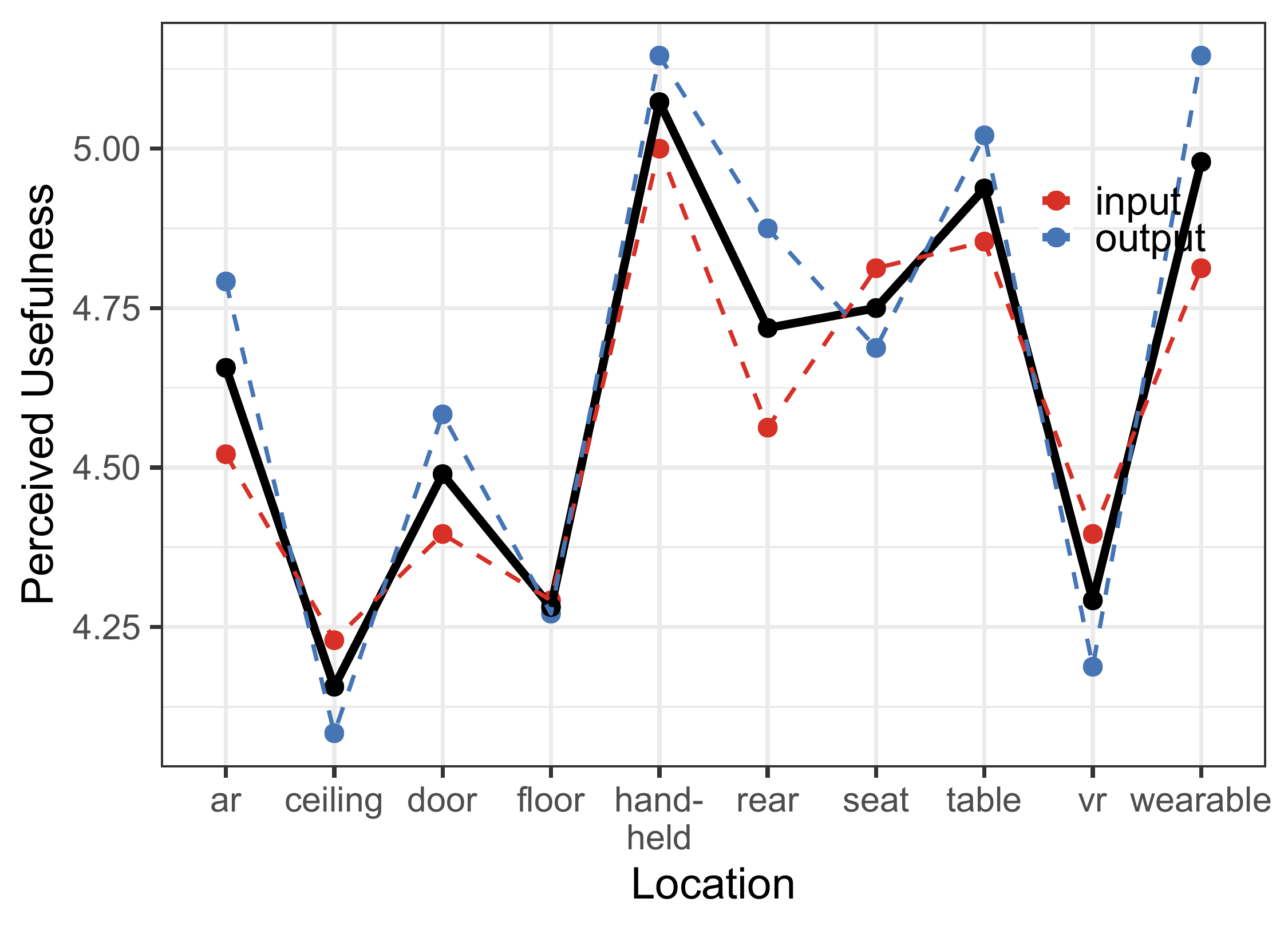}
    \caption{Effect on perceived usefulness.}
    \label{fig:pu_loc}
            \end{minipage}%
    \begin{minipage}{.33\textwidth}
    \centering
    \includegraphics[width=0.95\textwidth]{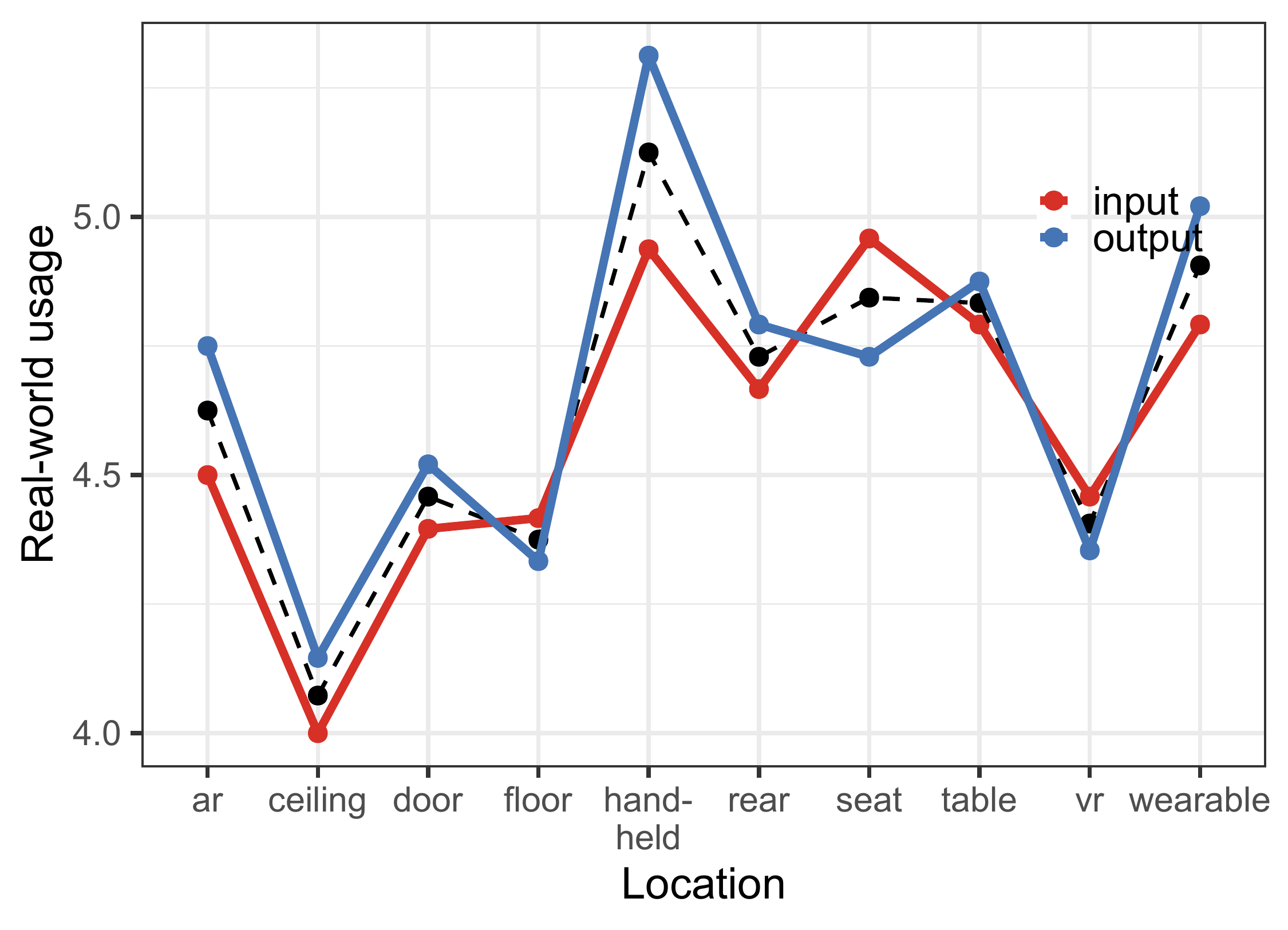}
    \caption{Effect on real-world usage.}
    \label{fig:ru_loc}
    \end{minipage}%
    \begin{minipage}{.33\textwidth}
        \centering
    \includegraphics[width=0.95\textwidth]{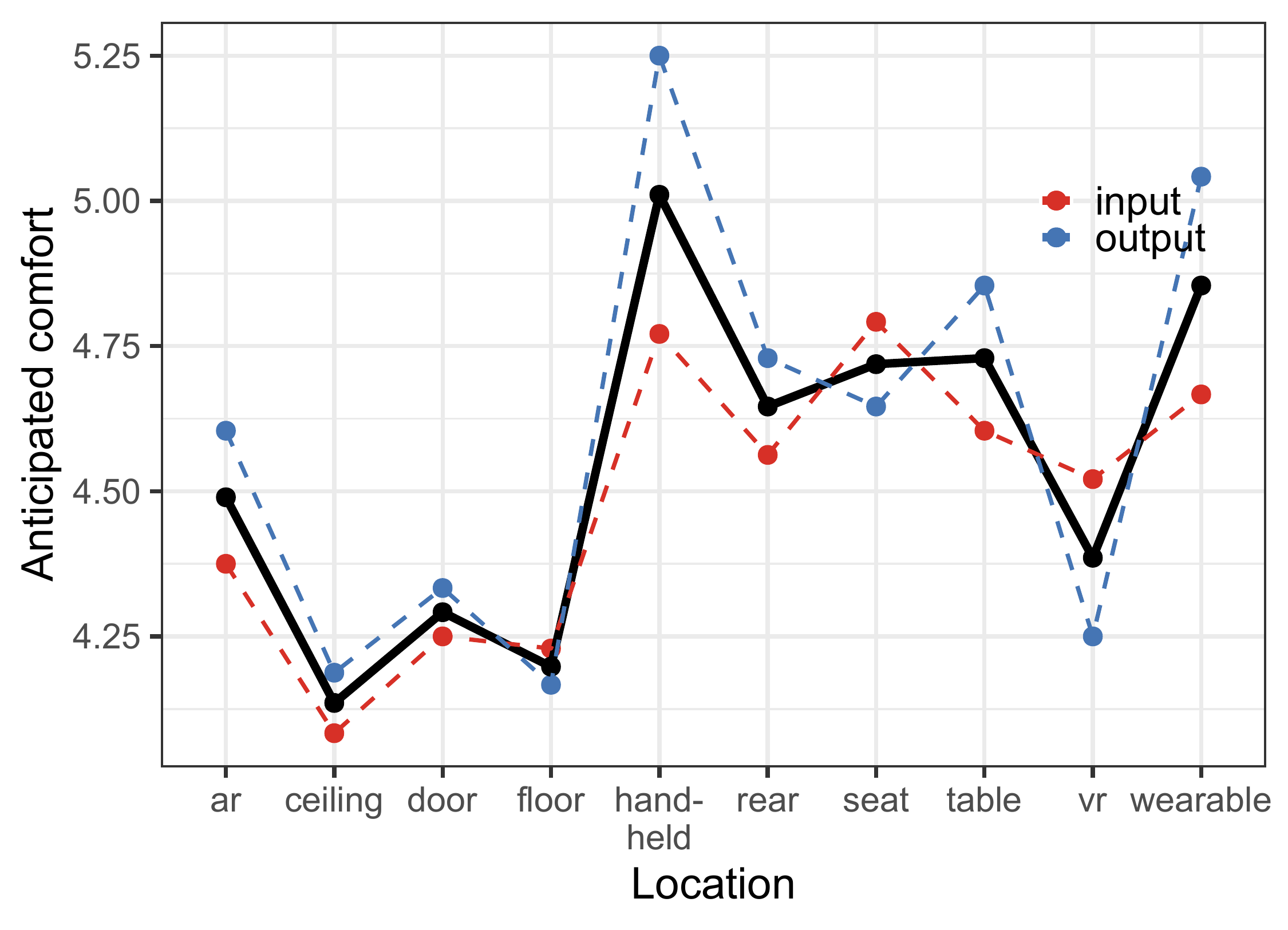}
    \caption{IE on anticipated comfort.}
    \label{fig:ac_loc}
        \end{minipage}%
\end{figure*}

\autoref{fig:pu_loc} shows the average ratings of the perceived usefulness of a \textit{location}.
The NPAV found a significant main effect of \textit{location} on perceived usefulness (\F{9}{423}{4.45}, \pminor{0.001}).
\autoref{fig:ru_loc} shows the average ratings of the perceived usefulness of a \textit{location}.
The NPAV found a significant main effect of \textit{location} on real-world usage (\F{9}{423}{3.69}, \pminor{0.001}).
\autoref{fig:ac_loc} shows the average ratings of the perceived usefulness of a \textit{location}.
The NPAV found a significant main effect of \textit{location} on anticipated comfort (\F{9}{423}{2.75}, \p{0.004}).

\subsubsection{Open Feedback}
Participants perceived the shown input and output modalities as interesting [P32] and comfortable [P19, P24, P25].
However, they also noted that, e.g., some inputs were especially invasive [P5], and the vehicle should not read minds [P11].
Besides, [P38] stated that except for speech output and vibration to acknowledge choices, all modalities are dangerous regardless if the user is a driver or passenger.
The shown interaction locations were perceived as interesting [P7], attractive [P25], and comfortable [P19] (e.g., for travel [P24]).
Still, [P5] noted that some locations generally seemed not comfortable, and even dangerous to interact with, in a moving vehicle.
[P11] thought that looking down for information (e.g., floor) is stressful, and [P38] would feel awkward using the majority of interaction locations in a car except leaning back and watching TV on the ceiling.

\section{Discussion}
In the following, we will discuss the results of the user study and our SLR and, based on these, outline future research interests in the in-vehicle interaction domain.

\subsection{Opportunities for Future In-Vehicle Interaction}
Our design space (see Section~\ref{extending_the_design_space}) revealed several underexplored opportunities for future in-vehicle interaction.
In this regard, insights from non-automotive HCI research may guide the creation of novel approaches.

Regarding \textbf{input modalities}, our design space revealed little to no utilization of thermal, olfactory, gustatory, cerebral, or cardiac modalities.
However, such implicit inputs might be helpful, e.g., for monitoring passengers, adapting outputs, or refining explicit interactions.
For example, the vehicle could measure the skin temperature employing wearable sensors similar to~\cite{welch_wearable_2018, mansor_body_2013} or without requiring invasive on-body sensors using thermal cameras~\cite{abdelrahman_cognitive_2017}, which might increase comfort and acceptance.
Besides, the vehicle may leverage olfactory inputs to distinguish passengers based on body odor~\cite{amores_fernandez_olfactory_2020} or restrict driving when alcohol breath is detected~\cite{romero2019alcohol}.
Gustatory input might also be possible by embedding an electronic tongue (e.g., see~\cite{lee2018bio}) into the vehicle that is capable of sensing flavors, e.g., picked up from the skin.
In contrast to gustatory input, which seems impractical, brain and heart activity is valuable in determining the passenger's state and, in the case of brain input, can even be used for active vehicle control~\cite{bi2018novel}.
However, there is more potential for using brain activity via brain-computer interfaces, e.g., demonstrated by~\cite{folgieri_bci-based_2012, anupama_brain_2012}.

Regarding \textbf{output modalities}, our design space revealed only a few approaches for vestibular, kinesthetic, and thermal output and none for electrodermal, gustatory, cerebral, or cardiac.
However, such output modalities may be useful as they address a dimension of human perception that differs from visual, auditory, or tactile stimuli, which other tasks might occupy (e.g., the driving task or an NDRT).
Moreover, there is a great potential for using them in multimodal interactions combined with visual, auditory, or tactile output.
While only two publications in our SLR considered vestibular output~\cite{winner_effects_2016,sun_shaping_2021}, there are various techniques to stimulate the vestibular sense, e.g., galvanic vestibular stimulation~\cite{fitzpatrick_probing_2004}, force simulation~\cite{hoppe_odins_2021}, vestibular stimuli~\cite{riecke_move_2012, colley_swivr_2021}, or motion cues~\cite{reymond_motion_2000}.
The vehicle can also apply such cues to provide kinesthetic output, e.g., via electric muscle stimulation~\cite{pfeiffer_let_2014}, exoskeletons and gloves~\cite{tzafestas_whole-hand_2003}, or a moving platform~\cite{zeng_analysis_2010}.
Electrodermal output can be conveyed by electric stimuli~\cite{pamungkas_tactile_2015}, such as slight electric shocks~\cite{mujibiya_haptic_2015} or electric arcs~\cite{spelmezan_sparkle_2016} to the skin.
While our SLR did not reveal approaches for gustatory output, there are HCI concepts that use, e.g., edible UIs~\cite{narumi_augmented_2011}, pseudo-gustatory displays~\cite{narumi_evaluating_2010}, or electrical tongue stimulation~\cite{ranasinghe_digital_2016}.
For this, the interior might be a suitable environment as edible assets can be stored.
Although we only found concepts that consider implicit brain and heart activity, the vehicle could also actively stimulate the brain, e.g., using a brain implant or a defibrillator.
However, technology is not yet advanced enough to enable safe and practical usage of such stimuli.
Therefore, future research must investigate the usability and user acceptance of experimental output modalities.

In contrast to input modalities, the \textbf{interaction location} is often critical for output modalities since the usability may depend on the location, e.g., as users need an unobstructed view on visual output or feel tactile output.
Our design space shows that publications mainly used the front as an output location, e.g., for displays~\cite{ulahannan_designing_2020} or vibration~\cite{ng_evaluation_2017-1}, while other interior locations are considered rarely, e.g., table, door, rear, floor, or ceiling.
However, those locations may increase in relevance regarding the transition to automated driving and NDRTs~\cite{detjen_wizard_2020, pfleging_ndrt_2016}, which should be comfortably available regardless of whether passengers are seated in front, rear, or swivel seats.
Therefore, we argue that automotive UI research should move away from a front-focused design for all input and output modalities in automated driving towards considering the whole vehicle as an interaction space.
Our extended design space can thereby guide the ideation process for such research concepts.

\subsection{User Acceptance of In-Vehicle Interaction Modalities and Locations}
The study results (see Section~\ref{study_results}) revealed differences between the acceptance of modalities, resembling findings from related work.
For example, visual output was perceived as more useful than visual input (gaze), which is in line with Akkil et al.~\cite{akkil_user_2016} who found that users perceive some gaze interactions as complex, strenuous, and slow.
This suggests that implicit visual inputs such as pupil diameter or blink rate should be employed primarily instead of straining explicit inputs (e.g., gaze).
Similarly, participants would rather use auditory output (sound) than auditory input (speech), which they might have perceived as impractical in some situations, e.g., surrounded by others~\cite{baier_not_2019}.
Hence, future in-vehicle UIs should provide alternative modalities for such situations.
Besides, tactile input via touch was generally more accepted than tactile output via vibration, which we mainly associate with the unfamiliar vibration location on the center stack.
This implies that acceptance of input/output modalities depends on their characteristics, e.g., the involved human sensors/actuators, the required effort, mental demand, or interior location.
Therefore, future in-vehicle UIs may employ different modalities depending on the context (e.g., touch for accurate selection instead of gaze or speech) or utilize multimodality to mitigate the drawbacks of one modality (e.g., using gesture to improve touch~\cite{ahmad_predictive_2018}, gaze, or speech input~\cite{aftab_multimodal_2019}).
Modalities such as olfactory, gustatory, or cerebral might be perceived as invasive, indicated by open feedback [P11, P38] and low usefulness and comfort ratings.
Hence, future vehicles should embed sensors unobtrusively (e.g., seat-embedded heart rate detection~\cite{wusk2018non}), and the sensed information should be made transparent to the respective users.
Besides, safety concerns may have influenced the ratings, as some participants perceived modalities [P38] and locations [P5] as dangerous to use in a vehicle regardless of being driver or passenger.
The users' safety needs~\cite{detjen_how_2021} require future interaction design to overcome such concerns by providing safe interactions in any driving situation, which might render modalities or locations unavailable (e.g., in manual or conditional automated driving).

Regarding nomadic locations, AR was generally more accepted than VR.
We assume that participants were concerned about VR motion sickness~\cite{mcgill_i_2017} and perceived the see-through AR as more useful for productivity, which users might prefer over gaming, e.g., see~\cite{pfleging_ndrt_2016}.
Also, participants may have expected isolation using an opaque VR HMD and worried about reduced fallback-readiness in a takeover (SAE~3~and~4).
Therefore, in-vehicle VR must not induce motion sickness (e.g., using visual motion cues~\cite{mcgill_i_2017}), provide applications other than entertainment, and consider passenger safety and trust for any SAE level.
Since handheld and wearable devices are already usable in current vehicles (e.g., Android Auto~\cite{android_auto} via smartphone or smartwatch) and have received high usefulness, real-world usage, and comfort ratings, future interaction design should first focus on such locations as acceptance already exists or can be quickly built.
Besides, there is potential for modality combinations in nomadic and anchored locations (e.g., using a smartphone while the seat is vibrating~\cite{petermeijer_take-over_2017}), which might assist in introducing unfamiliar anchored interaction locations.
The high perceived usefulness of rear, table, and seat indicates that users would accept interaction at locations other than the front.
However, as the ratings were lower for anticipated comfort and real-world usage, future design should aim to mitigate discomfort, e.g., by increasing the size of a rear display, using a tiltable table, or ergonomic seat armrest displays.
Other anchored locations, e.g., ceiling, door, and floor, were generally less accepted since the participants probably could not imagine any use case, which requires future interaction design to elicit suitable applications to gain user acceptance.

\subsection{Future Research Interests on In-Vehicle Interaction}
Emerging from the opportunities unveiled in our design space and further pointed out by the user study, achieving user acceptance of novel in-vehicle interaction concepts faces several challenges.
We derived the following research interests that can guide practitioners to overcome these.

\subsubsection{Technological Adavancement}
While novel modalities such as electrodermal, olfactory, gustatory, cerebral, and cardiac are interesting from a research perspective, technical challenges may hinder or even prevent their successful introduction in future vehicles.
For instance, technologies in early development stages, e.g., for cerebral, cardiac, or electrodermal modalities, may provide unreliable recognition of input~\cite{alfaras2020biosensing} and limited actuation of output, thereby negatively impacting usability.
Besides, experimental modalities such as olfactory input and gustatory input/output may be challenging to implement or impractical to operate in vehicles.
For example, technologies for an artificial tongue~\cite{lee2018bio} or nose~\cite{lee2012mimicking} exist but have not yet been transferred to a vehicle context.
Moreover, it is unclear whether all found modalities in this work can coexist in a vehicle.
We assume that manufacturers will use subsets of the presented possible vehicle sensors and actuators to suit different use cases.
The integration of multiple sensors enables the detection of erroneous data from one input modality using another~\cite{murali_intelligent_2021}.
Likewise, the vehicle can leverage different output modalities best suited to specific situations, e.g., seat vibration for subtle messages or loud sounds alerting the driver.
Furthermore, such cross-modal input/output usage can provide redundancy in interactions, increasing the reliability, especially when one vehicle sensor/actuator is unavailable~\cite{falco2019transfer}.
A safe and fast prototyping testbed for novel in-vehicle interactions are (open-source) VR driving simulators, such as~\cite{auto-wsd} or~\cite{low-cost-vr-sim}. Besides, low-cost vehicle motion simulators (e.g.,~\cite{colley_swivr_2021}) enable the testing of motion impacts on novel interactions without the need for an actual vehicle.


\subsubsection{Social Acceptance}
Due to our SLR's scope including passengers other than the driver, e.g., in the rear seat, there are social challenges, which persist for AVs, as they will be most likely used in social settings, such as leisure trips with friends and family~\cite{WILSON2022150}.
In this regard, the feasibility of in-vehicle modalities and locations may depend on the social context, i.e., sharing the vehicle with friends or strangers.
For instance, users are less sensitive to the presence of strangers when on a commute trip compared to a leisure-activity trip~\cite{lavieri2019modeling}, and they would feel more confident to privately use gesture~\cite{chittaro2010distinctive} or speech input~\cite{baier_not_2019}.
Likewise, in a social setting (e.g., car-sharing), we argue that users might prefer more subtle interactions via physiological input modalities, such as electrodermal, cerebral, cardiac, or thermal.
Besides, the output should be adapted to the social setting, e.g., an output modality (private audio message) might be intended for a specific passenger~\cite{kim2021please}, which excludes modalities (sound or displays) visible for everyone.
In this regard, the interaction location may be decisive.
We assume that using a touchscreen at the center table would be preferred when riding with friends than with strangers, or interacting with the rear location as a front-seat passenger might disturb rear-seat passengers.
Furthermore, there might be different "acceptance personalities"~\cite{detjen_how_2021} also visible in the open feedback where some participants had a more negative attitude towards interaction modalities or locations (e.g., [P5, P38]).
Hence, future in-vehicle interaction research should also investigate the social component of user acceptance.


\subsubsection{Explicit and Implicit In-Vehicle Interaction}
Implicit in-vehicle interaction enables passengers to produce input subconsciously~\cite{serim_implicit_2019} or perceive output from the vehicle implicitly embedded in output modalities.
Our SLR revealed various modalities feasible for implicit interactions, such as electrodermal, thermal, cerebral, and cardiac.
Other examples include driver fatigue detection~\cite{campagne2004correlation}, emotion recognition~\cite{zepf2020driver}, or body pose estimation~\cite{martin2012design}.
The advantages of implicit interactions are manifold.
For instance, the vehicle can tailor in-time airbag deployment, steering, brake, and crash avoidance if the passengers' body positions are estimated~\cite{murphy2010head,murali_intelligent_2021}.
AVs can detect driver attention in the driving task via implicit cues (e.g., emotion) to determine if a takeover can be achieved safely~\cite{zepf2020driver}.
Furthermore, the introduction of novel interaction locations, e.g., seat, door, or table, fosters new options for implicit interaction that utilize physiological vehicle sensors embedded in the seat~\cite{wusk2018non} or door handle~\cite{hernandez_autoemotive_2014}.
Besides, we found that modalities mainly used in explicit interactions, such as visual or auditory, are potentially usable in implicit interaction.
For example, the vehicle can use auditory output via adaptive music to subliminal affect the passengers~\cite{burnett_altering_2017} or alter the color of visual UIs to increase driving performance~\cite{hassib2019detecting}.
In contrast to implicit interaction, passengers can explicitly interact with the vehicle, e.g., in conversations and performing tasks.
Examples include voice commands~\cite{truschin_designing_2014}, gestures~\cite{may_designing_2017}, touch~\cite{kopinski_touch_2016}, and display-based modalities~\cite{broy_evaluating_2015}.
At times, implicit input can also assist while interacting explicitly, e.g., by guiding the interaction to the most appropriate modality (visual, auditory, or haptic) if the driver's intentions are recognized~\cite{murali_intelligent_2021}.
Such a combination of implicit and explicit interactions may be promising for future in-vehicle interaction design, e.g., regarding multimodal interactions.




\subsubsection{Perspectives on Multimodal In-Vehicle Interaction}

Our combination matrix for multimodal interaction (see~\autoref{tab:multimodal_interaction}) revealed research potential for thermal, electrodermal, and cerebral modalities, e.g., in combination with well-established input modalities, such as visual, auditory, and kinesthetic.
Besides, we found little research regarding multimodal output that considers vestibular or electrodermal output.
However, not every modality combination may be feasible, as one modality requires too much mental or physical workload to effectively use the other (e.g., using two explicit input modalities, gesture~$\times$~touch).
In this regard, combining an explicit modality such as touch, gesture, or speech with an implicit modality, e.g., EDA, might overcome this challenge.
Still, the location of each modality decisively influences the usability of such multimodal interaction.
For instance, sensing the passengers' EDA requires direct skin contact that might not be guaranteed when they lift (both) arms to perform a gesture.
Possible solutions may utilize fallback vehicle sensor locations, e.g., a nomadic wearable sensor (smartwatch), to compensate for an unavailable anchored sensor in the seat armrest.
Therefore, we propose a new perspective on multimodal in-vehicle interaction that includes the location to specify the design of possible multimodal interactions, e.g., kinesthetic input~\textit{at}~center stack~$\times$~electrodermal input~\textit{at}~seat or visual output~\textit{at}~door~$\times$~visual output~\textit{at}~HMD.


\subsubsection{Explainable In-Vehicle Artificial Intelligence}
Vehicles can leverage artificial intelligence (AI) to enable adaptive or proactive implicit, explicit, and multimodal interactions.
In our SLR, we found AI-based approaches, for example, in conversational speech systems~\cite{lin_adasa_2018, kim_predicting_2019}.
Besides, using physiological input modalities, such as electrodermal, cerebral, or cardiac activity, mostly requires AI to synthesize the data and make correct inferences to provide a suitable output to passengers.
However, to achieve user acceptance, passengers should presumably understand the decision-making of AI, especially in case of wrong decisions or faulty output.
Similarly, driving automation systems require adaptive UIs to foster trust and understanding~\cite{wintersberger_explainable_2020}.
Therefore, future in-vehicle interaction research should consider explainable AI-based implicit and multimodal interactions that may further support the successful introduction of novel modalities (e.g., thermal, cerebral, or cardiac).
Solutions may include interfaces that explain the current AI decision or monitoring systems that automatically detect unreasonable outputs and provide explanations (e.g., see~\cite{gilpin_monitor_2018}).

\subsection{Limitations and Future Work}
We limited the SLR database search to query only abstracts in a set of selected venues to keep the SLR focused and the resulting set of publications manageable, which had potentially excluded relevant research.
However, we assume that our results sufficiently cover the domain of in-vehicle interaction to answer our RQs.

Our user study's presentation of self-made images might have influenced participants' understanding of the shown interactions.
However, we deem this negligible as we intended to gain meta-level insights on the topic and limit the study complexity.
Still, future work should evaluate possible interaction modalities and locations using video or interactive simulations, e.g., VR or a real-world setting.
Besides, in the study, we only showed one example for a modality category for internal consistency (see~\autoref{fig:input_output_study_concepts}).
However, there might be examples of the same modality category, which may be more accepted.
Therefore, future work may present more than one example per modality, e.g., tactile output via vibration, ultrasound, and cutaneous push.
Similarly, the perceived usefulness, usability, and comfort of input at different interaction locations may be influenced by showing only the gesture input modality.
Hence, future work should evaluate different modalities for each location to elicit strengths and weaknesses adequately to determine the feasibility of input and output modalities.
Another limitation is that the study results refer to US residents only. However, there might be a regional bias, and the user acceptance of information technology may depend on culture (e.g., see~\cite{tech_accept_cultural_diffs}). Especially in the EU region, we could imagine more pessimistic results. In the future, surveys should target additional regions and compare cultural differences.

Our design space asks for usage as an ideation tool.
Therefore, future work should consider the unaccounted aspects in this design space primarily.
The presented research interests can further guide this process.

\section{Conclusion}
This work presents a comprehensive overview of existing literature on in-vehicle interaction regarding SAE levels 0-5 by performing an SLR on 327 publications selected from 2534 candidates.
We found that existing literature mainly covers established input and output modalities, such as visual, auditory, kinesthetic, tactile, and vehicle interior locations, such as the front.
However, little research was related to other modalities and interior locations. 
Our SLR revealed multimodal interaction as a relevant topic for future in-vehicle interactions.
Therefore, we created a combination matrix for multimodal interaction, enabling practitioners to retrieve existing approaches and elicit novel modality combinations systematically.
To classify the existing literature and outline research gaps and future approaches, we presented a design space for in-vehicle interaction with dimensions \textit{location} and \textit{modality}.
Our design space extends previous design spaces~\cite{kern_design_2009,detjen_how_2021} to update in-vehicle interaction research, considering a more extensive set of possible human sensors and actuators and novel interaction locations throughout the vehicle interior.
To gain initial insights on the user acceptance of possible concepts for input and output modalities and interaction locations, we conducted an image-based online study (\N{48}).
The shown concepts were deduced from the gaps unveiled in our design space and existing literature.
The results revealed general user acceptance of novel input and output modalities and locations while also showing differences between modalities and locations, encouraging future work to consider a modality's characteristics and interior location to achieve high usability.
We argue that the SLR results, the combination matrix of multimodal interactions, the design space, and insights from our user study advise future work to evaluate key design decisions, exploit trends, and explore new areas in the domain of in-vehicle interaction.

\begin{acks}
The authors thank all study participants. This work was supported by the project 'SEMULIN' (\textbf{se}lbstunterstützende, \textbf{mu}ltimodale \textbf{In}teraktion) funded by the Federal Ministry for Economic Affairs and Energy (BMWi).
\end{acks}
\bibliographystyle{ACM-Reference-Format}
\bibliography{GeneralReferences}
\appendix


\section{Input and Output Modality Keywords}
\begin{table*}[ht!]
\centering
\caption{Keywords for input and output modalities that are categorized by the respective human sensors/actuators. \textit{Blue}: keywords obtained from a preliminary literature review in the automotive domain. \textit{Brown}: keywords obtained from a Google Scholar literature search for input/output modalities, including non-automotive research. \textit{Olive}: keywords obtained in a brainstorm session among the authors.}
\resizebox{\textwidth}{!}{%
\begin{tabular}{lll|l|l|}
\cline{4-5}
 &  &  & \textbf{Input Modality (generated by Human Actuators)} & \textbf{Output Modality (sensed by Human Sensors)} \\ \hline
\multicolumn{3}{|l|}{\textbf{Visual}} & \begin{tabular}[c]{@{}l@{}}\color{blue}Gaze, Eye Movement\\ {\color{blue}Eye Blink}, \color{brown}Blink Rate \cite{chen_using_2014}, Squinting \cite{vasisht_human_2019},\\ \color{brown}Eyebrow Raises \cite{grauman_communication_2003}\\ \color{brown}Pupillary Behavior \cite{babiker_pupillary_2013}, Pupil Diameter \cite{babiker_pupillary_2013,chen_using_2014}\end{tabular} & \begin{tabular}[c]{@{}l@{}}\color{blue}LED, LED Strip, Ambient Light, Peripheral Light Glasses\\ {\color{blue}Display, Head-Up-Display}, \color{olive}Screen, Monitor\\ {\color{blue}AR}, {\color{brown}Windshield Display \cite{hauslschmid_augmenting_2015}}, \color{olive}Mixed Reality\\ \color{blue}VR\\ \color{blue}Robotic Companion, Co-driver, Robotic Gestures\\ \color{olive}Shape, Texture\end{tabular} \\ \hline
\multicolumn{3}{|l|}{\textbf{Auditory}} & \begin{tabular}[c]{@{}l@{}}{\color{blue}Speech, Voice, Verbal}, \color{brown}Vocal Affect\\ {\color{brown}(Rate, Pitch, Intensity, Quality) \cite{chen_joint_2000,zupan_importance_2009}}, \color{brown}Natural Language\\ \color{brown}Non-Speech Sounds (Whistling, Humming, ...) \cite{sporka_non-speech_2006,sporka_longitudinal_2007},\\ \color{brown}Acoustic Gesture (Tone and Pitch) \cite{sporka_non-speech_2006,sporka_longitudinal_2007}\end{tabular} & \begin{tabular}[c]{@{}l@{}}{\color{blue}Auditory Cues, Earcons, Sounds}, \color{brown}Sonification \cite{hermann_taxonomy_2008},\\ \color{brown}Auditory Icons \cite{hermann_taxonomy_2008}, Spatial Sound \cite{spors_spatial_2013}\\ \color{blue}Speech-based Assistant, Conversational Agent, Voice-Command,\\ {\color{brown}Synthetic Speech \cite{chen_joint_2000}}, \color{olive}Natural Language\\ \color{blue}Music\end{tabular} \\ \hline
\multicolumn{1}{|l|}{\multirow{22}{*}{\rotatebox[origin=c]{90}{\textbf{Haptic}}}} & \multicolumn{2}{l|}{\textbf{Vestibular}} & \begin{tabular}[c]{@{}l@{}}\color{brown}Proxemics \cite{marquardt_proxemic_2015}\\ \color{brown}Balance \cite{doyle_base_2010}, Vestibular Behavior \cite{kim_virtual_2018}\\ \color{olive}Accelerating Body Motions\end{tabular} & \begin{tabular}[c]{@{}l@{}}{\color{blue}Vehicle Roll Motion}, \color{brown}Vestibular Stimuli \cite{riecke_move_2012},\\ \color{brown}Motion Cues \cite{reymond_motion_2000}, Induced Self-Motion \cite{hoppe_odins_2021},\\ \color{brown}Perception of Self-Motion \cite{rietzler_vrspinning_2018},\\ \color{brown}Force Simulation \cite{hoppe_odins_2021}\\ \color{brown}Proxemics \cite{marquardt_proxemic_2015}\\ \color{brown}Galvanic Vestibular Stimulation \cite{fitzpatrick_probing_2004}\end{tabular} \\ \cline{2-5} 
\multicolumn{1}{|l|}{} & \multicolumn{2}{l|}{\textbf{Kinaesthetic}} & \begin{tabular}[c]{@{}l@{}}\color{blue}Lever, Joystick\\ {\color{blue}Gesture, Hand-, Head-}, \color{brown}Finger-\cite{crowley_finger_1995}, Arm-\cite{ebert_sensx_2017}, Leg-\cite{ebert_sensx_2017},\\ \color{brown}Feet-Movement \cite{velloso_feet_2015}\\ {\color{blue}Body Posture, Body Motion}, \color{brown}Walking \cite{bhandari_legomotion_2017},\\ \color{brown}Jumping \cite{wolf_jumpvr_2020}\\ \color{olive}Emotion, Facial Expression\\ \color{brown}Muscle Activity \cite{rietzler_virtual_2019}\end{tabular} & \begin{tabular}[c]{@{}l@{}}\color{brown}Electric Muscle Stimulation \cite{pfeiffer_let_2014},\\ \color{brown}Functional Electrical Stimulation \cite{nishida_wearable_2015}\\ \color{brown}Kinesthetic Feedback,\\ \color{brown}Exosceleton force-feedback glove \cite{tzafestas_whole-hand_2003}\\ \color{brown}Moving Platform/Plate \cite{zeng_analysis_2010}\end{tabular} \\ \cline{2-5} 
\multicolumn{1}{|l|}{} & \multicolumn{1}{l|}{\multirow{12}{*}{\rotatebox[origin=c]{90}{\textbf{Cutaneous}}}} & \textbf{Electrodermal} & \begin{tabular}[c]{@{}l@{}}\color{blue}EDA\\ \color{blue}Facial Features\end{tabular} & \begin{tabular}[c]{@{}l@{}}\color{brown}Electric Shocks \cite{mujibiya_haptic_2015},\\ \color{brown}Electric Arcs \cite{spelmezan_sparkle_2016},\\ \color{brown}Electric Stimuli \cite{pamungkas_tactile_2015}\end{tabular} \\ \cline{3-5} 
\multicolumn{1}{|l|}{} & \multicolumn{1}{l|}{} & \textbf{Tactile} & \begin{tabular}[c]{@{}l@{}}\color{blue}Touch,\\ \color{brown}Pressure \cite{stewart_characteristics_2010}\end{tabular} & \begin{tabular}[c]{@{}l@{}}{\color{blue}Shape Changing}, \color{brown}Tactile Click Feedback \cite{jung_pinpad_2017},\\ \color{brown}Texture Modulation \cite{jung_pinpad_2017},\\ \color{brown}Pin-array Displays \cite{jung_pinpad_2017}\\ \color{blue}Ultrasound Haptic\\ \color{blue}Vibration, Vibro-tactile\\ \color{brown}Tactons \cite{brewster_tactons_2004}\\ \color{brown}Liquid \cite{richter_liquitouch_2013}\\ \color{brown}Air Jet \cite{suzuki_air_2005}\end{tabular} \\ \cline{3-5} 
\multicolumn{1}{|l|}{} & \multicolumn{1}{l|}{} & \textbf{Thermal} & \begin{tabular}[c]{@{}l@{}}\color{brown}Skin Temperature \cite{welch_wearable_2018},\\ \color{brown}Body Temperature \cite{mansor_body_2013}\end{tabular} & \begin{tabular}[c]{@{}l@{}}\color{blue}Changes in Temperature, Thermal Feedback\\ \color{brown}Liquid \cite{richter_liquitouch_2013}\end{tabular} \\ \cline{3-5} 
\multicolumn{1}{|l|}{} & \multicolumn{1}{l|}{} & \textbf{Pain (Nociception)} & \textit{-} & \textit{-} \\ \hline
\multicolumn{3}{|l|}{\textbf{Olfactory}} & \color{brown}Body Odor \cite{amores_fernandez_olfactory_2020}, Breath Smell \cite{amores_fernandez_olfactory_2020} & \color{blue}Scent \\ \hline
\multicolumn{3}{|l|}{\textbf{Gustatory}} & \textit{-} & \begin{tabular}[c]{@{}l@{}}\color{brown}Edible User Interface, Edible Marker \cite{narumi_augmented_2011}\\ \color{brown}Pseudo-Gustatory Display \cite{narumi_evaluating_2010}\\ \color{brown}Electrial stimulation on the tongue \cite{ranasinghe_digital_2016}\end{tabular} \\ \hline
\multicolumn{3}{|l|}{\textbf{Cerebral}} & \begin{tabular}[c]{@{}l@{}}{\color{blue}Cortical Activity}, \color{brown}Slow Cortical Potentials \cite{hinterberger_multimodal_2004},\\ \color{brown}EEG or fNIRS \cite{folgieri_bci-based_2012}, Brain Signals \cite{anupama_brain_2012},\\ \color{olive}Brain-Computer Interface\end{tabular} & \color{olive}Brain-Computer Interface, Brain Chip \\ \hline
\multicolumn{3}{|l|}{\textbf{Cardiac}} & \begin{tabular}[c]{@{}l@{}}{\color{blue}Heart Rate Variability, Heart Rate}, \color{brown}Blood Pressure \cite{welch_wearable_2018},\\ \color{olive}Cardiovascular\end{tabular} & \color{olive}Defibrillator \\ \hline
\multicolumn{3}{|l|}{\textbf{Nomadic Device}} & \begin{tabular}[c]{@{}l@{}}{\color{blue}Smartphone, Tablet}, \color{olive}Laptop\\ \color{olive}Smartwatch, On-body Device, Wearable\end{tabular} & \begin{tabular}[c]{@{}l@{}}{\color{blue}Smartphone, Tablet}, \color{olive}Laptop\\ \color{olive}Smartwatch, On-body Device, Wearable\end{tabular} \\ \hline
\end{tabular}%
}
\label{tab:keyword_list}
\end{table*}

\section{Search Queries}
\begin{table}[ht!]
\centering
\caption{Search queries based on Boolean algebra used in the SLR covering four contexts relevant to our research topic including the results count for each queried database (ACM DL, IEEE Xplore, and ScienceDirect).}
\resizebox{\textwidth}{!}{%
\begin{tabular}{|l|l|l|l|l|l|l|l|l|l|}
\hline
\cellcolor[HTML]{FCE5CD}\textbf{\begin{tabular}[c]{@{}l@{}}(1) Automotive\\ Context\end{tabular}} & \cellcolor[HTML]{FFF2CC}\textbf{\begin{tabular}[c]{@{}l@{}}(2) Interaction\\ Context\end{tabular}} & \multicolumn{3}{l|}{\cellcolor[HTML]{D9EAD3}\textbf{\begin{tabular}[c]{@{}l@{}}(3) Sensor/Actuator\\ Context\end{tabular}}} & \cellcolor[HTML]{C9DAF8}\textbf{\begin{tabular}[c]{@{}l@{}}(4) Input and Output Modalities\\ Context\end{tabular}} & \textbf{Resulting Query (1 AND 2 AND (3 OR 4))} & \multicolumn{1}{c|}{\textbf{ACM}} & \multicolumn{1}{c|}{\textbf{IEEE}} & \multicolumn{1}{c|}{\textbf{\begin{tabular}[c]{@{}c@{}}Science\\ Direct\end{tabular}}} \\ \hline
 &  & \multicolumn{3}{l|}{\begin{tabular}[c]{@{}l@{}}("Visual" OR\\ "Eye" OR\end{tabular}} & \begin{tabular}[c]{@{}l@{}}"Gaze" OR "Eyebrow" OR\\ "Pupillary" OR "Pupil" OR\\ "LED" OR "Display" OR\\ "Screen" OR "Monitor" OR\\ "Head-Up-Display" OR\\ "Augmented Reality" OR\\ "Virtual Reality" OR\\ "Mixed Reality" OR "Robotic" OR\\ "Shape" OR "Texture")\end{tabular} & \begin{tabular}[c]{@{}l@{}}(vehicle? OR car? OR driver? OR driving OR "in-vehicle") AND\\ (interaction? OR interface?) AND\\ (visual OR eye OR gaze OR eyebrow? OR pupil* OR "LED" OR\\ display? OR screen? OR monitor? OR "head-up-display" OR\\ "augmented reality" OR "virtual reality" OR "mixed reality" OR\\ robotic OR shape? OR texture?)\end{tabular} & 614 & 395 & 233 \\ \cline{3-10} 
 &  & \multicolumn{3}{l|}{\begin{tabular}[c]{@{}l@{}}("Auditory" OR\\ "Audio" OR\\ "Ear" OR\end{tabular}} & \begin{tabular}[c]{@{}l@{}}"Speech" OR "Voice" OR\\ "Verbal" OR "Vocal" OR\\ "Non-Speech" OR "Sound" OR\\ "Whistling" OR "Humming" OR\\ "Acoustic" OR "Earcon" OR\\ "Conversational" OR "Music")\end{tabular} & \begin{tabular}[c]{@{}l@{}}(vehicle? OR car? OR driver? OR driving OR "in-vehicle") AND\\ (interaction? OR interface?) AND\\ (auditory OR audio OR ear? OR speech OR voice OR verbal OR\\ vocal OR "non-speech" OR sound? OR whistling OR humming OR\\ acoustic OR earcon? OR conversational OR music)\end{tabular} & 285 & 136 & 92 \\ \cline{3-10} 
 &  & \multicolumn{3}{l|}{\begin{tabular}[c]{@{}l@{}}("Vestibular"OR\\ "Equilibrium" OR\end{tabular}} & \begin{tabular}[c]{@{}l@{}}"Balance" OR "Body Acceleration" OR\\ "Vehicle Roll Motion" OR\\ "Motion Cue" OR "Self-Motion")\end{tabular} & \begin{tabular}[c]{@{}l@{}}(vehicle? OR car? OR driver? OR driving OR "in-vehicle") AND\\ (interaction? OR interface?) AND\\ (vestibular OR equilibrium OR balance OR "body acceleration" OR\\ "vehicle roll motion" OR "motion cue" OR "self-motion")\end{tabular} & 19 & 89 & 1 \\ \cline{3-10} 
 &  &  & \multicolumn{2}{l|}{\begin{tabular}[c]{@{}l@{}}"Kinaesthetic" OR\\ "Kinesthetic" OR\end{tabular}} & \begin{tabular}[c]{@{}l@{}}"Lever" OR "Joystick" OR\\ "Button" OR "Gesture" OR\\ "Movement" OR "Moving" OR\\ "Motion" OR "Posture" OR\\ "Pose" OR "Emotion" OR\\ "Facial" OR "Muscle" OR\\ "Functional Electrical Stimulation" OR\\ "Exosceleton" OR "Proxemic")\end{tabular} & \begin{tabular}[c]{@{}l@{}}(vehicle? OR car? OR driver? OR driving OR "in-vehicle") AND\\ (interaction? OR interface?) AND\\ (haptic? OR somatosens*)\\ \\ (vehicle? OR car? OR driver? OR driving OR in-vehicle) AND\\ (interaction? OR interface?) AND\\ (kinaesthetic? OR kinesthetic? OR lever? OR joystick? OR\\ button? OR gesture? OR movement OR moving OR motion? OR\\ posture? OR pose? OR emotion? OR facial OR muscle? OR\\ "functional electrical stimulation" OR exosceleton OR proxemic?)\end{tabular} & 494 & 509 & 123 \\ \cline{4-10} 
 &  &  &  &  & \begin{tabular}[c]{@{}l@{}}"Electrodermal" OR "Electrical" OR\\ "Facial Features")\end{tabular} & \begin{tabular}[c]{@{}l@{}}(vehicle? OR car? OR driver? OR driving OR "in-vehicle") AND\\ (interaction? OR interface?) AND\\ (cutaneous OR skin)\\ \\ (vehicle? OR car? OR driver? OR driving OR in-vehicle) AND\\ (interaction? OR interface?) AND\\ (electrodermal? OR electric* OR face)\end{tabular} & 97 & 207 & 8 \\ \cline{5-10} 
 &  &  &  & \begin{tabular}[c]{@{}l@{}}"Tactile" OR\\ "Tactition" OR\end{tabular} & \begin{tabular}[c]{@{}l@{}}"Touch" OR "Pressure" OR\\ "Shape" OR "Texture" OR\\ "Pin-Array" OR "Ultrasound"\\ OR "Vibration" OR "vibro-tactile" OR\\ "Liquid" OR "Air")\end{tabular} & \begin{tabular}[c]{@{}l@{}}(vehicle? OR car? OR driver? OR driving OR "in-vehicle") AND\\ (interaction? OR interface?) AND\\ (tacti* OR touch* OR pressure OR "pin-array" OR\\ ultrasound OR vibrat* OR "vibro-tactile")\end{tabular} & 169 & 106 & 22 \\ \cline{5-10} 
 &  &  &  & \begin{tabular}[c]{@{}l@{}}"Thermal" OR\\ "Temperature" OR\end{tabular} & "Body" OR "Heat" OR "Liquid") & \begin{tabular}[c]{@{}l@{}}(vehicle? OR car? OR driver? OR driving OR "in-vehicle") AND\\ (interaction? OR interface?) AND\\ (thermal? OR temperature? OR body OR heat OR liquid OR air)\end{tabular} & 133 & 140 & 0 \\ \cline{5-10} 
 &  & \multirow{-22}{*}{\begin{tabular}[c]{@{}l@{}}("Haptic" OR\\ "Somatosensory" OR\end{tabular}} & \multirow{-18}{*}{\begin{tabular}[c]{@{}l@{}}"Cutaneous" OR\\ "Skin" OR\end{tabular}} & \begin{tabular}[c]{@{}l@{}}"Pain" OR\\ "Nociception")\end{tabular} & \textit{-} & \begin{tabular}[c]{@{}l@{}}(vehicle? OR car? OR driver? OR driving OR "in-vehicle") AND\\ (interaction? OR interface?) AND\\ (pain OR nociception)\end{tabular} & 2 & 1 & 2 \\ \cline{3-10} 
 &  & \multicolumn{3}{l|}{\begin{tabular}[c]{@{}l@{}}("Olfactory" OR\\ "Olfaction" OR\\ "Smell" OR\end{tabular}} & \begin{tabular}[c]{@{}l@{}}"Odor" OR "Breath"\\ OR "Scent" OR "Nose")\end{tabular} & \begin{tabular}[c]{@{}l@{}}(vehicle? OR car? OR driver? OR driving OR "in-vehicle") AND\\ (interaction? OR interface?) AND\\ (olfact* OR smell* OR odor* OR breath* OR scent* OR nose)\end{tabular} & 52 & 4 & 1 \\ \cline{3-10} 
 &  & \multicolumn{3}{l|}{\begin{tabular}[c]{@{}l@{}}("Gustatory" OR\\ "Gustation" OR\\ "Taste" OR\end{tabular}} & \begin{tabular}[c]{@{}l@{}}"Edible" OR\\ "Pseudo-Gustatory" OR\\ "Tongue")\end{tabular} & \begin{tabular}[c]{@{}l@{}}(vehicle? OR car? OR driver? OR driving OR "in-vehicle") AND\\ (interaction? OR interface?) AND\\ (gustat* OR taste* OR edible* OR "pseudo-gustatory" OR tongue?)\end{tabular} & 1 & 1 & 0 \\ \cline{3-10} 
 &  & \multicolumn{3}{l|}{\begin{tabular}[c]{@{}l@{}}("Brain" OR\\ "Brain-Computer Interface" OR\end{tabular}} & "Cortical") & \begin{tabular}[c]{@{}l@{}}(vehicle? OR car? OR driver? OR driving OR "in-vehicle") AND\\ (interaction? OR interface?) AND\\ (brain OR "brain-computer interface" OR cortic* OR "EEG")\end{tabular} & 18 & 49 & 3 \\ \cline{3-10} 
 &  & \multicolumn{3}{l|}{\begin{tabular}[c]{@{}l@{}}("Heart" OR\\ "Cardiac" OR\\ "Cardio" OR\end{tabular}} & "Blood" OR "Defibrillator") & \begin{tabular}[c]{@{}l@{}}(vehicle? OR car? OR driver? OR driving OR "in-vehicle") AND\\ (interaction? OR interface?) AND\\ (heart OR cardi* OR blood OR defibrillator?)\end{tabular} & 20 & 19 & 8 \\ \cline{3-10} 
\multirow{-60}{*}{\begin{tabular}[c]{@{}l@{}}("vehicle" OR\\ "car" OR\\ "driver" OR\\ "driving" OR\\ "in-vehicle") AND\end{tabular}} & \multirow{-60}{*}{\begin{tabular}[c]{@{}l@{}}("Interaction" OR\\ "Interface") AND\end{tabular}} & \multicolumn{3}{l|}{\begin{tabular}[c]{@{}l@{}}("Nomadic Device" OR\\ "Portable Device" OR\\ "Hand-held Device"\end{tabular}} & \begin{tabular}[c]{@{}l@{}}"Smartphone" OR "Tablet"\\ OR "Laptop" OR\\ "Smartwatch" OR "On-Body" OR\\ "Wearable")\end{tabular} & \begin{tabular}[c]{@{}l@{}}(vehicle? OR car? OR driver? OR driving OR "in-vehicle") AND\\ (interaction? OR interface?) AND\\ ("nomadic device" OR "portable device" OR\\ "hand-held device" OR smartphone? OR tablet? OR\\ laptop? OR smartwatch* OR "on-body" OR wearable?)\end{tabular} & 144 & 55 & 32 \\ \hline
\end{tabular}%
}
\label{tab:search_queries}
\end{table}

\newpage
\section{Result Counts of Queried Venues}
\begin{table*}[ht!]
\centering
\caption{Selected venues and number of publications found after applying the SLR search queries.}
\resizebox{\textwidth}{!}{%
\begin{tabular}{l|c}
\hline
\textbf{Conference / Venue} & \textbf{Number of publications} \\ \hline
ACM Conference on Human Factors in Computing Systems (CHI) & \textbf{40} \\
ACM Conference on Computer-Supported Cooperative Work \& Social Computing (CSCW) & \textbf{0} \\
ACM Conference on Pervasive and Ubiquitous Computing (UbiComp) & \textbf{2} \\
ACM/IEEE International Conference on Human Robot Interaction (HRI) & \textbf{1} \\
ACM Symposium on User Interface Software and Technology (UIST) & \textbf{3} \\
ACM Transactions on Computer-Human Interaction (TOCHI) & \textbf{1} \\
ACM International Conference on Multimodal Interaction (ICMI-MLMI) & \textbf{7} \\
ACM Designing Interactive Systems Conference (DIS) & \textbf{6} \\
ACM International Conference on Intelligent User Interfaces (IUI) & \textbf{8} \\
ACM International Conference on Human-Computer Interaction with Mobile Devices and Services (MobileHCI) & \textbf{9} \\
ACM International Conference on Tangible, Embedded, and Embodied Interaction (TEI) & \textbf{1} \\
ACM Conference on Automotive User Interfaces and Interactive Vehicular Applications (AutoUI) & \textbf{122} \\
IEEE Transactions on Affective Computing & \textbf{0} \\
IEEE Transactions on Human-Machine Systems & \textbf{2} \\
IEEE Transactions on Haptics & \textbf{2} \\
IEEE Conference on Virtual Reality and 3D User Interfaces (VR) & \textbf{1} \\
IEEE Transactions on Intelligent Transportation Systems (ITS) & \textbf{12} \\
IEEE Transactions on Vehicular Technology & \textbf{1} \\
IEEE Transactions on Intelligent Vehicles & \textbf{0} \\
IEEE Intelligent Vehicles Symposium (IV) & \textbf{24} \\
IEEE International Conference on Intelligent Transportation Systems (ITSC) & \textbf{18} \\
IEEE Access & \textbf{8} \\
International Journal of Human-Computer Studies & \textbf{5} \\
Transportation Research Part F: Traffic Psychology and Behaviour & \textbf{32} \\
Applied Ergonomics & \textbf{22} \\
Robotics and automated Systems & \textbf{0} \\ \hline
\textbf{Combined} & \textbf{327}
\end{tabular}%
}
\label{tab:selected_venues}
\end{table*}

\newpage
\section{Concept Images of In-Vehicle Interactions}
\begin{figure*}[ht!]
        \centering
        \includegraphics[width=.89\textwidth]{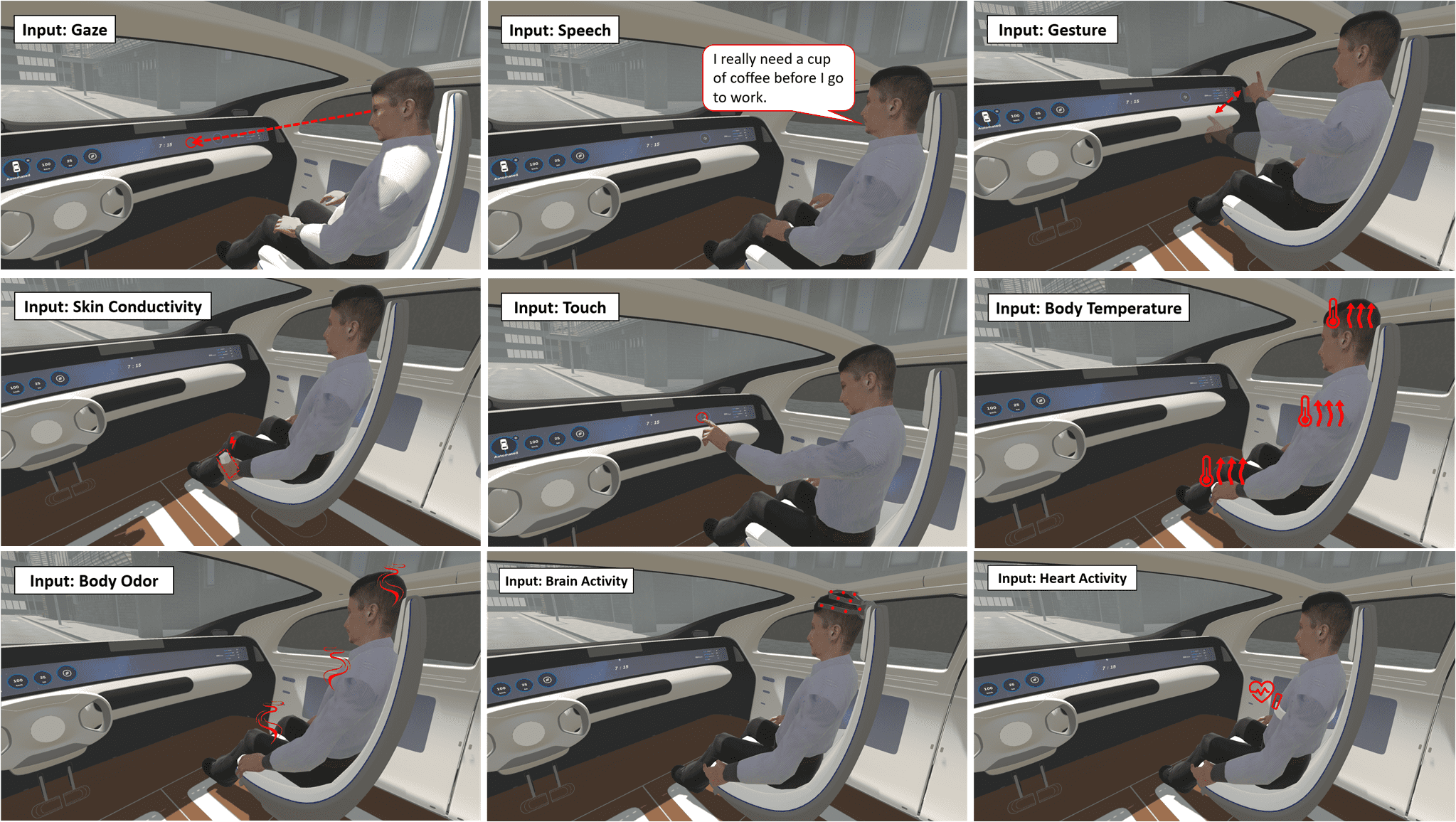}
    \caption{Concept images for in-vehicle input modalities used in our online study.}
\label{fig:study_images_full_input}
\end{figure*}
\begin{figure*}[ht!]
        \centering
        \includegraphics[width=.89\textwidth]{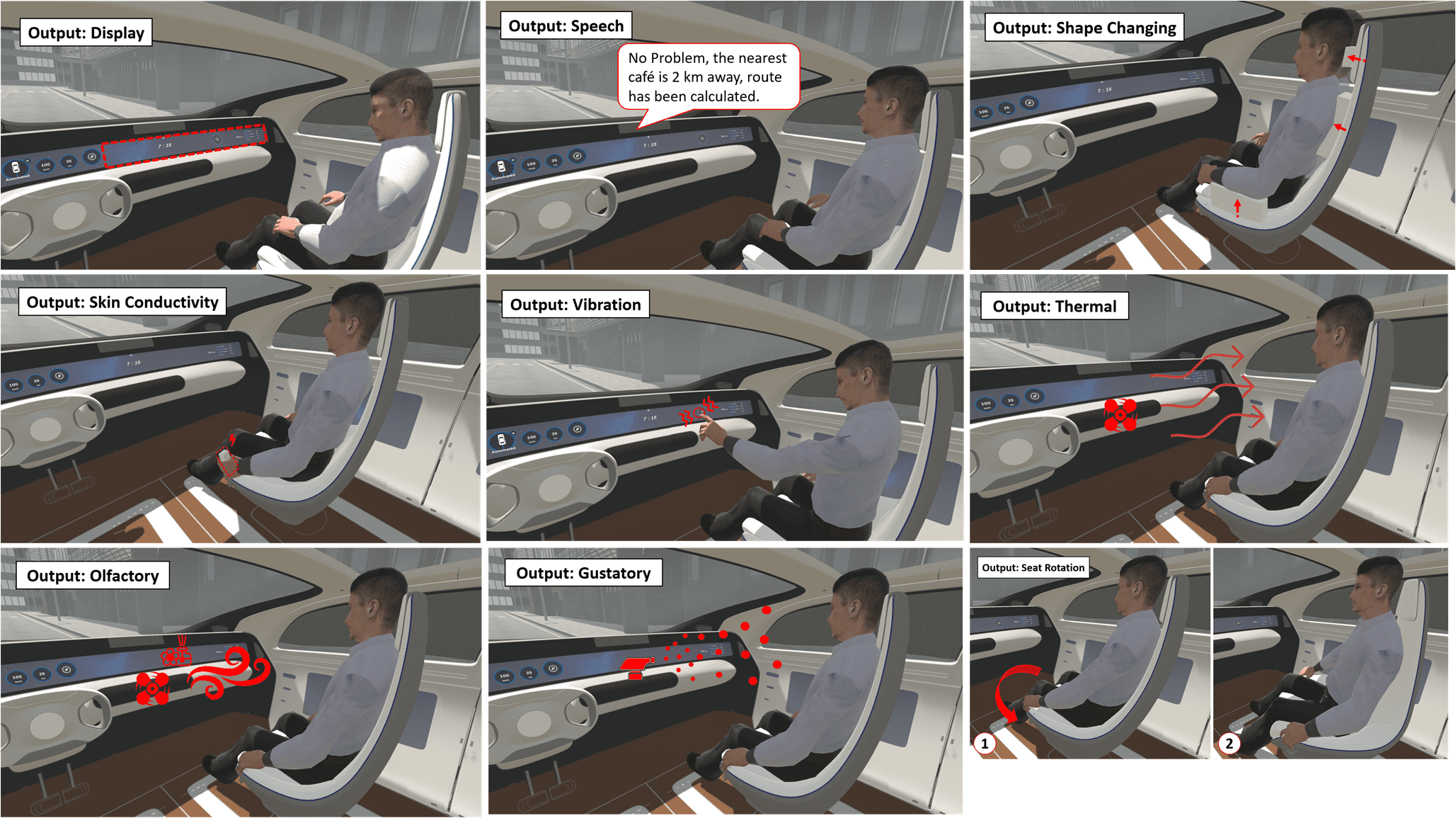}
    \caption{Concept images for in-vehicle output modalities used in our online study.}
\label{fig:study_images_full_output}
\end{figure*}
\begin{figure*}[ht!]
        \centering
        \includegraphics[width=\textwidth]{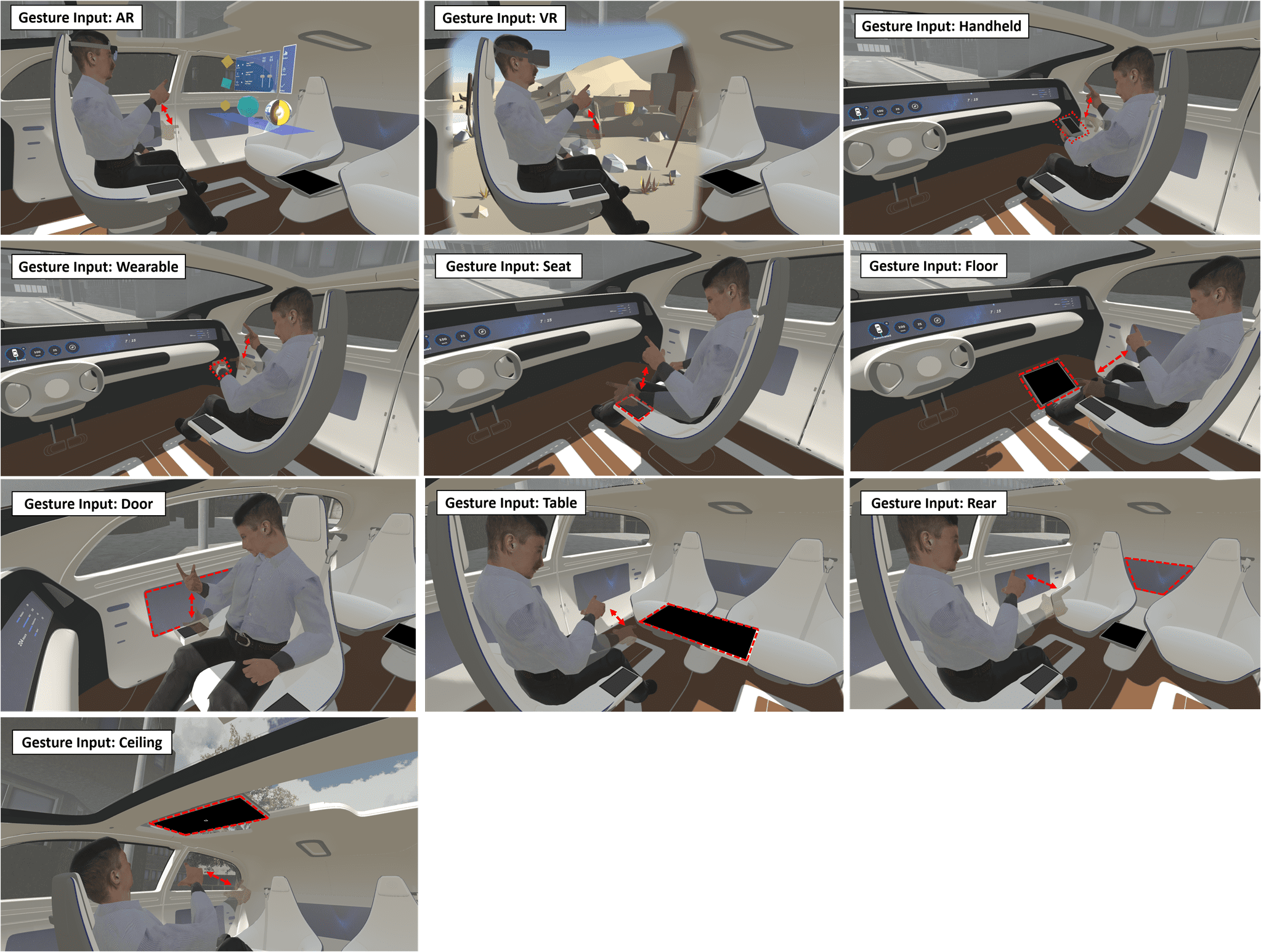}
    \caption{Concept images for input at nomadic (1-4) and anchored (5-10) in-vehicle locations used in our online study.}
\label{fig:study_images_full_location_in}
\end{figure*}
\begin{figure*}[ht!]
        \centering
        \includegraphics[width=\textwidth]{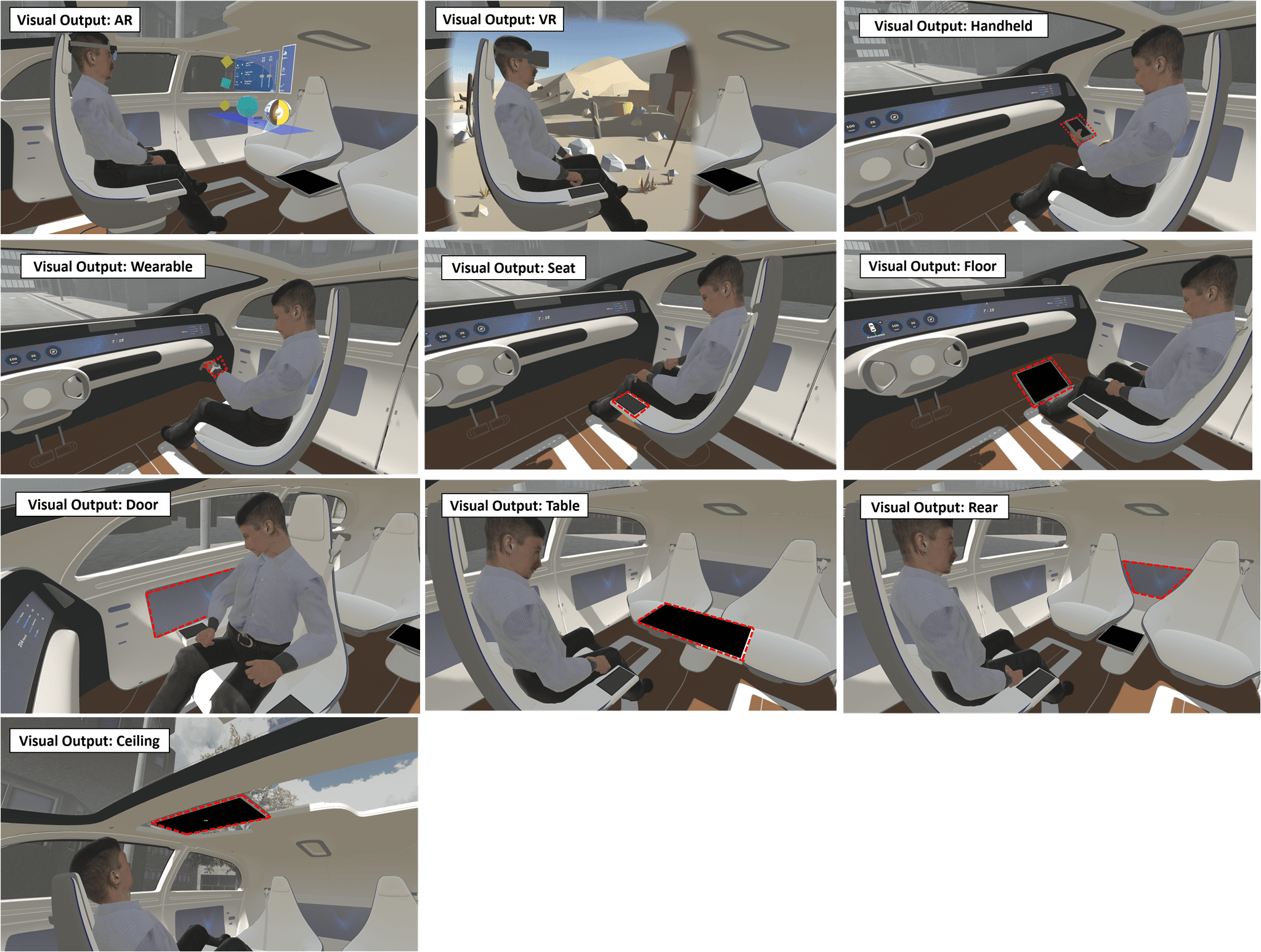}
    \caption{Concept images for output at nomadic (1-4) and anchored (5-10) in-vehicle locations used in our online study.}
\label{fig:study_images_full_location_out}
\end{figure*}




\end{document}